\documentclass[%
aip,
 amsmath,amssymb,
 reprint,%
]{revtex4-2}

\usepackage{graphicx}% Include figure files
\graphicspath{ {./Figures/} {./}  }
\usepackage{dcolumn}% Align table columns on decimal point
\usepackage{bm}% bold math
\usepackage{dirtytalk}
\usepackage{xcolor}
\usepackage{hyperref}
\hypersetup{
    colorlinks=true,
    linkcolor=blue,
    filecolor=magenta,      
    urlcolor=cyan,
}
\usepackage{soul} % for \hl
\usepackage{ulem} % for \sout
\usepackage{cancel} % for \xcancel in equations
\usepackage[utf8]{inputenc}
\usepackage[T1]{fontenc}
\usepackage{mathptmx}
\usepackage{etoolbox}
\usepackage[]{natbib}

\newcommand{\beq}{\begin{equation}}
\newcommand{\eeq}{\end{equation}}
\newcommand{\bea}{\begin{eqnarray}}
\newcommand{\eea}{\end{eqnarray}}
\newcommand{\p}{\partial}
\newcommand{\mb}{\mathbf}
\newcommand{\D}{\Delta}

\begin{document}

%\preprint{AIP/123-QED}

\title{
Transport of Particles in Strongly Turbulent 3D Magnetized Plasmas}% Force line breaks with \\
%\thanks{Footnote to title of article.}

\author{Heinz Isliker  and Loukas Vlahos\\Department of Physics,
              Aristotle University, 54124 Thessaloniki, Greece}
              \email{isliker@astro.auth.gr; vlahos@astro.auth.gr}
              %\homepage{http://www.astro.auth.gr/~vlahos.}
%\author{}%
 %\email{Second.Author@institution.edu.}
%\affiliation{ Department of Physics,
              %Aristotle University, 54124 Thessaloniki, Greece}%\\This line break forced with \textbackslash\textbackslash
%

\date{\today}% It is always \today, today,

\begin{abstract}
In this review, we examine particle transport in strongly turbulent three-dimensional (3D) magnetized plasmas, characterized by intense (large-amplitude) magnetic field ($B$) fluctuations $\delta B$ ($\delta B/B>1$). Such environments naturally give rise to a network of coherent structures (CoSs), including current sheets, filaments, shocks, switchbacks, and significant magnetic perturbations, which critically influence particle dynamics at the kinetic level. Within this turbulent regime, two fundamental particle energization mechanisms emerge, stochastic acceleration and systematic acceleration. Systematic acceleration within open turbulent volumes promotes the development of power-law tails in energy distributions, a hallmark of many astrophysical plasmas. Our analysis distinguishes the roles of two electric fields: the perpendicular (or convective) fields $(\mathbf{E}_{\perp}\sim - \mathbf{V}\times \mathbf{B})$, which drive stochastic heating via interactions with randomly moving scatterers, and the parallel electric fields $(\mathbf{E}_{||} \sim \mathbf{J}\cdot (\mathbf{B}/|\mathbf{B}|)$, which enable systematic particle acceleration in regions of strong currents. Combined with accurate estimates of particle escape times in finite volumes, the interplay of these mechanisms leads to the formation of Kappa distributions. These distributions are frequently observed in strongly turbulent environments across laboratory, space, and astrophysical plasmas. The transport properties differ significantly between the two energization modes. Stochastic energization follows Gaussian statistics and can be effectively described by the Fokker-Planck equation. In contrast, systematic acceleration exhibits Lévy flight statistics, necessitating a fractional transport equation for an accurate description. Furthermore, the fractal  spatial distribution of CoSs introduces deviations from traditional transport models, influencing e.g.\  particle escape times. Systematic acceleration is most efficient during the early, high-energy phases of turbulence, while stochastic heating becomes dominant during the later stages, contributing to gradual particle energization. These mechanisms are ubiquitous across various plasma environments, including the edges of Tokamaks, the solar corona, the turbulent solar wind, the Earth's magnetotail, astrophysical jets, and supernova remnants, where Kappa distributions are routinely observed. This review underscores the gaps in our understanding of particle interactions with diverse CoSs beyond classical reconnection sites, emphasizing their critical role in accurately modelling particle dynamics in turbulent plasmas.
\end{abstract}

%Use showkeys class option if keyword
\maketitle

%\tableofcontents

\section{Introduction \label{Intro}}
In physics, a well-established and highly effective approach to tackling complex problems involves decomposing them into smaller, more manageable components, often assuming that the system under study is close to equilibrium. A prominent model for ``turbulent" heating and particle acceleration is grounded in three foundational assumptions: (1) the existence of a spectrum of low-amplitude waves (i.e.\ weak or wave turbulence), (2) the use of the quasilinear approximation to estimate transport coefficients, and (3) the application of the Fokker-Planck equation to describe particle transport. This wave-particle interaction mechanism, central to second-order Fermi acceleration \cite{Fermi49}, is widely recognized as “stochastic turbulent acceleration” \cite{Schhlickeiser02Book, Petrosian12, Zank14book, Blasi23}.

Another widely studied energization mechanism is magnetic reconnection, predicated on forming isolated, two-dimensional current sheets driven by turbulent flows. This process has been extensively explored for its role in generating magnetic islands that contribute to particle acceleration (e.g., \citet{Drake06}). A third mechanism, shock acceleration, occurs when turbulence-driven flows encounter a shock, facilitating energy transfer from the upstream flow to downstream magnetic fluctuations. Known as first-order Fermi acceleration \cite{Fermi54, Longair11}, this process is regarded as the most efficient systematic acceleration mechanism in space and astrophysical plasmas.

These three mechanisms are traditionally investigated as distinct processes, raising critical questions: (1) Which is the most efficient mechanism? (2) How closely do the resulting high-energy particle distributions align with observational data? (3) Which mechanism best reflects the dynamics of large-scale environments? Textbooks and reviews on particle acceleration in astrophysical, space, and laboratory plasmas often address these mechanisms independently \cite{Melrose2009, Marcowith20}. While such analyses -- frequently based on 2D periodic numerical simulations -- offer valuable insights into small-scale dynamics, they face significant challenges when extrapolating to the complex, large-scale, 3D environments typical of astrophysical or laboratory plasmas \cite{Marcowith20}. A vital limitation of these simulations is their difficulty in accurately estimating particle escape times in extensive, 3D, open turbulent systems. This limitation underscores the unresolved and crucial question of how transport in energy and spatial transport are coupled in the context of particle energization.

In highly unstable or explosive laboratory, space, or astrophysical plasmas, large-amplitude magnetized turbulence (where $\delta B/B \approx 1$, with $\delta B$ representing magnetic fluctuations and $B$ the ambient magnetic field) drives particle heating and acceleration, resulting in high-energy power-law distributions. Such phenomena are observed in diverse systems, including solar flares \cite{Lin03}, the solar wind, the magnetosheath \cite{Xu23}, Earth's magnetotail \cite{Ergun20}, astrophysical jets, and accretion disks.

The interplay between magnetic reconnection and large-amplitude magnetized turbulence has been explored since the 1980s \cite{Matthaeus86}, with subsequent studies elaborating on this relationship \cite{Biskamp89}. Recent reviews continue to examine how intense turbulence can host reconnecting current sheets and, conversely, how reconnection can generate turbulence \cite{Matthaeus11, Cargill12, Lazarian12, Karimabadi13a, Karimabadi2013c, Lazarian20, MatthaeusREv2021}. Similarly, the interaction between shocks and turbulent reconnection has emerged as an area of interest \cite{Karimabadi2014, Trotta21, Trotta22}. In the solar atmosphere, turbulence driven by convective-zone dynamics often results in the spontaneous formation of strong 3D turbulence, as examined in several studies \cite{Parker83, Parker88, Galsgaard96, Galsgaard97a, Gasgaard97b, Rappazzo13, Einaudi21}.

\begin{figure}[!ht]
\centering \includegraphics[width=0.8\columnwidth]{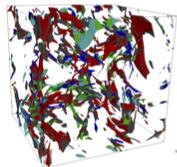}
\caption{A large-amplitude magnetic disturbance generates a variety of CoSs within a 3D strongly turbulent plasma. Reproduced with permission from Thibaud et al., Astr.\ and Astroph.\ {\bf 664}, A193 (2022), Copyright 2022 ESO. }\label{TurReconnection} 
\end{figure}

The formation and evolution of a network of Coherent Structures (CoSs) --- such as magnetic disturbances, current sheets, shock waves, vortices, and magnetic discontinuities --- across many scales in a large-scale, strongly turbulent magnetized environment far from equilibrium present a complex, multi-scale challenge. Before addressing particle transport in such turbulent environments, it is crucial first to understand how CoSs form and evolve (see Fig.\ \ref{TurReconnection}). 

A different approach from the traditional frameworks is followed in this review,  by examining large-amplitude MHD waves, current sheets, and shocks \textit{together} as mechanisms driving CoS development, as recently proposed by \citet{Vlahos23}. Within strong 3D turbulence, CoSs dissipate energy at the kinetic level, leading to two main particle energization processes: (1) stochastic and (2) systematic acceleration, rooted in the foundational concepts introduced by \citet{Fermi49, Fermi54}. This perspective treats strong turbulence as inherently integrating all three acceleration mechanisms, emphasizing their interaction as fundamental to particle heating and acceleration.

This review is structured as follows: Section II explores the limitations of the Fokker-Planck equation and introduces a novel framework for particle transport through CoSs distributed over fractal sets in strongly turbulent 3D environments (fractional transport). Section III examines stochastic and systematic particle acceleration processes emerging from CoSs, focusing on the synergy between these mechanisms. Section IV presents particle heating and acceleration findings through MHD and kinetic simulations. Finally, Section V summarizes the key insights of this review and outlines a comprehensive perspective on particle transport in three dimensions in strongly turbulent, magnetized plasmas.

\section{Normal and Anomalous Transport}
\subsection{Normal transport and the Fokker-Planck Equation}\label{normalDiff1}

	The classical random walk serves as a paradigm for normal transport phenomena, where particles move randomly in discrete steps. The Fokker-Planck equation is a partial differential equation that can be derived from the random walk model as an approximate description of the evolution of the probability density function of the particle position. This derivation is valid in the limit of a large number of steps with small individual displacements, providing a continuous approximation of the discrete random walk process. In the following, we will demonstrate this derivation and explore its implications.

\subsubsection{Classical random walk and normal transport
\label{formalrw}}

\begin{figure}[ht]
  % Requires \usepackage{graphicx}\
  \centering
  \includegraphics[width=8cm]{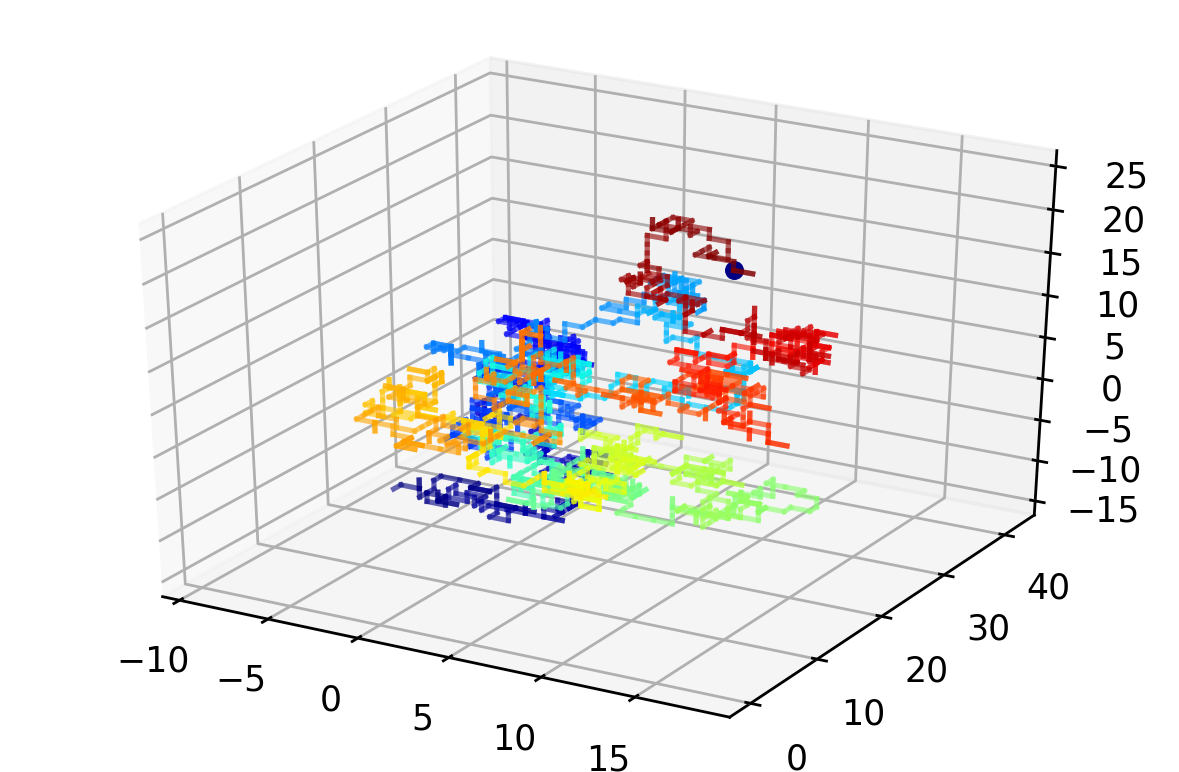}\\
  \caption{Random walk in 3D space of several particles, marked with different colors,
as a function of time.}\label{rw}
\end{figure}

We can define the classical random walk problem as follows:
We consider the position $\mathbf r$ of a particle in 1, 2, or 3D space,
and we assume that this position changes in a series of repeated random steps $\Delta \mathbf r$ (see Fig. \ref{rw}).
%\begin{figure}[ht]
%\centering
%\includegraphics[width=4cm]{rw21.eps}
%\includegraphics[width=4cm]{rw3.eps}\\
% \caption{Random walks in two and three dimensions}\label{rw23}
%\end{figure}

The
time $\Delta t$ elapsing between two subsequent steps is assumed to be constant,
time plays a dummy role, it is a simple counter.
The position $\mathbf r_n$ of a particle after $n$ steps, corresponding to
time $t_n=n\Delta t$, is
\beq
\mathbf r_n = \Delta \mathbf r_n + \Delta \mathbf r_{n-1} + \Delta \mathbf r_{n-2} + \, ...
\, + \Delta \mathbf r_1 + \mathbf r_0 ,
\label{rnclass}
\eeq
where $\mathbf r_0$ is the initial position, and $\Delta \mathbf r_{i}$ is the $i$th
step (or increment or displacement). The position $\mathbf r_n$ as well as the
increments $\Delta \mathbf r_{i}$ are all random variables. To specify the problem
completely, we have to prescribe the probability distribution
$q_{\Delta\mathbf r}(\Delta\mathbf r)$
for the increments $\Delta \mathbf r_{i}$, which
yields the probability for the particle to make a certain
step $\Delta \mathbf r_{i}$ (with given length and direction).
Writing $q_{\Delta\mathbf r}(\Delta\mathbf r)$ in this way, we have
assumed that all the increments have the same probability
distribution and
that the increments are independent of each other (the value that $\Delta \mathbf r_{i}$
takes in any realization is completely independent of the value
taken in the previous step by $\Delta \mathbf r_{i-1}$).
Generalizations to time-dependent increment distributions or correlated
increments are, of course, possible.

Since $\mathbf r_n$ is a random variable, the solution we are looking for in the random
walk problem is in the form of the probability distribution $P(\mathbf r,t_n)$, which yields
the probability for a particle to be at position $\mathbf r$ at time $t=t_n\equiv n\Delta t$.

The mean square displacement is an important quantity that allows to characterize the nature of transport, namely to distinguish between normal and anomalous (including ballistic) transport. In order to derive it for the random walk, we can just
square Eq.\ (\ref{rnclass}), and,
rearranging the terms, we find for $\mathbf r_0=0$
\beq
\langle \mathbf r_n^{\,2} \rangle
= \sum_{j=1,\,(k=j)}^n \langle \D \mathbf r_j^{\,2} \rangle
  + \sum_{j,k=1,\, k\ne j}^n \langle \D \mathbf r_j \D \mathbf r_k \rangle .
\label{vecrsquare}
\eeq
The first term on the right-hand side is just a sum of the variances
$\sigma^2_{\Delta\mathbf r,\,j}$ of $q_{\Delta\mathbf r}(\Delta\mathbf r)$, since
by definition 
\beq \label{variance}
\sigma^2_{\Delta\mathbf r} :=
\langle \D\mathbf r_j^{\,2} \rangle =
\int \Delta\mathbf r^{\, 2} q_{\Delta\mathbf r}(\Delta\mathbf r) \,d\Delta\mathbf r
\eeq
if the mean value of
$q_{\Delta\mathbf r}(\Delta\mathbf r)$ is zero. The second term on the
right hand side is the covariance $\textrm{cov}(\Delta\mathbf r_j,\Delta\mathbf r_k)$
of the random walk steps, and it is zero if the steps a particle takes
are independent of each other. We thus can write Eq.\ (\ref{vecrsquare})
as
\beq
\langle \mathbf r_n^{\,2} \rangle
= \sum_{j=1,\,(k=j)}^n  \sigma^2_{\Delta\mathbf r,\,j}
  + \sum_{j,k=1,\, k\ne j}^n \textrm{cov}(\Delta\mathbf r_j,\Delta\mathbf r_k).
\label{vecrsquare2}
\eeq
In the case with
zero covariance and finite variance $\sigma^2_{\Delta\mathbf r,\,j} =: \lambda_{sc}^2$,
the mean square displacement is 
\beq
\langle \mathbf{r}^{\,2} \rangle = n \lambda_{sc}^2.
\label{normaltr}
\eeq
The standard deviation $\lambda_{sc}$ of the walk-steps can be interpreted here as a mean free path, since, as just stated, the mean value of $q_{\Delta\mathbf r}(\Delta\mathbf r)$ is assumed to be zero. It can be estimated with a simple model in an ideal gas in equilibrium. Assuming that a particle is moving inside a
gas with a mean speed $\langle v \rangle,$ the distance traveled between two successive collisions is
$\lambda_{sc} = \langle v \rangle \tau$, where $\tau$ is the mean collision time, which plays the role of the so-far unspecified time-step $\Delta t$.
We may thus conclude that the number of steps a particle executes
inside the plasma  during a time $t$ is
$n=t/\tau$,
%where $t$ is the time the particle travels inside the gas
and 
the mean squared
distance it travels is
\begin{equation}\label{dif}
    \langle\mathbf{r}^{\,2}\rangle = n \lambda_{sc}^2 = (t/\tau) (\langle v \rangle \tau) \lambda_{sc} 
    = [\langle v\rangle \lambda_{sc}] t \equiv [\lambda_{sc}^2/\tau] t,
\end{equation}
 or
\begin{equation}\label{dif1}
	\langle \mathbf{r}^{\,2} \rangle = 
	%[\langle v \rangle \lambda_{sc}] t = 
	2 D t  ,
\end{equation}
where  $D:= \langle v \rangle \lambda_{sc}/2 \equiv \lambda_{sc}^2/2\tau$ is called the diffusion coefficient,
which is a useful parameter to characterize particle transport
in the normal case.
{\it Important here is to note the linear scaling relation between the mean square displacement} $\langle \mb r^2 \rangle$
and time $t$.

\subsubsection{Einstein's Formalism for the Classical Random Walk and the
Diffusion Equation\label{einsform}}

 The approach to treating normal transport that we present here was introduced initially 
by Bachelier\cite{Bachelier1900} and later by Einstein \cite{Einstein1905}.
 The starting point is the classical random walk, and we consider the 1-dimensional case for simplicity.
We assume that the solution of the random walk
problem is in the form of the probability distribution $P(z,t)$ for
a particle at time $t$ to be at position $z$. Let us assume that we would
know the distribution $P(z,t-\Delta t)$ one time-step $\Delta t$ earlier
(we remind that $\Delta t$ is assumed constant). If particles are conserved,
the relation
\beq
P(z,t) = P(z-\Delta z,t-\Delta t) \, q_{\Delta z}(\Delta z)
\label{eins1}
\eeq
must hold, with $q_{\Delta z}$ the distribution of random walk steps.
Eq.\ (\ref{eins1}) states that the probability to be at time $t$ at position
$z$ equals the probability to have been at position $z-\Delta z$ at time
$t-\Delta t$ and to have made a step of length $\Delta z$ during time $\Delta t$. We still have
to sum over all possible $\Delta z$, which leads to the Einstein (or Bachelier)
diffusion equation,
\beq
 P(z,t) = \int_\infty^\infty P(z-\Delta z,t-\Delta t) \,q_{\Delta z}(\Delta z)\,d\Delta z .
\label{eins2}
\eeq
This is an integral equation that determines the solution $P(z,t)$ of the
random walk problem as defined in Sec.\ \ref{formalrw}. The power of this
equation will become clear below when we will show ways to treat cases of
anomalous transport. Here, we still focus on normal transport. As will become
clear later, it is a characteristic of normal transport that the
particles take only small steps $\Delta z$ compared to the system size.
This implies that
$q_{\Delta z}(\Delta z)$ is non-zero only for small $\Delta z$,
the integral in Eq.\ (\ref{eins2}) is only over a small $\D z$-range, and we
can expand $P(z-\Delta z,t-\Delta t)$ in $z$ and $t$ (also
$\Delta t$ is small),
\bea
P(z-\Delta z,t-\Delta t) &=& P(z,t) - \Delta t\,\partial_t P(z,t) - \Delta z\,\partial_z P(z,t)\nonumber\\
&+& \frac{1}{2} \Delta z^2 \, \partial_z^2 P(z,t)
\label{expP}
\eea
Inserting Eq.\ (\ref{expP}) into Eq.\ (\ref{eins2}), and since (i)
 $P(z,t)$, its derivatives, and $\Delta t$ are not affected by the integration, (ii) $q_{\Delta z}$ is normalized 
($\int q_{\Delta z}(\Delta z)\,d\Delta z = 1$), (iii) $q_{\Delta z}$ is assumed
to have mean
value zero ($\int \Delta z \,q_{\Delta z}(\Delta z)\,d\Delta z = 0$), and (iv) using
the definition of the variance $\sigma^2_{\Delta z}$
($\int \Delta z^2 \,q_{\Delta z}(\Delta z)\,d\Delta z = \sigma^2_{\Delta z}$), we find
\beq
P(z,t) = P(z,t) -\Delta t \,\partial_t P(z,t)
+ \frac{1}{2} \sigma^2_{\Delta z} \,\partial^2_z P(z,t) ,
\eeq
or
\beq
\frac{\partial P(z,t)}{\partial t} = \frac{\sigma^2_{\Delta z}}{2\Delta t} \frac{\partial^2 P(z,t)}{\partial z^2} ,
\label{einsdiff}
\eeq
i.e.\ we recover the simple diffusion equation,
with  diffusion coefficient $D_z=\frac{\sigma^2_{\Delta z}}{2\Delta t}$.
In infinite space, and if all particles start initially from $z=0$,
the solution of Eq. (\ref{einsdiff}) is
\begin{equation}\label{SolDiff}
    P(z,t)=\frac{N_0}{\sqrt{4 \pi D_z t}}e^{-z^2/4D_zt}  ,
\end{equation}
where $N_0$ is the total number of particles inside the volume under
consideration. The solution is identical to a Gaussian distribution
with mean zero and variance $2D_zt$. The mean square displacement can be calculated as the variance of $P$, 
%Using
%Eq. (\ref{solDiff})
\begin{equation}\label{normalDiff}
    \langle z^2(t)\rangle = \int z^2 P(z,t) \, dz = 2D_zt ,
\end{equation}
which shows a scaling with time that is identical to the one obtained earlier for the 3D case in Sect.\ \ref{formalrw}, Eq.\ (\ref{dif1}). 
With $\mb r = (x,y,z)$,
	and assuming that the background scattering medium (e.g.\ a gas) is in equilibrium and isotropic, we expect that $\langle x^2 \rangle = \langle y^2 \rangle = \langle z^2 \rangle = \langle \mathbf{r}^{\,2} \rangle /3$, so that the 1D and 3D diffusion coefficients are related as
	$D_z = D/3$.

{\it Transport obeying Eq.\ (\ref{normalDiff}) is called \textbf{normal}
and is characteristic of systems in equilibrium or very close to equilibrium.}

\subsubsection{Fokker-Planck Equation\label{fokkpla}}

The Fokker-Planck (FP) equation (or Kolmogorov forward equation)
is a more general transport  equation than the simple
equation introduced in
Sec.\ \ref{einsform}.
We again start with a description of transport in terms of a random
walk, as in Sec.\ \ref{formalrw},
we though relax two assumptions made there: (i) We assume now
that the mean value $\mu_{\Delta z}$ of the random walk steps
%$\int \Delta z \,q_{\Delta z}(\Delta z)\,d\Delta z = \mu_{\D z}$
can be different from zero,
which corresponds to a systematic motion of the particles
in the direction of the sign of $\mu_{\Delta  z}$, and (ii) we assume that
both the mean and the variance can be spatially dependent,
$\mu_{\Delta z}=\mu_{\Delta  z}(z)$ and
$\sigma_{\Delta  z}^2=\sigma_{\Delta  z}^2(z)$, which means
that the distribution of increments depends on the spatial location,
i.e.\ it is of the form $q_{\Delta z,z}(\Delta z,z)$. To be compatible
with these
assumptions, Eq.\ (\ref{eins2}) must be rewritten in a slightly
more general form,
\beq 
P(z,t) = \int_\infty^\infty P(z-\Delta z,t-\Delta t)q_{\Delta z,z}(\Delta z,z-\Delta z)\,d\Delta z  ,
\label{chapkol}
\eeq
which is the {\it Chapman-Kolmogorov equation}, and where
now $q_{\Delta z, z}(\Delta z,z)$ is the probability
density for being at position $z$ and making a step $\Delta z$
in time $\Delta t$.
The FP equation
can be derived in a way similar to the one presented
%used to derive Eq.\ (\ref{einsdiff})
in Sec.\ \ref{einsform}:
%Repeating then the derivation of Sec.\ \ref{einsform},
We expand the integrand of Eq.\ (\ref{chapkol})
in a Taylor-series in terms of $z$, so that
$P(z,t) = \int_\infty^\infty A B \,d\Delta  z$,
with
\bea
 A&=&P(z,t)-\p_t P(z,t)\Delta t-\p_z P(z,t)\Delta  z\\&+& \frac{1}{2}\p_z^2 P(z,t)\Delta z^2 + ... ,
\label{Aeq}
\eea
where we have also expanded to first order in $t$ and
which is, of course, the same as Eq. (\ref{expP}),
and newly, we have 
\bea
B&=&q_{\Delta z,z}(\Delta  z,z)-\p_z q_{\Delta z,z}(\Delta z,z)\Delta z \\&+&\frac{1}{2}\p_z^2
q_{\Delta z,z}(\Delta z,z)\Delta  z^2 + ...
\label{Beq}
\eea
(note that the Taylor expansion is concerning the second
argument of $q_{\Delta z,z}$, we expand only with respect to $z$,
not though with respect to $\Delta  z$).
In multiplying and evaluating the integrals, we use the normalization
of $q_{\Delta  z,z}$ ($\int q_{\Delta  z,z}(\Delta z,z) d \Delta z)=1$),
the definition of
the mean value 
($\mu_{\Delta z}(z) :=  \int \Delta z q_{\Delta z,z}(\Delta  z,z)\,d\Delta  z$) and of the
second moment
($\langle \Delta  z^2\rangle (z) := \int \Delta z^2 q_{\Delta  z,z}(\Delta  z,z)\,d\Delta  z$),
and expressions
like $\int \Delta  z \, \p_z q_{\Delta  z,z}(\Delta  z,z)\,d\Delta  z $ are considered to equal
$\p_z \int \Delta  z  q_{\Delta  z,z}(\Delta  z,z)\,d\Delta  z\equiv \p_z \mu_{\Delta  z}(z)$, so that,
on keeping all terms up to the second order in $\D z$, we find
the {\it Fokker-Planck equation},
% We get one term more on the right
%hand side, and we find
\beq
\frac{\partial P(z,t)}{\partial t}  = -\frac{\partial [F(z)P(z,t)]}{\partial z}+\frac{\partial^2 [D(z)P(z,t)]}{\partial z^2} ,
\label{FokkerP}
\eeq
with %$n$ the particle density,
$F(z)\equiv\mu_{\Delta z}(z)/\Delta t$ a drift velocity,
and $D(z)\equiv \langle \Delta z^2\rangle (z)/2\Delta t$
the diffusion coefficient (for a 3D formulation, see Gardiner
\citep{Gardiner2004}).
The basic difference between the FP equation and the
simple transport equation in Eq.\ (\ref{einsdiff}) is the appearance
of a drift term, and that both the drift velocity and the diffusion
coefficient are allowed to be spatially dependent.
These differences allow the FP equation to model
more complex diffusive behavior.
The FP equation can also be applied e.g.\ to velocity space,
or to position and velocity space together. 
%\heinz{
In scenarios where the surrounding magnetic field significantly exceeds the magnetic fluctuations ($\delta B \ll B_0$, weak turbulence), the transport and acceleration processes are affected by changes in the pitch angle, which can also be included in the FP equation. Further insights into the Fokker-Planck equation, particularly the significance of the pitch angle, are detailed in \citet{Schhlickeiser02Book}.
%}

The FP equation has the advantage of being a deterministic differential equations
that allows describing the evolution of stochastic systems, as long as
the diffusivities ($D(z)$) and drift velocities ($V(z)$)  are known, and as long as the
conditions for its applicability are met; see the remarks below.
From its derivation, it is clear that the FP
equation is suited only for systems close to equilibrium,
with just small deviations of  particles from equilibrium, or, in the
random walk sense, with just small steps of the particles performing the
random walk, exactly as it holds for the simple transport equation
%for in the derivation of Eq.\ (\ref{einsdiff})
in Sect.\ \ref{einsform}.
%It is beyond the scope of the present tutorial to formally introduce
%the Fokker Planck equation,

A further natural generalization for a transport
equation in the approach followed here would be
not to stop the Taylor expansions in Eqs.\ (\ref{Aeq}) and (\ref{Beq})
at second order in $z$,
but to keep all terms, which would lead to the so-called
{\it Kramers-Moyal expansion}. More details about the Fokker-Planck equation can be found in the
literature \citep{Risken96,  Gardiner2004}.

The above analysis assumes that the particle scattering is passive and no energy is gained or lost. The transport of particles in energy space is more complex since transport in space and energy are coupled \cite{Bouchet04}. By neglecting though spatial transport and focusing on stochastic random walk in energy space, the evolution of the distribution $P(W,t)$ of particles in energy $W$   obeys a Fokker-Planck equation in  the form
\beq
\frac{\partial P(W,t)}{\partial t}  = -\frac{\partial [F(W)P(W,t)]} {\partial W} + \frac{\partial^2 [D(W)P(W,t)]}{\partial W^2} ,
\label{FokkerPVel}
\eeq
where the transport coefficients are defined as follows: $F(W,t)=<\Delta W(t)>/\Delta t$ is the convection coefficient which describes the systematic energy transfer to the particles, and $D(W,t)=<\Delta W^2>/ 2 \Delta t$ is the diffusion term describing the stochastic interaction of particles with scatterers. If a system under study is finite and the average time particles remain inside its volume is $t_{esc}(W)$, then the FP equation includes an additional loss term,
\bea
\frac{\partial P(W,t)}{\partial t}  &=&-\frac{\partial [F(W)P(W,t)]} {\partial W}+\frac{\partial^2 [D(W)P(W,t)]}{\partial W^2} \nonumber\\&-&\frac{P(W,t)}{t_{esc}(W)} ,
\label{FokkerPVelLoss}
\eea
see e.g.\ \citet{Schhlickeiser02Book}.

\subsubsection{Why Normal Diffusion Should Be the Usual Case\label{CLT}}

The appearance of normal diffusion in many natural phenomena close
to equilibrium
and the particular Gaussian form of
the solution of the diffusion equation
can  be understood from
probability theory. The Central Limit Theorem (CLT) states that if a statistical
quantity (random variable) is the sum of many other statistical quantities,
such as the position of a random walker after $n$ steps according to
Eq.\ (\ref{rnclass}), and if (i) all the steps $\D z_i$ have finite mean $\mu_{\D z}$
and variance $\sigma_{\Delta  z}^2$, (ii) all the steps $\Delta z_i$ are mutually independent, and
(iii) the number $n$ of steps $\Delta  z$ is large, then, independent of the
distribution of the steps $\Delta  z_i$, the distribution $P(z_n,t_n)$ of $z_n$ is a Gaussian. In
particular, if $\mu_{\Delta  z}= 0$, $z_0=0$ and all the $\D z_i$ have the same variance, then
\beq
P(z_n,t_n) = \frac{1}{2\pi n \sigma_{\D z}^2} e^{-\frac{z^2}{2n\sigma_{\D z}^2}} ,
\label{CLTG}
\eeq
with variance $\sigma_{z_n}^2 = n\sigma_{\Delta  z}^2$, or, if we set $n=t_n/\D t$ (constant time-step), then
$\sigma_{z_n}^2 = t_n\sigma_{\D z}^2/\Delta  t$.
%Eq.\ (\ref{CLTG}) corresponds to the solution Eq.\ (\ref{nztG}) of the simple
%diffusion equation 
The mean
square displacement equals, per definition, the variance,
\beq
\langle z_n(t_n)^2\rangle = \int z_n^2 P(z_n,t_n) \,dz_n=\sigma_{z_n}^2 = [\sigma_{\Delta  z}^2/\Delta t] t_n
\eeq
(for $z_0=0$), and diffusion is thus always normal in cases where the CLT
applies.

%{\bf
 In sum, the CLT predicts quantities that are the result
of many small-scale interactions to be distributed according
to a Gaussian,  as we found it here on the example  of the classical random walker in Sec.\ \ref{einsform}.
Stated differently, we may say that
the appearance of non-Gaussian distributions
is something unexpected and unusual, according to the CLT.
%, not normal, or, as it is usually termed, anomalous,
%, which we will deal with in the next section.
We mention
that also the equilibrium velocity distributions of gas or fluid particles
obey the CLT, the velocity components,
say $v_x$, $v_y$, $v_z$, follow Gaussian distributions,
and  as as result, the velocity magnitude
$v=(v_x^2+v_y^2+v_z^2)^{1/2}$ exhibits a Maxwellian distribution. Again
then, the appearance of non-Maxwellian velocity distributions is unexpected
on the base of the CLT.

\subsection{Anomalous Transport and Fractional Transport Equation\label{anodi}}

\subsubsection{The Scaling of "Anomalous" Trajectories\label{anomdef}}

Normal transport
%we have met so far were all normal ones (
%where the system is close to equilibrium,
has as
basic characteristic the linear scaling of the mean square displacement
of particles with time, $\langle r^2 \rangle\sim Dt$.
%, just in the
%case of small times, we met a scaling with $t^2$.
As we will show in the following sections, the energization of particles by CoSs is not always weak, and particles may also be trapped for a relatively long time inside CoSs. This  leads to
deviations from normal transport, in that transport is either
faster or slower, which is termed anomalous transport. A useful
characterization of the transport process is again through the scaling
of the mean square displacement with time, where though now we are
looking for a more general scaling of the form
\beq
\langle r^2(t) \rangle \sim t^\gamma  .
\label{asca}
\eeq
Transport is then classified through the scaling index $\gamma$. The
case $\gamma=1$ is normal transport, and all other cases
are termed anomalous transport. The cases $\gamma>1$ form the family of super-diffusive
processes, including the particular case $\gamma=2$, which is called ballistic
transport, and the cases $\gamma<1$ are the sub-diffusive processes. If the
trajectories of a sufficient number of particles inside a %non-equilibrium
system are known, either through experimental observation or through theoretical modeling, then plotting $\log \langle r^2\rangle$ vs $\log t$ is a 
way to determine the type of transport occurring in a given
%nonlinear
system.

\begin{figure}[ht]
\centering
  \includegraphics[width=4cm]{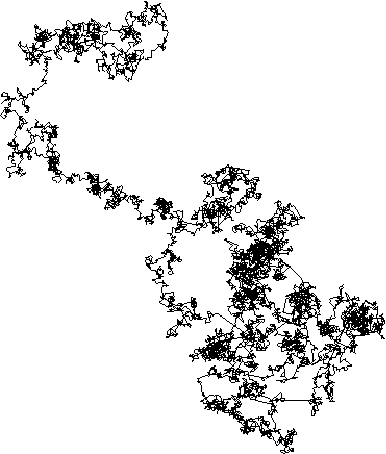}
  \includegraphics[width=4cm]{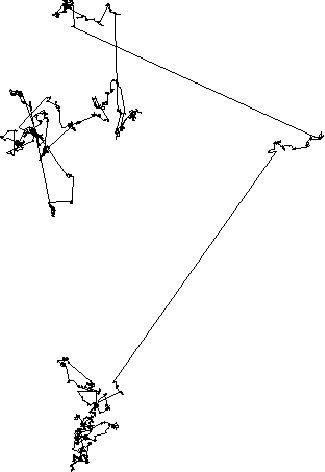}
  \caption{Left: Random walk in a dynamical system close to equilibrium
(normal transport); right: Random walk in a dynamical
system far from equilibrium (anomalous transport).}
\label{levy}
\end{figure}

As an illustration, let us consider particles that are moving with
constant velocity $v$ and undergo no collisions and experience no
friction forces. It then
obviously holds that $r=vt$ for each individual particle, so that $\langle r^2(t)\rangle\sim t^2$.
Free particles are thus super-diffusive in the terminology used here,
which is also the origin of the name ballistic for the case $\gamma=2$.
Accelerated particles would even be transported faster. The difference between
normal and anomalous transport is  illustrated in Fig. \ref{levy},
where in the case of anomalous transport
long ``flights" are followed by efficient ``trapping" of particles in
localized spatial regions,
in contrast to the more homogeneous
picture of normal transport.

It is to be noted that anomalous transport manifests itself not only
in the scaling of the mean square displacement in Eq.\ (\ref{asca}) with $\gamma\ne 1$ (which
experimentally may also be difficult to measure), but also in 'strange'
and 'anomalous' phenomena such as 'uphill' transport, where particles
or heat are transported in the direction of a higher concentration (against a 'driving gradient'), or the appearance
of non-Maxwellian distributed particle velocities (see Sec.\ \ref{CLT}),
very often of power-law
shape, which is very common in high-energy space, astrophysical, and laboratory plasmas,
etc.

\subsubsection{Continuous Time Random Walk}
\label{CTRW_Levy_real}

%\subsubsection{Definition\label{subsub1}}

Given the experimental ubiquity of anomalous transport phenomena, the
question arises of how to model such phenomena. One way of tackling the problem
is through the random walk formalism. So far, we have used the random
walk to model classical diffusion, and in Sec.\ \ref{einsform}, it had
been shown how the random walk is related to a simple transport equation
if the steps the particles take on their walk are small compared with the size of the system under consideration. One way to model
anomalous transport is by relaxing the latter condition and allowing the
particles to also take large steps, where 'large' in a finite system means
large up to system size, and in infinite systems it means that the steps
are potentially unbounded in length. Useful in this context is the family of Levy
distributions \cite{Shlesinger87, Blumen89} as step-size distributions $q_{\Delta z}$. They are
defined in closed form in Fourier space (see Sec.\ \ref{symmLevy} below),
and they have the property that
\beq
q^{L,\alpha}_{\Delta z}(\Delta z) \sim \vert \Delta z\vert^{-1-\alpha},
\ \ \ \textrm{for}\ \,\vert\Delta z\vert\ \,\textrm{large},\ \ \, 0 < \alpha < 2,
\eeq
so that there is always a small, though finite, probability for an arbitrarily large
step size. The Levy distributions all have infinite variance,
$\sigma_{L,\alpha}^2=\int\Delta z^2 q^{L,\alpha}_{\Delta z}(\Delta z) \,d\Delta z=\infty$,
which makes their direct use as a step size distribution in the
classical random walk of Sec.\ \ref{formalrw} and Eq.\ (\ref{rnclass})
impossible:
%Eq.\ (\ref{vecrsquare2})
Consider the case of a random walk in 1-D, with the position of the
random walker after $n$ steps given by the 1-D version of Eq.\ (\ref{rnclass}),
%$z_n=\D z_{n-1} + ... + \D z_1 + z_0$,
and the mean square displacement (for $z_0=0$)
given by Eq.\ (\ref{vecrsquare2}).
%$z_n^2 = \sum_i \D z_{i}^2 + \sum_{i,j,i\ne j} \D z_i \D z_j$.
%Taking the
%mean, we note that $\langle \D z_i \D z_j \rangle = 0$ if we assume
%the steps to be independent of each other, and $\langle \D z_{i}^2 \rangle$
%is just the variance $\sigma_{L,\alpha}^2$, so that for the mean
%square displacement we have $\langle z_n^2\rangle = n \sigma_{L,\alpha}^2$,
%, which is infinite, already after the first step.
Let us assume that
the steps are independent of each other so that the covariances are zero
%, and $\langle \D z_{i}^2 \rangle$
%is just the variance $\sigma_{L,\alpha}^2$,
and the mean
square displacement is
$\langle z_n^2\rangle = n \sigma_{L,\alpha}^2$,
which is infinite, already after the first step.

A way out of the problem is to release time from its dummy role and make
it a variable that evolves dynamically, as does the walker's position.
In this way,
infinite steps in space can be accompanied by an infinite time for
the step to be completed, and the variance of the random walk, i.e., its
mean square displacement, remains finite. The extension of the random
walk to include the timing is called Continuous Time Random Walk (CTRW).
Its formal definition consists again of Eq.\ (\ref{rnclass}) for the evolution 
of the position $\mathbf r_n$ of a particle in $n$ steps, as described
in Sec. \ref{formalrw}, and, moreover, the time at which the $n$th step of
the walk takes
place is now also random (a random variable), and it evolves according
to
\beq
t_n = \Delta t_n + \Delta t_{n-1} + \Delta t_{n-2} + \, ...
                                             \, + \Delta t_1 + t_0,
\label{tctrw}
\eeq
where $t_0$ is the initial time, and the $\Delta t_i$ are random
temporal increments. To complete the definition of the CTRW, we need
also to  define the probability distribution of the $\Delta t_i$, i.e., we
must specify the probability for the $i$th step to last a time $\Delta t_i$.

For the most general case of a CTRW, one would have to specify the joint
	probability distribution $q_{\D z,\D t}(\D z,\D t)$ for the spatial and
	temporal increments.
In practice, two simpler cases are usually considered in order to keep the technical problems
for an analytical treatment at a manageable level. (i) In the waiting model, the steps in position
and time are independent, and one specifies two probabilities,
one for $\D\mathbf r$, the $q_{\D\mathbf r}$ already introduced, and one
for $\D t$, say $q_{\D t}$. Here, then, $\D t$ is interpreted as a waiting time,
the particle waits at its current position until the time $\D t$ has elapsed,
and then it performs a spatial step $\D \mathbf r$ during which no time is
consumed
\cite{Montroll65}.
(ii) In the velocity model, the time $\D t$ is interpreted as the traveling
time of the particle, $\D t=\vert \D \mathbf r\vert /v$, where $v$ is an
assumed constant
velocity, so that
the distribution of increments is
$q_{\D z,\D t} = \delta(\D t - \vert \D \mathbf r\vert/v)q_{\D \mathbf r}(\D \mathbf r)$
\cite{Shlesinger87}. 
Despite of the explicit use of a velocity, the model does obviously not include an actual velocity dynamics. 
	
We note that the waiting model is easier to treat analytically than the velocity model. In
Sec.\ \ref{velo}, we extend the CTRW to a model that also includes the velocity as a dynamic random variable, 
which though is prone to sole numerical treatment.

\subsubsection{The CTRW Equations}

The CTRW equations can be understood as a generalization of the Einstein
%or Bachelier
equation, Eq.\ (\ref{eins2}), or the Chapman-Kolmogorov equation,
Eq.\ (\ref{chapkol}). It is helpful to introduce the concept of the
turning points, which are the points at which a particle arrives and starts
a new random walk step. The evolution equation of the distribution of turning points
$Q(z,t)$ (here in 1-D) follows basically from particle conservation,
\bea
Q(z,t) &=& \int d\D z\int_0^t d\D t Q(z-\D z,t-\D t) q_{\D z,\D t}(\D z,\D t)
\nonumber\\
&& +\, \delta(t) P(z,t=0) + S(z,t),
\label{Qeq}
\eea
where the first term on the right side describes a completed random walk step, including
stepping in space and in time, the second term takes the initial condition
$P(z,t=0)$ into account, and the third term $S$ is a source term
\cite{Zumofen93}.

The expression for $P(z,t)$, the probability for the walker to be at position $z$ at time $t$,
is different for the waiting
and for the velocity model, respectively.
In the case of the waiting model, where
$q_{\D z,\D t}(\D z,\D t)=q_{\D t}(\D t)q_{\D z}(\D z)$, we have
\beq
P_W(z,t) = \int_0^t d\D t\, Q(z,t-\D t) \Phi_W(\D t)   ,
\label{PWeq}
\eeq
with $\Phi_W(\D t):=\int_{\D t}^\infty dt^\prime q_{\D t}(t^\prime)$ the probability
to wait at least a time $\Delta t$ \cite{Zumofen93}.

In the velocity model, where
$q_{\D z,\D t}(\D z,\D t) =
\delta(\D t - \vert \D z\vert/v) \,q_{\D z}(\D z)$,
$P(z,t)$ takes the form
\beq
P_V(z,t) = \int_{-vt}^{vt} d\D z \int_0^t d\D t\, Q(z-\D z,t-\D t) \, 
\Phi_V(\D z,\D t)
\label{PVeq}
\eeq
with
\bea
& &\Phi_V(\D z,\D t) = \frac{1}{2} \delta(\vert \D z\vert -v \D t) \nonumber \\&&
\int_{\vert\D z\vert}^\infty dz^\prime
\int_{\D t}^\infty dt^\prime
\delta(t^\prime-\vert z^\prime\vert/v) q_{\D z}(z^\prime)
\eea
the probability of making a step of length at least $\vert \D z\vert$ and of
duration at least $\D t$ \cite{Shlesinger87, Zumofen93}.

Both, the expression for $P_W$ and $P_V$ determine the probability for
seeing the particle when moving in between two turning points, taking into
account only the part of the random walk in which the particle consumes time.

The kind of transport that the CTRW formalism yields depends on the distribution
of step increments. If the increments are small, then the treatment of Sec.\ \ref{einsform}
can be applied again, transport is normal, and again, a simple transport equation
can be derived. If the increments are not small, then super- as well as sub-diffusion
can result, depending on the concrete choice of increment distributions. For instance,
small spatial steps in combination with Levy-distributed, long waiting times will
yield sub-diffusion in the waiting model. An essential property of the CTRW equations
is that they are non-local in space and time (also termed non-Markovian), being in the form of integral equations over the spatial and temporal domain.
Anomalous transport phenomena in the CTRW approach are thus considered non-local
processes, and with that, they are far from equilibrium processes.

\subsubsection{Treating the CTRW Equations}

%\subsubsubsection{Remarks on the solution of the CTRW equations}

A standard way to treat the CTRW equations is by transforming them
to Fourier (F)
%($z\to k$)
and Laplace (L) space,
%($t\to s$),
whereby
the convolution theorems of the two transforms are used. We will
illustrate this procedure on the example of the waiting model
below in Sec.\ \ref{solvwait}.

The CTRW equations are though not always of a convolution type, e.g., \
the velocity model does not have a convolution structure anymore due to the
appearance of time in the $\D z$-integration limits (see Eq.\ (\ref{PVeq}), and also in the expression
for the turning points $Q$, Eq.\ (\ref{Qeq}), time appears in the $\D z$-integration limits 
in this case), %of the velocity model),
so that Fourier Laplace
methods are not directly applicable anymore, and other methods
are needed.

FL transforms, if applicable, usually do not allow the calculation of the
probability $P(z,t)$ in closed
analytical form, but rather some asymptotic properties of it,
such as the mean square displacement at large times
\cite{Klafter87,Blumen89}.
On the other hand, FL transforms, if applicable, allow transforming
the CTRW equations into other kinds of equations,
e.g., \  into one instead of the two
integral equations \cite{Klafter87,Blumen89},
into an integrodifferential equation
\cite{Klafter87}, or even into
a fractional transport equation,
which has the form of a generalized diffusion
equations with fractional
derivatives that are generalized, non-local differentiation operators.
In Sec.\ \ref{fracdiffeq}, we will show how a fractional diffusion
equation arises naturally in the context of the waiting model
\cite{Metzler00,Metzler04}.

Another standard way of treating the CTRW equations is with Monte Carlo simulations
\cite{Vlahos04}, or else,
the equations can be solved numerically with an appropriate method
\cite{Isliker03}.

\subsubsection{Fourier and Laplace transforms of probability densities}

For any probability density function (pdf) such as $q_{\D z}$, we can define the Fourier transform
as ($z\to k$)
\beq
\hat{q}_{\D z}(k) = \int e^{-ik\D z}q_{\D z}(\D z) \, d\D z,
\eeq
which is often called the characteristic function of $q_{\D z}$.
Considering then the expression
\bea
i^n \p_k^n \hat{q}_{\D z}(k) \Big\vert_{k=0}
&=& i^n \int (-i\D z)^n e^{-ik\D z}q_{\D z}(\D z) \, d\D z \Big\vert_{k=0} \nonumber \\
&=& \int \D z^n q_{\D z}(\D z) \, d\D z  ,
\eea
for $n=0,1,2,3...$,
we see that, because $q_{\D z}$ is a pdf, the
last expression is the expectation value $\langle \D z^n \rangle$ of $\D z^n$,
the so-called $n$th moment, and
we have
\beq
i^n \p_k^n \hat{q}_{\D z} (0)
=\langle \D z^n \rangle   .
\label{dermo}
\eeq
In particular, it  holds that $\langle \D z^0 \rangle= 1$, since
$q_{\D z}$ is a pdf that is normalized to one
($\langle \D z^0 \rangle=\int q_{\D z}(\D z) \, d\D z=1$). Furthermore,
$\langle \D z^1 \rangle=\int \D z q_{\D z}(\D z) \, d\D z$
is the mean value of $q_{\D z}$.

In the use of Fourier and Laplace transforms for solving the CTRW
equations, we will concentrate on the asymptotic, large $|z|$ % , large $t$
regime,
%That this is valid only for the asymptotic regime where $z^2$ is large
%follows from the fact that we let
which corresponds to small values of $k$ (small
wave numbers correspond to large length scales or wave lengths).
We thus can make a
%The relations in Eq.\ (\ref{dermo})
%are useful in two respects:
%First, consider the
Taylor expansion of $\hat{q}_{\D z}(k)$ around
$k=0$ and keep only a few low-order terms,
\beq
\hat{q}_{\D z}(k) = \hat{q}_{\D z}(0) + \p_{k} \hat{q}_{\D z}(0)k
                 + \frac{1}{2} \p_{k}^2 \hat{q}_{\D z}(0) k^2 + ... \ .
\eeq
With Eq.\ (\ref{dermo}),
the derivatives can be replaced with the moments,
\beq
\hat{q}_{\D z}(k) = 1 - i \langle \D z \rangle  k
                 - \frac{1}{2} \langle \D z^2 \rangle k^2 + ...
\label{TaylorF}
\eeq
(with $\langle \D z^0 \rangle = 1$). The
%and we have expressed the
Taylor expansion of a pdf in terms of the moments is
practical because
the moments are the natural characteristics of a pdf.
% in that they are related to the mean value, variance, etc. of a distribution.
Often, the distribution of spatial increments is assumed to be
symmetric around $z=0$, so that $\langle \D z \rangle = 0$, and the
Taylor expansion is written in this case as
\beq
\hat{q}_{\D z}(k) = 1
                 - \frac{1}{2} \langle \D z^2 \rangle k^2 + ...
\label{TaylorF2}
\eeq

Temporal distributions, such as the time-step distribution
$q_{\D t}(\D t)$ in the waiting model,
have the characteristic to be
'one-sided', i.e.\ they are defined and used only for $t\geq 0$, so that
it is more
appropriate to use Laplace transforms in this case, defined as
\beq
\tilde{q}_{\D t}(s) = \int_0^\infty e^{-s\D t} q_{\D t}(\D t)\, d\D t  .
\eeq
Straightforward calculations and the use of the respective
definitions leads to the analogue of Eq.\ (\ref{dermo}) for
Laplace transforms,
\beq
(-\p_s)^n \tilde{q}_{\D t}(s) \Big\vert_{s=0}
=\int_0^\infty \D t^n q_{\D t}(\D t)\, d\D t = \langle \D t^n \rangle ,
\label{dermoL}
\eeq
where the $\langle \D t^n \rangle$ are again the moments.
As with respect to $z$, we will focus on the asymptotic, large $t$ regime,
which corresponds to small values of $s$,
and we make a Taylor-expansion of the Laplace transform around
$s=0$, replacing the derivatives by the
moments through Eq.\ (\ref{dermoL}),
\beq
\tilde{q}_{\D t}(s) = 1 - \langle \D t \rangle s
%                   + \frac{1}{2} \langle \D t^2 \rangle s^2
+ ...  ,
\label{TaylorL}
\eeq
in complete analogy to Eq.\ (\ref{TaylorF}) ($\langle \D t^0 \rangle=1$ is
the normalization of $q_{\D t}$).
%Similarly, if we consider a small $s$ values,
%then it determines diffusion at large times.
The Taylor expansions in Eqs.\ (\ref{TaylorF}), (\ref{TaylorF2}), and (\ref{TaylorL})
can, of course, only be used if the involved moments are finite.

\subsubsection{The symmetric and the one-sided Levy distributions\label{symmLevy}}

The {\it symmetric Levy distributions} are defined in Fourier space as
\beq
\hat{q}^{L,\alpha}_{\Delta z}(k) = \exp(-a\vert k\vert^\alpha) ,
\label{levydis}
\eeq
with $0<\alpha\leq 2$. It is not possible to express them in closed form
in real space, with two exceptions,
the case $\alpha=2$ is the usual Gaussian distribution
(the Fourier back-transform of a Gaussian is a Gaussian), and the
case $\alpha=1$ is known as the Cauchy distribution
(see Chapters 4 and 5 in \citet{Hughes1995}). As mentioned
already, the Levy distributions for $\alpha<2$
all have infinite
variance, and for $\alpha\leq 1$, they even have an infinite mean value,
so that the expansion in the form of Eq.\ (\ref{TaylorF2}) is not applicable.
We can though directly expand the exponential in Eq.\ (\ref{levydis}) and
find the small $k$ expansion as
\beq
\hat{q}^{L,\alpha}_{\Delta z}(k) = 1 -a\vert k\vert^\alpha
%                                    + \frac{1}{2} a^2\vert k\vert^{2\alpha}
+ ... ,
\label{LexpaF}
\eeq
The case $\alpha=2$ corresponds obviously to the classical case
of Eq.\ (\ref{TaylorF2})
with finite second moment ($a\equiv (1/2)\langle \D z^2\rangle$) and zero
mean   (we are considering only the symmetric Levy-distributions).

If one-sided distributions with infinite variance are needed, then they can also be defined
via Fourier space as a specific asymmetric Levy distribution,
or, more convenient for our purposes,
as shown in \citet{Hughes1995} (Chap.\ 4.3.2), they can be defined
in Laplace space as
\beq
\tilde{q}^{L1,\beta}(s) = \exp(-b s^\beta) ,
\label{Llap}
\eeq
with $b$ strictly positive, and where now $0<\beta\leq 1$. These
one-sided Levy distributions decay as $t^{-1-\beta}$ for $t\to\infty$,
and they have a small $s$ expansion
\beq
\tilde{q}^{L1,\beta}(s) = 1- b s^\beta
%+ \frac{1}{2} b^2 s^{2 \beta}
+ ...
\label{LexpaL}
\eeq
Note that for $\beta=1$ we recover the finite mean case of
Eq.\ (\ref{TaylorL}), with $b=\langle \D t\rangle$,
and the back-transform of Eq.\ (\ref{Llap}) in this case
yields the distribution $\delta(\D t- b)$, i.e.\ the time-steps
are constant and equal to $b$ ($\Delta t\equiv \langle \D t\rangle\equiv b$).

In the following, we will use the (small $k$) Fourier expansion for $\hat{q}_{\D z}(k)$
in the form,
\beq
\hat{q}_{\D z}(k) = 1 -a\vert k\vert^\alpha + ... ,
\label{qexpaF}
\eeq
which for $\alpha<2$ corresponds to the Levy case, Eq.\ (\ref{LexpaF}),
and for $\alpha=2$ it recovers the normal, finite variance case of
Eq.\ (\ref{TaylorF2}), with $a=(1/2)\langle \D z^2\rangle$.
Correspondingly, the (small $s$) Laplace expansion for $\tilde{q}_{\D t}(s)$ will be used
in the form
\beq
\tilde{q}_{\D t}(s) = 1- b s^\beta  + ... ,
\label{qexpaL}
\eeq
which for $\beta<1$ yields the Levy distributions, Eq.\ (\ref{LexpaL}),
and for $\beta=1$ the normal, finite mean case of Eq.\ (\ref{TaylorL}),
with $b=\langle \D t\rangle$.

\subsubsection{Solving the CTRW equations with Fourier and Laplace transforms\label{solvwait}}

In this section, we will illustrate the use of the
Fourier and Laplace (F-L) transform to solve the CTRW
equation on the example of the waiting model, Eqs.\ (\ref{Qeq})
and (\ref{PWeq}).
%A standard way to treat the CTRW equations is through Fourier
%and Laplace (F-L) transform.
The use of the respective convolution theorems, namely
\beq
\int f(x-y)g(y) \,dy \to \hat{f}(k)\,\hat{g}(k)
\eeq
for Fourier transforms, and
\beq
\int_0^t \phi(t-\tau)\psi(\tau) \,d\tau \to \tilde{\phi}(s)\,\tilde{\psi}(s)
\eeq
for Laplace transforms (with $f,\,g,\,\phi$, and $\psi$ any
transformable functions),
allows in the
case of the waiting model to determine
the solution in Fourier Laplace space: For the initial condition
$P(z,t=0)=\delta(t)\delta(z)$ and in the absence of any source ($S=0$),
Eq.\ (\ref{Qeq}) turns into
$\tilde{\hat{Q}}(k,s)=\tilde{\hat{Q}}(k,s)\,\hat{q}_{\D z}(k)\,\tilde{q}_{\D t}(s)+1$, and
Eq.\ (\ref{PWeq}) takes the form
$\tilde{\hat{P}}_W(k,s)=\tilde{\hat{Q}}(k,s)\tilde{\Phi}_W(s)$,
and we can eliminate $\tilde{\hat{Q}}$ and solve for $\tilde{\hat{P}}$. Noting
further that $\Phi_W(t)=\int_t^\infty q_{\D t}(\D t)\,d\D t
=1-\int^t_\infty q_{\D t}(\D t)\,d\D t$, so that
$\tilde{\Phi}_W(s)=(1-\tilde{q}_{\D t}(s))/s$, we find
\beq
\tilde{\hat{P}}_W(k,s) = \frac{1-\tilde{q}_{\D t}(s)}{s\left[1-\hat{q}_{\D z}(k)
\tilde{q}_{\D t}(s)\right]}  ,
\label{mowei}
\eeq
which is known as the {\it Montroll-Weiss equation} \cite{Montroll65,Zumofen93,Klafter87}.

Looking for asymptotic solutions, we insert the general form
of the transformed temporal and spatial step distribution,
Eqs.\ (\ref{qexpaF}) and (\ref{qexpaL}), respectively,
into Eq.\ (\ref{mowei}), which yields
\beq
\tilde{\hat{P}}_W(k,s) = \frac{b s^{\beta-1}}
{b s^\beta +a\vert k\vert^\alpha}  .
\label{PWsmall}
\eeq
Unfortunately, it is not possible to
Fourier and Laplace back-transform $\tilde{\hat{P}}_W(k,s)$ analytically.
We can though use Eq.\ (\ref{dermo}) for $n=2$, namely
\beq
\langle z^2(s) \rangle
=
- \p_k^2 \tilde{\hat{P}}_W(k=0,s) ,
\label{msdfou}
\eeq
to determine the mean square
displacement in the asymptotic
regime
(note that we set $k=0$ at the end, which clearly is in the large $|z|$
regime).
%That this is valid only for the asymptotic regime where $z^2$ is large
%follows from the fact that we let $k\to 0$, i.e.\ we consider small
% wave numbers that correspond to large length scales (wave lengths).
%Similarly, if we consider Eq.\ (\ref{msdfou}) only at small $s$ values,
%then it determines diffusion at large times.
Inserting $\tilde{\hat{P}}_W(k,s)$ from Eq.\ (\ref{PWsmall})
into Eq.\ (\ref{msdfou}),
without yet setting $k=0$, we find
\beq
\langle z^2(s)\rangle = -\frac{2a^2\alpha^2 \vert k\vert^{2\alpha-2}}{bs^{2\beta+1}}
+\frac{a \alpha(\alpha-1) \vert k\vert^{\alpha-2}}{bs^{\beta+1}}
\label{aux1}
\eeq
The first term on the right side diverges for $\alpha<1$, and the second term
diverges for $\alpha<2$, so that $\langle z^2(s)\rangle$ is infinite in these
cases. This divergence must be interpreted in the sense that the transport
process is very efficient, so that in the asymptotic
regime, $P_W$ has
already developed  so fat wings at large $|z|$ (power-law tails)
that the variance, and with that, the mean square displacement,
of $P_W$ is infinite, $P_W$ has already become a Levy type
distribution \cite{Klafter87,Balescu07a}.
Of course, with the formalism we apply, we cannot say anything
about the transient phase, before the asymptotic regime
is reached.
%As noted by \citet{Klafter87} \citep[see also][]{Balescu07},
%Eq.\ (\ref{PWsmall}) with
%$\alpha$ and $\beta$ taking arbitrary values in their allowed
%ranges leads to extremely efficient diffusion, with
%$\hat{P}_W$ developing very fast power-law tails, so the
%the variance of $P_W$, and with that the mean square displacement
%are infinite like they are for any Levy distribution.
We just note here that the velocity model (Eq.\ (\ref{PVeq})) in this regard
is not so over-efficient, it allows
super-diffusion with a more gradual build-up of the fat wings of the
distribution $P_V$ \cite{Klafter87}.

Less efficient diffusion can only be achieved
in the frame of the waiting model for $\alpha=2$, i.e.\ for
normal, Gaussian distributed spatial steps (see Eq.\ (\ref{qexpaF})). In this case,
Eq.\ (\ref{aux1}) takes the form
%(see Eqs.\ (\ref{TaylorF}) and (\ref{LexpaF})).
%Applying now the recipe of Eq.\ (\ref{msdfou}) for $\alpha=2$,
%we find
\beq
\langle z^2(s)\rangle = \frac{\langle\D z^2\rangle}{bs^{\beta+1}}
\eeq
for $k\to 0$ ($a=(1/2)\langle\D z^2\rangle$ for $\alpha=2$, see Eq.\ \ref{TaylorF2}).
This expression is valid for small $s$,
%as follows from the derivation of Eq.\ (\ref{msdfou}),
and with the help of the
Tauberian theorems, which relate the power-law scaling
of a Laplace transform at small $s$ to the scaling in original
space for large $t$ \cite{Hughes1995,FellerBook71}, it follows that
\beq
\langle z^2(s)\rangle\rangle \sim t^\beta .
\eeq
With our restriction $0<\beta\leq 1$, diffusion is always
of sub-diffusive character, and for $\beta=1$ it is normal,
as expected, since we have, in this case, waiting times with finite
mean and variance (see
Eq.\ (\ref{qexpaL})).

\subsubsection{Including Velocity Space Dynamics\label{velo}}

 In applications to turbulent systems, and  very prominently to turbulent or driven
plasma systems, it  is often not enough to monitor the position and the timing of
a particle, since its velocity may drastically change, e.g.\  in interactions with
local electric fields generated by turbulence.
For these cases, the standard CTRW can be extended to include, besides the
position space and the temporal dynamics, also the velocity space dynamics, which
%, which yields a more
%realistic model
%for particle dynamics and
allows to study anomalous diffusive behavior also in energy space.

To formally define the extended CTRW that also includes momentum space,
we keep Eq.\ (\ref{rnclass}) and Eq.\ (\ref{tctrw}) for position and
time evolution as they
are, and newly, the momentum (or velocity) also becomes a random,
dynamic variable,
with temporal evolution of the form
\beq
\mathbf p_n = \Delta \mathbf p_n + \Delta \mathbf p_{n-1} + \Delta \mathbf p_{n-2} + \, ...
                                             \, + \Delta \mathbf p_1 + \mathbf p_0 ,
\label{pext}
\eeq
with $p_0$ the initial momentum, and the $\D \mathbf p_i$ the momentum increments.
Again,
one has to specify a functional form for the distribution of momentum
increments
$q_{\D \mathbf p}(\D \mathbf p)$ in order to specify the random walk problem
completely (in the most generic case, one would have to define the joint probability distribution of increments $q_{\D\mb p,\D\mb r,\D t}(\D p,\D z,\D t)$, since the increments can be mutually dependent). The solution of the extended CTRW is in the form of
the distribution $P(\mathbf r,\mathbf p,t)$ for a particle at time $t$ to be
at position $\mathbf r$ and to have momentum $\mathbf p$.

Monte-Carlo simulations can treat the extended CTRW,
as done e.g.\ in \citet{Vlahos04}, \citet{vlahos16}, \citet{Isliker17},  \citet{Pisokas16}, \citet{Garrel18}, \citet{Pisokas18}, \citet{Sioulas20}, \citet{Sioulas20b}, and \citet{Sioulas22c}.
%or 
In 
%\citet{Isliker03} 
\citet{Isliker2007}   a set of equations for the extended CTRW
has been introduced, which is a generalization of
Eq.\ (\ref{Qeq}) and Eq.\ (\ref{PVeq}), and a way to solve
the equations numerically is presented.

\subsubsection{From random walk to fractional diffusion equations}

The purpose of this section is to show how fractional transport equations
naturally arise in the context of random walk models.
The starting point here is the CTRW equations for the waiting
model, Eqs.\ (\ref{Qeq}) and (\ref{PWeq}), which in Fourier Laplace
space take the form of Eq.\ (\ref{mowei}), and on inserting
the small $k$ and small $s$ expansion of the step-size and waiting-time
distributions, Eqs.\ (\ref{qexpaF}) and (\ref{qexpaL}), respectively,
the waiting CTRW equation
takes the form of Eq.\ (\ref{PWsmall}), with $\alpha\leq2$ and $\beta \leq 1$.
Multiplying Eq.\ (\ref{PWsmall}) by the numerator on its right side, we find
\beq
\tilde{\hat{P}}_W(k,s)(b s^\beta +a\vert k\vert^\alpha) = b s^{\beta-1}  ,
\label{PWsmall2}
\eeq
and rearranging, we can bring the equation for $\tilde{\hat{P}}_W$ to the form
\beq
s^\beta \tilde{\hat{P}}_W(k,s) - s^{\beta-1}
= -\frac{a}{b} \vert k\vert^\alpha \tilde{\hat{P}}_W(k,s)  .
\label{PWsmall3}
\eeq

It is illustrative first to consider the case of normal transport,
with $\beta=1$ and $\alpha=2$,
where according
to Eqs.\  (\ref{qexpaF}) and (\ref{qexpaL}) we have $a=(1/2)\langle \D z^2\rangle$ and
$b=\langle \D t\rangle$, so that
\beq
s \tilde{\hat{P}}_W(k,s) - s^{0}  = -\frac{\langle \D z^2\rangle}{2\langle \D t\rangle}
\vert k\vert^2 \tilde{\hat{P}}_W(k,s)
\label{PWsmall4}
\eeq
Now recall how a first-order temporal derivative of a function $\psi$ is expressed in Laplace space,
\beq
\frac{d}{dt} \psi(z) \to s \tilde{\psi}(s) - s^{0}\psi(0) ,
%- s^{n-2} \frac{d}{dz} \psi(0) - ...- s^{0} \frac{d^{n-1}}{dz^{n-1}} \psi(0)
\label{derivL}
\eeq
and how a spatial derivative of a function $f$ translates to Fourier space,
\beq
\frac{d^n}{dz^n} f(z) \to (-ik)^n \hat{f}(k)
\label{derivF}
\eeq
Obviously, for $P_W(z,t=0)=\delta(z)$, Eq.\ (\ref{PWsmall4}) can be
back-transformed as
\beq
\p_t P_w(z,t) = \frac{\langle \D z^2\rangle}{2\langle \D t\rangle}
                 \p^2_z P_W(z,t),
\eeq
so that we just recover the simple transport equation of the normal
diffusive case in Eq.\ (\ref{einsdiff}).

\subsubsection{Fractional derivatives}

Fractional derivatives generalize the usual derivatives
of $n$th order to general non-integer orders. Several definitions exist,
and in original space ($z$ or $t$) they are a combination of usual derivatives
of integer order and integrals over space or time. The latter property makes
them non-local operators so that fractional differential
equations are non-local, as are the CTRW integral equations.
For the following, we need to define the Riemann-Liouville left-fractional
derivative of order $\alpha$,
\beq
{}_a D^\alpha_{z} f(z)
:=
\frac{1}{\Gamma(n-\alpha)}
\frac{d^n}{dz^n} \int_a^z \frac{f(z^\prime)}{(z-z^\prime)^{\alpha+1-n}}
\,dz^\prime  ,
\eeq
with $\Gamma$ the usual Gamma-function, $a$ a constant, $n$ an integer
such that $n-1\leq\alpha<n$, $\alpha$ a positive real number,
and $f$ any suitable function.
Correspondingly, the Riemann-Liouville right-fractional
derivative of order $\alpha$ is defined as
\beq
{}_z D^\alpha_{b} f(z)
:=
\frac{(-1)^n}{\Gamma(n-\alpha)}
\frac{d^n}{dz^n} \int_z^b \frac{f(z^\prime)}{(z^\prime-z)^{\alpha+1-n}} \,dz^\prime ,
\eeq
with $b$ a constant. It is useful to combine these two asymmetric definitions into a
a new, symmetric fractional derivative, the so-called
Riesz fractional derivative,
\beq
D^\alpha_{|z|} f(z)=
-\frac{1}{2\cos(\pi\alpha/2)}
\left({}_{-\infty} D^\alpha_{z}  + {}_z D^\alpha_{\infty} \right) f(z)   ..
\eeq
The Riesz fractional derivative has the interesting property that its representation in Fourier
space is
\beq
{}^{(R)}D^\alpha_{\vert z \vert} f(z) \to -\vert k\vert^\alpha \hat{f}(k) .
\label{RieszF}
\eeq
Comparison of this simple expression with Eq.\ (\ref{derivF})
makes it obvious that the Riesz derivative is a natural generalization
of the usual derivative with now non-integer $\alpha$.

To treat time, a different variant of fractional derivative
is useful, the
Caputo fractional derivative of order $\beta$,
\beq \label{Caputo}
{}^{(C)}D_t^\beta \psi(t)
:=
\frac{1}{\Gamma(n-\beta)}
 \int_0^t \frac{1}{(t-t^\prime)^{\beta+1-n}} \frac{d^n}{dt^{\prime n}} \psi(t^\prime)\,dt^\prime ,
\eeq
with $n$ an integer such that $n-1\leq\beta<n$, and
$\psi$ any appropriate function.
The Caputo derivative
translates to Laplace space as
\beq
{}^{(C)}D_t^\beta \psi(t) \to s^\beta \tilde{\psi}(s) - s^{\beta-1}\psi(0) ,
%- s^{\beta-2} \frac{d}{dz} \psi(0) - ...- s^{\beta-n} \frac{d^{n-1}}{dz^{n-1}} \psi(0)
\label{CapF}
\eeq
for $0<\beta\leq 1$,
which is again a natural generalization of Eq.\ (\ref{derivL}) for
the usual derivatives, with now non-integer $\beta$ (the Caputo derivative
is also defined for $\beta\geq1$, with Eq.\ (\ref{CapF}) taking a more
general form).

Further details about fractional derivatives can be found e.g.\
in \citet{Kilbas2006}, \citet{Podlubny2009}, or in the extended Appendix of Balescu \cite{Balescu07a}.

\subsubsection{Fractional Transport Equation (FTE)\label{fracdiffeq}}

Turning now back to Eq.\ (\ref{PWsmall3}), the CTRW equation in F-L-space in the 'fluid' limit (large $z$ and $t$), we obviously can identify the
fractional Riesz and Caputo derivatives in their simple Fourier and
Laplace transformed form, Eqs. (\ref{RieszF}) and (\ref{CapF}), respectively,
and write
\beq
{}^{(C)}D_t^\beta P_W(z,t) = \frac{a}{b}\,\,\,
                  {}^{(R)}D^\alpha_{\vert z \vert} P_W(z,t)
\label{fraceq}
\eeq
From this derivation, it is clear that the order of the fractional derivatives,
$\alpha$ and $\beta$, are determined by the index of the
step-size ($q_{\D z}$) and the waiting time ($q_{\D t}$)
Levy distributions, respectively.
It is also clear that Eq. (\ref{fraceq}) is just an alternative way of writing
Eq.\ (\ref{PWsmall3}) or (\ref{PWsmall}), and as such it is the asymptotic,
large $|z|$, large $t$ version of the CTRW equations
(\ref{Qeq}) and (\ref{PWeq}).
It allows though to apply different mathematical tools for its
analysis that have been
developed specially for fractional differential equations.

%As an example, we
%the Laplace back transform of Eq.\ (\ref{PWsmall})

As an example, we may consider the case $\beta=1$ and $0<\alpha\leq 2$,
where the transport equation is fractional just in the spatial part,
\beq
\p_t P_w(z,t) = \frac{a}{b}\,\,\,
                  {}^{(R)}D^\alpha_{\vert z \vert} P_W(z,t)
\label{fraceqexam}
\eeq
In Fourier Laplace space, this equation takes the form
\beq
\hat{P}_W(k,s) = \frac{b}
{b s +a\vert k\vert^\alpha}  ,
\label{PWexam}
\eeq
which, on applying the inverse Laplace transform, yields
\beq
\hat{P}_W(k,t) = \exp\left(- \frac{a}{b}\, |k|^\alpha t\right) ,
\eeq
which is the Fourier transform of a symmetric Levy-distribution,
with time $t$ being part of the parameter (see Eq.\ (\ref{levydis})),
and with index $\alpha$ equal to the
one of the spatial step distribution $q_{\D z}$.
Thus, for $\alpha<2$, the
the solution has power-law tails and the mean square displacement
(or variance or second moment)
is infinite, as we found it in Eq.\ (\ref{aux1}).
For $\alpha=2$, the solution $P_W(z,t)$ is a Gaussian
(the Fourier back-transform of a Gaussian is a Gaussian),
and we have normal transport.

%\heinz{
Both continuous time random walk (CTRW) models, the one with finite velocity (Eq.\ (\ref{PVeq})) and the waiting model (Eq. (\ref{PWeq})), are obviously causal, despite their non-locality, since the involved temporal integrals extend only over the past, so only the past determines the future. This property --- causality despite non-locality --- is inherited by the fractional transport equation (FTE), which is derived from the waiting model. Correspondingly, the derivatives in the general FTE (Eq.\ (\ref{fraceq})) are fractional and therewith non-local in space and time. The temporal derivative is though in the form of the Caputo fractional derivative (Eq.\ (\ref{Caputo})), which is non-local only in the backward time direction, ensuring causality.
%} 

\subsubsection{Other ways to model anomalous transport}

Escande and Sattin \cite{Escande07} review and discuss under what
circumstances the Fokker-Planck equation (Eq.\ (\ref{FokkerP}))
can model anomalous transport.
They note that the FP equation, which is a local model, is
able to model anomalous transport behavior
in cases where there is a non-zero drift velocity,  $F(z)\ne 0$,
anomalous transport is thus based here on systematic drift effects.

Klafter et al. \cite{Klafter87} briefly review attempts to use the Fokker-Planck
equation with zero drift velocity but spatially or temporally dependent
transport coefficient. To account for anomalous
transport, they find that the spatial or temporal dependence of the diffusion coefficient
must be chosen in very particular functional forms, which are difficult
to interpret physically. 

%\heinz{
Spatially varying diffusion coefficients are routinely utilized in astrophysics, notably to model TeV halos surrounding pulsars, where suppressed cosmic-ray diffusion around pulsars results in localized gamma-ray emission observed at TeV energies \cite{Mukhopadhyay22}. 
%}

Lenzi at al. \cite{Lenzi03} shortly discuss
non-linear transport equations, which have the form of a simple
transport equation as in Eq.\ (\ref{einsdiff}), with $P$ though raised
on one side of the equation to some power $\nu$, 
%\heinz{
$\p_t P = \p^2_z P^\nu$, with usually $\nu \geq 2$, which is known as the porous media equation. In \citet{Frank05}, an introduction to the theory of nonlinear Fokker-Planck equations is provided, discussing key attributes of both transient and enduring solutions. The text highlights stability analyses of stationary solutions by employing self-consistency equations, linear stability analysis, and Lyapunov's direct method. Additionally, it explores Langevin equations. Moreover, the book surveys various applications in domains such as condensed matter physics, the physics of porous media and liquid crystals, accelerator physics, neurophysics, social sciences, population dynamics, and computational physics.
%}

%subsection{Transport in a Fractal Environment}

\section{Stochastic and systematic energization}

Before examining the mechanisms of particle heating and acceleration within strongly turbulent, magnetized plasmas where Coherent Structures (CoSs) spontaneously form and evolve in large-scale space and astrophysical environments \cite{Vlahos23}, it is essential to emphasize several key points about the energization processes: (1) they are fundamentally {\bf electromagnetic}, (2) they involve the {\bf intermittent formation of CoSs} that act as ``scattering centers" within the turbulent plasma, and (3) the turbulent 3D plasma volume is {\bf open}, allowing energized particles to escape with a characteristic time, $t_{esc}$. 

 The electromagnetic forces can be much more efficient and act on much faster time-scales than collisions. For longer times, collisions may become effective, too, only at low energies though, typically affecting the heating process but not much the acceleration phenomena, above all if the electromagnetic effects are strong. Astrophysical plasmas are highly conductive when viewed on large scales, meaning that any large-scale electric field adheres to the condition of ideal magnetohydrodynamics (MHD), which states that the plasma’s irregular bulk motion induces the electric field
\beq \label{Econv}
\mathbf{E}^c_{\perp} = -{\mathbf{V}} \times \mathbf{B} ,
\eeq
where $\mathbf{V}$ is the plasma's bulk velocity, and $\mb B$ is the magnetic field. Within a structured, turbulent plasma, the CoSs' random motions create a distribution of convective electric fields that interact stochastically with particles.

%\heinz{
Turbulent regions are often marked by small-scale features like reconnecting current sheets and current filaments, which may exhibit finite resistivity $(\eta)$, and which also permit considerable current densities ($\vec{J} \sim \nabla \times \vec{B}$) and pressure discontinuities, leading in turn to the non-ideal electric field components of the generalized Ohm's law,
\beq \label{Eres}
\mathbf{E}^{ni} = \mathbf{E}^{ni}_{||} + \mathbf{E}^{ni}_{\perp} = \eta \mathbf{J} + \frac{1}{en_e} \mb{J} \times \mb{B} - \frac{1}{en_e} \nabla P_e .
\eeq 
In this expression for the non-ideal electric field $\mathbf{E}^{ni}$, $\eta \mb{J}$ signifies the resistive term, $(1/en_e) \mb{J} \times \mb{B}$ describes the Hall effect, and $(1/en_e) \nabla P_e$ addresses the effect of the electron pressure $P_e$. Here, $e$ denotes the electron charge and $n_e$ the electron density. In most magnetohydrodynamic studies, only the resistive term seems to be taken into account, at least in the astrophysical context. The total electric field $\mathbf{E}$ incorporates both ideal and non-ideal magnetohydrodynamic elements, most prominently the resistive term, and it is given by $\mathbf{E} = \mathbf{E}^c_{\perp} + \mathbf{E}^{ni}_{\perp} + \mathbf{E}^{ni}_{||}$.
%}

The equation of motion for a charged particle determines the rate of change in kinetic energy, 
$W = (\gamma - 1) mc^2$, as
\beq \label{enrgygain}
\frac{dW}{dt}= q(\mathbf{E} \cdot \mathbf{v}) ,
\eeq
where $q$ is the particle charge, $\mathbf{v}$ is the particle velocity, $\mb p=\gamma m \mb v$  the particle momentum ($\gamma=1/\sqrt{1-v^2/c^2}$), 
governed by
\beq \label{particleMotion}
\frac{d\mathbf{p}}{dt}=q(\mathbf{E}+\frac{1}{c}(\mathbf{v}\times \mathbf{B}) .
\eeq

The role of moving CoSs, or magnetized "clouds", in particle acceleration was first recognized by Fermi \cite{Fermi49, Fermi54}, who introduced two foundational models of cosmic rays acceleration that continue to underpin current research:
\begin{itemize}
    \item {\bf Stochastic Acceleration:} Particles gain or lose energy incrementally through interactions with randomly positioned electric fields in the turbulent plasma.
   \item {\bf Systematic Acceleration:} Particles gain energy more systematically as they interact directly with CoSs.
\end{itemize}
In the stochastic model, particles undergo small random energy changes during their encounters with CoSs. In the systematic model, all particles gain energy in more significant and sometimes substantial  amounts when interacting with CoSs. These mechanisms often coexist in turbulent, magnetized plasmas, as particles may undergo stochastic acceleration by large-scale magnetic fluctuations, eddies, or colliding filaments and systematic acceleration when crossing CSs or shocks.

The spatial transport and diffusion of energized particles within a finite turbulent volume significantly influences the energy-dependent escape time, $t_{esc}(W)$, because interactions with CoSs impact both the particles' trajectories and energies. Certain CoSs may confine particles temporarily before they are released and continue to subsequent scatterers, which affects both their spatial propagation and energy evolution.
%\subsection

\subsection{Stochastic energization  of particles by waves and CoSs in turbulent plasma}

\subsubsection{Fermi's initial ideas on "stochastic'' energization}

Fermi proposed in 1949 \cite{Fermi49} the first mechanism for particle heating and acceleration in turbulent space and laboratory plasma.   

\begin{figure}[h]
\centering
\includegraphics[width=0.90\columnwidth]{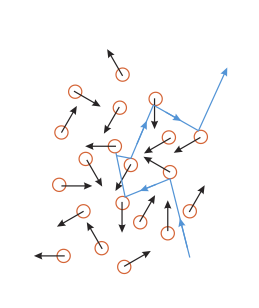}
\caption{Illustration of stochastic Fermi energization for a particle  scattered e.g.\ by CoSs. The scattering centers are in red and the particle trajectory in blue color. }
\label{FermiStoch}
\end{figure}
As illustrated in Fig.\ \ref{FermiStoch}, the basic idea of stochastic energization involves scattering centers (red) that scatter a light particle in collisions that are elastic in the rest frame of the scattering center (the trajectory of the scattered particle is shown in blue). In stochastic Fermi acceleration, the scattering centers move randomly with velocity  $\mathbf{V}$ and are {\bf uniformly} distributed in the turbulent volume. The energy change for a particle in a scattering event can be estimated by the equation \cite{Fermi49}
\beq \label{Fermi1}
\Delta W=W_{f} - W_{i}= 2\Gamma^2 \left( \frac{V^2}{c^2}-\frac{\mathbf{V}\cdot \mathbf{v}}{c^2} \right) W_{i}.
\eeq
%\heinz{
A particle gains energy in a `head-on’ collision with a scattering center, $\mathbf{V}\cdot \mathbf{v} <0$, and it loses energy in an overtaking collision,  $\mathbf{V}\cdot \mathbf{v} >0$ (see details in \citet{Achterberg08} and \citet{Longair11}).
%}

In the astrophysical applications of stochastic energization, the scattering centers are  CoSs (e.g.\ sub-Alfvenic large scale fluctuations, nonreconnecting current sheets, eddies with randomly distributed small scale electric fields, etc), where the Lorentz force associated with the CoSs is responsible for the deflection and energization of a charged particle.

Eq.\ (\ref{Fermi1}) is an order of magnitude estimate of the energy change. In particular, if the motion of the scatterers is nonrelativistic ($|\mathbf{V}|<<c$), and the particle velocity is much larger ($|\mathbf{v}|>> |\mathbf{V}|$), the typical {\bf stochastic} energy change will be 
\beq \label{StocFermi1}
\frac{|\Delta W|}{W} \sim \frac{vV}{c^2} .
\eeq
Since, on the average, the head-on collisions dominate the overtaking ones off the CoSs,
the estimated {\bf average} rate of {\bf systematic} kinetic energy gain by relativistic charged particles $(v \sim c)$  is 
\beq \label{FSenergtgain}
\frac{\langle \Delta W \rangle}{t_{sc}}=\frac{8c}{3\lambda_{sc}} \left(\frac{V}{c} \right )^2 W = \frac{W}{t_{acc}} 
\eeq
(see the details in \citet{Achterberg08}),
where $t_{acc}=3c\lambda_{sc}/8V^2$ is the typical acceleration time, $\lambda_{sc}$ is the mean free path between collisions, and $t_{sc}=\lambda_{sc}/c$ the mean travel time between subsequent collisions. The mean energy of the particles thus grows exponentially as
\beq \label{Fer2Acc}
\langle W\rangle (t)=W(0)e^{t/t_{acc}}.
\eeq
Eq.\ (\ref{FSenergtgain}) actually is the definition of the convective transport coefficient $F_W$ in the FP equation (Eq.\ (\ref{FokkerPVelLoss})), so that we have
\beq
F_W = \frac{W}{t_{acc}} .
\label{F_W}
\eeq
The coefficient $F_W$ is a measure of the degree of systematic energy gain, and, in terms of random walk, it represents a systematic trend that is superimposed upon the erratic diffusive motion in energy space.

The mean energy change per collision is a factor $\sim V/c$ smaller than the magnitude of
the typical energy change in an individual collision, see Eq.\ (\ref{StocFermi1}). In fact, to leading order in $V/c$ one
has \cite{Achterberg08}
\beq \label{SFenergygain1}
\frac{\langle \Delta W\rangle}{W} = \frac{8}{3} \left ( \frac{V}{c} \right )^2,\  \frac{\langle \Delta W^2\rangle}{W^2}=\frac{4}{3} \left ( \frac{V}{c} \right )^2 .
\eeq 
This means
that the particles will not only increase their mean energy systematically but will also rapidly disperse
in energy. This dispersion can be thought of as a random walk along the energy axis,
where a particle randomly takes forward and backward steps in energy, with the average step
size equal to
\beq\label{SFengai2}
\Delta 
W_{rms}\sim (\langle \Delta W^2\rangle)^{1/2} ,
\eeq
and with the probability of a forward or backward step  equal. Such a random
walk can be characterized by an energy diffusion coefficient $D_W$ that is defined as
\beq \label{SFDiff}
D_W:= \frac{\langle(\Delta W)^2\rangle}{2 t_{sc}}=\left (\frac{2V^2}{3c \lambda_{sc}} \right ) W^2.
\eeq
Again, $t_{sc} = \lambda_{sc}/c$ is the mean time between two subsequent collisions with CoSs for a relativistic particle. The diffusive term represents the stochastic part of the so-called second-order Fermi energization and plays a crucial role in the stochastic heating of a strongly turbulent plasma. 

 In summary, the interaction of particles with CoSs, as  described by Eq.\ (\ref{Fermi1}), represents a {\bf synergy of systematic and stochastic energization mechanisms}.  Another interesting point in  this approach is the choice of the mean free path $\lambda_{sc}$ as constant, which means that the scatterers are uniformly distributed in space. This is a strong assumption for a strongly turbulent plasma (see Fig.\ \ref{TurReconnection}), which though is not very realistic. We will return to this issue in a later section.

Since the particles execute a random walk in energy space with small steps, see Eq.\ (\ref{SFenergygain1}), we can use the Fokker-Planck equation (Eq.\ (\ref{FokkerPVelLoss})) to evaluate the evolution of the energy distribution of the particles $(P(W,t) )$ inside a turbulent volume.

\subsubsection{How can a power-law energy distribution be built?}
\label{FP_power_law}

Using the Fokker Planck equation for a leaky box, Eq.\ (\ref{FokkerPVelLoss}), and focusing on the systematic energy gain, assuming for now that the diffusion term for particles undergoing many collisions with CoSs is not important (i.e.\ keeping only the convective term, Eq.\ (\ref{F_W})), the FP equation reduces to 
\bea \label{FP2}
\frac{\partial P(W,t)}{\partial t} &=& -\frac{\partial}{\partial W} \left [\left (\frac{W}{t_{acc}} \right )P(W,t)\right ]  \nonumber\\ &-&  \frac{P(W,t)}{t_{esc}(W)} .
\eea
In other words, we have ignored here the stochastic energy transport, which, as we will show next, plays a crucial role in heating the non-relativistic plasma.  The steady-state solution of Eq.\ (\ref{FP2}) is  
\beq \label{highenergyslope}
P(W) \sim \left (\frac{W}{W_0} \right )^{-(1+t_{acc}/t_{esc})}.
\eeq

Fermi also assumed without proof that $t_{acc} \sim t_{esc}$ to reach the well-known result for the cosmic ray energy distribution function ($P(W) \sim W^{-2}$). If the spatial transport inside the turbulent volume is locked with the acceleration time, then the mystery of the near constancy of the power law index of the cosmic rays will be resolved. We will return to this issue later in this article.

Fermi's approach to building a power law energy distribution in a leaky acceleration box can be generalized to other systematic particle accelerators, as shown in the following sections.
%todo: check again after reading everything

\citet{Parker58}, \citet{Ramaty79}, and \citet{Miller90} analyzed the interaction of electrons and ions with large-amplitude magnetic perturbations, which they assumed to be hard spheres in order to obtain analytical results. They kept in their analysis both, the systematic and the diffusion term, and 
the energy distribution is obtained from the Fokker–Planck equation. For the low-energy particles ($W << mc^2$), the solution can be approximated with the function \cite{Miller90}
\beq \label{SFHS4}
P(W) \sim K_2 \left ( 2 \sqrt{\frac{3p}{mc (t_{esc}/t_{acc})}} \right ),
\eeq 
where $K_2$ is the Bessel function of second kind, and $p$ is the momentum of the particles. For relativistic particles $(W >> mc^2)$, the solution is
\beq \label{SFHS5}
P(W) \sim W^{1/2-(1/2)(9+12/(t_{esc}/t_{acc}))^{1/2}}.
\eeq
Assuming that  $t_{acc} \sim t_{esc}$, $P(W) \sim W^{-1.8}.$
Therefore, the results reported by Fermi in his original paper are modified for non-relativistic or relativistic particles when the analysis is based on the assumption of hard spheres. 

\citet{Lemoine25} formulated a theory of stochastic Fermi acceleration that encompasses all types of nonresonant acceleration within ideal electric fields and can be applied in general scenarios.

We  now discuss several simplifications Fermi  made to achieve his famous result: 
(1) The estimate of $t_{esc}$ depends on the transport properties of the energized particle inside the acceleration volume, which are very complex.  
    (2) The size of the acceleration volume influences both the acceleration  and the escape time, e.g., for an infinite turbulent volume, the escape time is also infinite, and the index of the high energy slope in Eq.\ (\ref{highenergyslope}) will be $\sim -1$. 
    (3) The interaction of particles with CoSs is transient in a finite 3D volume. The system (CoSs and particles) may not reach the steady state distribution assumed in Eq.\ (\ref{highenergyslope}). Strong turbulence will dissipate energy and heat, as well as accelerate particles, on a fast timescale. Weak wave turbulence will gradually interact with particles on a slower time scale. 
    (4)  The efficiency of the energization of low energy particles $(v << c)$ is more complex, and this issue is not included in the analysis presented by Fermi in his initial article\cite{Fermi49} (see more details in \citet{Parker58, Ramaty79, Miller90, Mertsch11}).
    (5) The CoSs are distributed homogeneously in the  models mentioned here, and $\lambda_{sc}$ is assumed constant. It is though well known that in strongly turbulent plasmas, the CoSs are fractally distributed inside the turbulent volumes.  

For all these reasons, more careful numerical work on stochastic Fermi energization is necessary, which we will present and discuss in the following subsections.

\subsubsection{Numerical Studies on Stochastic Fermi Energization within a uniform Network of CoSs}\label{stochasticFermi2}

\citet{Pisokas16} investigate stochastic Fermi energization using a three-dimensional grid of size $N \times N \times N$ with a linear length $L$. The grid has a spacing of $\ell = L/(N-1)$, and each grid point is designated as either {\bf active} (i.e., a scatterer) or {\bf inactive} (i.e., no scatterer). Only a small fraction $R = N_{\rm sc}/N^3$ of the grid points are active, allowing us to define the scatterer density as $n_{\rm sc} = R \times N^3/L^3$. The mean free path between scatterers is then given by $\lambda_{\rm sc} = \ell / R$.

Particles (such as electrons or ions) move through this grid and interact with the active points, where they experience energy changes according to the scatterers' characteristics. After an interaction, each particle moves randomly with its newly assigned velocity $v$, continuing until it encounters another scatterer or leaves the grid. The grid width $\ell$ represents the minimum distance between two scatterers, and the time interval between successive scatterings is given by $\Delta t = s/v$, where $s$ is the distance traveled, which is always an integer multiple of $\ell$. At time $t = 0$, all particles are placed at random positions on the grid, with initial velocities selected randomly from a Maxwellian distribution of temperature $T$.

The standard stochastic Fermi energization process is used, as described in the previous section (see Eq. (\ref{Fermi1})), with the velocity of the scatterers being assumed to equal the average Alfv\'en speed in the environment.

The parameters employed in these simulations correspond to conditions in the low solar corona. The magnetic field strength is $B = 100$ G, with a plasma density of $n_0 = 10^9\mathrm{cm}^{-3}$ and an ambient temperature of around $100\,$eV, which equals the initial mean energy of the particles. The Alfv\'en speed in this environment is $V_A \simeq 7 \times 10^8\,\mathrm{cm/s}$, comparable to the thermal speed $V_e$ of the electrons. The energy gain per scattering is approximately $(\Delta W / W) \approx (V_A / c)^2 \sim 10^{-4}$ (see Eq. (\ref{SFenergygain1})), and the length $L$ of the simulation box is $10^{10}\mathrm{cm}$. \citet{Pisokas16} consider the grid to be open, allowing particles to escape from the acceleration region when they reach any grid boundary at $t = t_{\rm esc}$. The escape time $t_{\rm esc}$ differs for each particle, depending on when it reaches a boundary.

In their simulations, they assume that $R = 10\%$ of the grid points in the $601^3$ grid are active, meaning that the scattering sites are sparsely distributed throughout the acceleration region.

\begin{figure}[!ht]
\includegraphics[width=8cm]{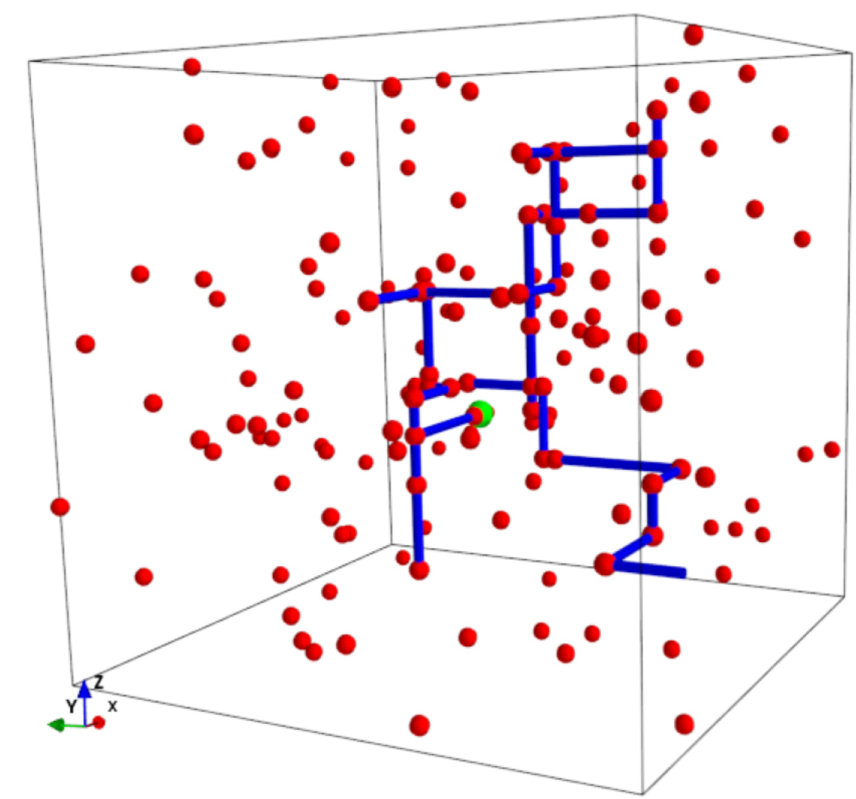}	  
	\caption{Trajectory of a typical particle (blue tube) inside a grid with linear dimension $L={10}^{10} \mathrm{cm}$. Spheres in red mark CoSs. The particle starts at a random grid point (green sphere) and moves along a straight path on the grid until it meets a CoS.  At a CoS, it changes direction randomly, its energy changes according to the CoS's nature, and it continues this process until it exits the simulation box. Reproduced with permission from Pisokas et al., Astrop. J. {\bf 835}, 214 (2017), Copyright 2017 AAS.}
    \label{o2}
\end{figure}
 A typical trajectory of a particle inside the simulation box is displayed in Fig.\ \ref{o2}; the particles move along the grid on straight lines until they encounter a scatterer, which affects their energy and direction of motion (see Eq.\ (\ref{Fermi1})). The motion of the particles is typical for a stochastic system with random-walk-like energy gain and loss before exiting the simulation box. The theoretical mean free path is calculated as $\lambda_{\rm sc} = \ell/R \simeq 1.7 \cdot 10^8 \mathrm{cm}$, which coincides with the value estimated numerically by tracing particles inside the simulation box.

\begin{figure}[!ht]
\includegraphics[width=8cm]{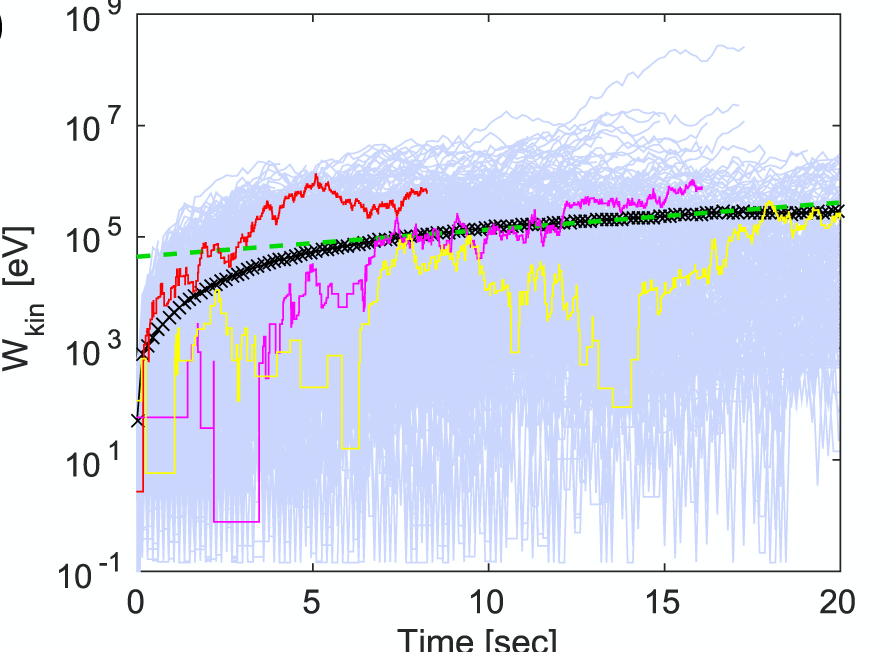}\\
\includegraphics[width=8cm]{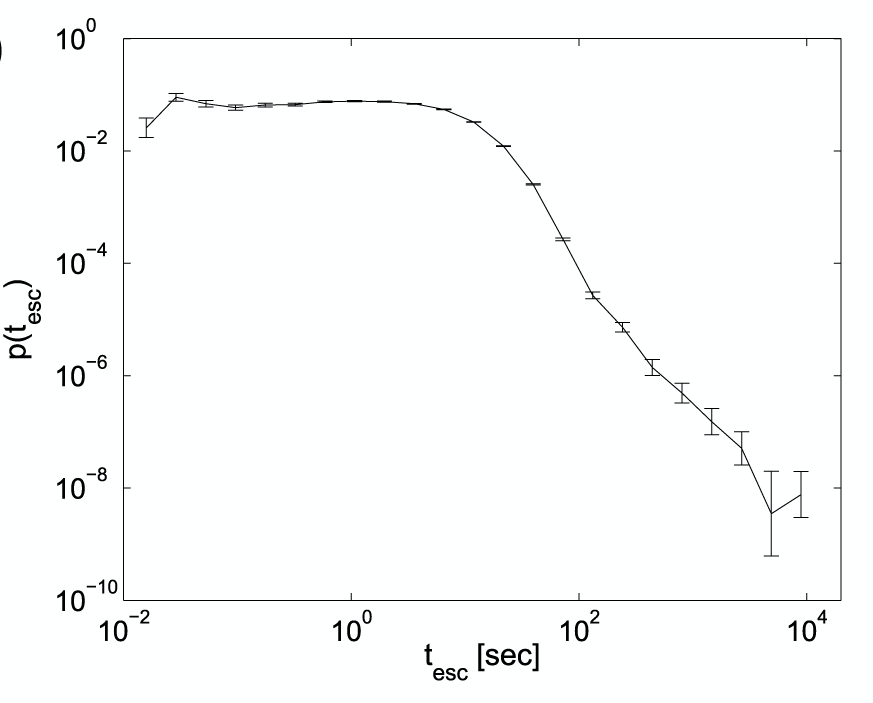}	%   
	\caption{(a) Kinetic energy of the electrons remaining inside the box as a function of time (blue), their mean energy (black) with an exponential fit (green), and the kinetic energy of three typical electrons.  (b) Distribution of the escape time of the electrons. Reproduced with permission from Pisokas et al., Astrop. J. {\bf 835}, 214 (2017), Copyrigt 2017 AAS. \cite{Pisokas16}. }\label{pisf21}
\end{figure}

The temporal evolution of the mean kinetic energy of the electrons that remain inside the simulation box, on using the parameters stated above, is presented in Fig.\ \ref{pisf21}a, along with the kinetic energy evolution of some typical energetic electrons. 
The mean energy increases exponentially (after an initial period of a few seconds), as expected from the analysis performed initially by Fermi (see Eq.\ (\ref{Fer2Acc})). 
Using the analytical expression  given in Eq.\ (\ref{FSenergtgain}), ${t}_{{\mathrm{acc}},{\mathrm{th}}}=(3{\lambda}_{\mathrm{sc}}c)/{8V}_{A}^{2})\approx 4\, \mathrm{s}$. 
We can also estimate the acceleration time from our simulation by making an exponential fit to the asymptotic exponential form of the mean kinetic energy, as predicted by Eq.\ (\ref{Fer2Acc}), which yields an acceleration time of ${t}_{{\mathrm{acc}},{\mathrm{num}}}\approx 9 \mathrm{s}$, a value similar to the analytical estimate, and which we will use as an estimate of $t_{acc}$ below.

Fig.\ \ref{pisf21}b presents the escape-time distribution for all electrons (i.e., the time at which they have reached any boundary); it starts as a uniform distribution at low values and turns over to a power-law distribution at large values. Here, we use the median value, $t_{esc} \sim 8s$, to estimate a characteristic escape time from the system. If we assume that the particle is executing a classical random walk with thermal speed $V_e$ (see Eq.\ (\ref{dif1})), then $t_{esc} \sim L^2 / (\langle v \rangle \lambda_{sc}) \ \sim L^2 / (V_e \lambda_{sc}) \sim 840\, \mathrm{s}$, which is much larger than the numerical estimate. In other words, coupling acceleration and space transport in stochastic Fermi energization reduces the escape time (see similar results in \citet{Zang23}).

\begin{figure}
\includegraphics[width=8cm]{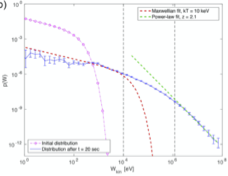}
	\caption{Energy distribution at $t = 0 s$ and
$t = 20 s$ (stabilized) for the electrons remaining inside the box. Stochastic
Fermi acceleration from the large-scale magnetic disturbances formed inside
a strongly turbulent plasma, as seen in the simulation box above, heats and
accelerates the particles from the thermal pool. Reproduced with permission from Pisokas et al., Astrop. J. {\bf 835}, 214 (2017), Copyrigt 2017 AAS.}

\label{pisf3}
\end{figure}

The energy distribution function of the electrons inside the box represents a hot plasma with a mean temperature of approximately $100 \mathrm{keV}$, along with a power-law tail (see Fig.\ \ref{pisf3}). After about $20 \mathrm{s}$, the power-law index is $k \approx 2.1$, and the power-law tail extends up to $100 \mathrm{MeV}$. Using the values for $t_{acc}$ and $t_{esc}$ estimated earlier, the simple expression in Eq.\ (\ref{highenergyslope}) provides an estimate for the slope of the tail as $k=1+{t}_{acc}/{t}_{esc}\approx 1+9/8\approx 2.1$, which matches the numerical result derived from the simulation. Applying the hard-sphere model (see Eq.\ (\ref{SFHS5})), a similar power-law index is obtained, with $k \approx 1.9$.

A key aspect of understanding stochastic Fermi acceleration lies in accurately estimating the transport coefficients and  the escape time for non-trapped particles. In their paper, \citet{Pisokas16} carefully attempt to calculate both. The energy diffusion coefficient can be estimated directly from the particle dynamics, using the relevant relation
\beq \label{DifW}\
D_W(t)=\frac{\left<(W(t+\Delta t)- W(t))^2\right>_W}{2\Delta t} ,
\eeq
and the energy convection coefficient, representing systematic acceleration, is given as
\beq \label{SystW}
F_W (t)=\frac{\left <W(t+\Delta t)-W(t)\right>_W}{2\Delta t} .
\eeq
Here, ${\langle \ldots \rangle }_{W}$ denotes the conditional average that $W(t) = W$, which is applied to determine the functional dependence of the transport coefficients on the energy $W$ (see, e.g., \citet{Ragwitz2001}).  $\Delta t$ must be a small time interval, which should just be large enough so that most particles show measurable changes of the energy over the time interval $ \Delta t$ (theoretically the limit ${\rm{\Delta }}t\to 0$ would apply).
\begin{figure}
\includegraphics[width=8cm]{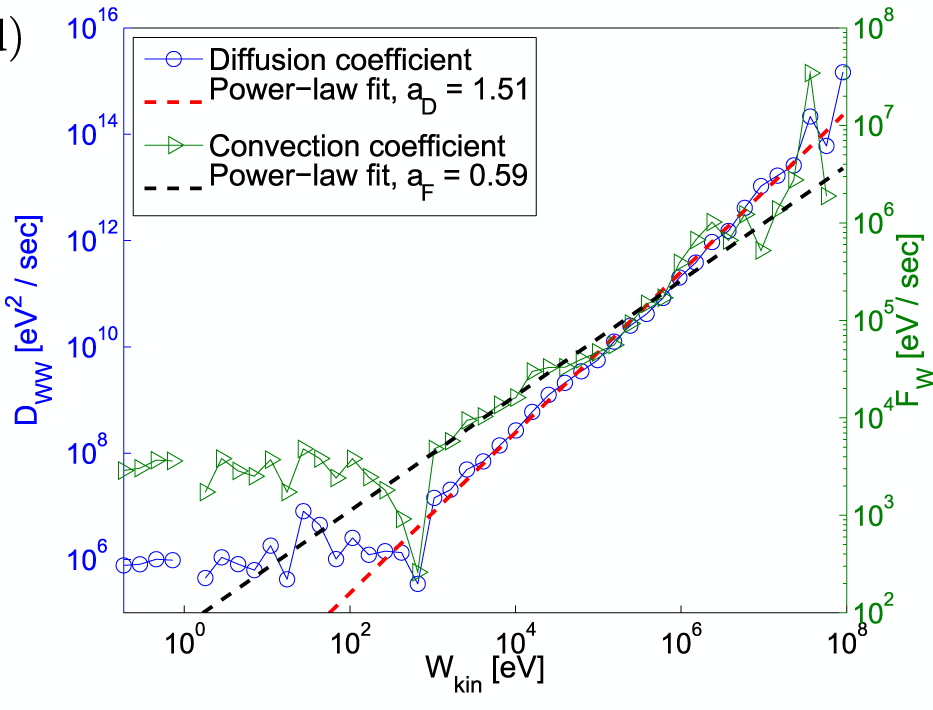}
	\caption{ Energy diffusion and convection coefficients as functions of the kinetic energy at time $t=20\,{\rm{s}}$. Reproduced with permission from Pisokas et al., Astrop. J. {\bf 835}, 214 (2017), Copyrigt 2017 AAS.  }\label{pisf4}
\end{figure}

In Fig.\ \ref{pisf4}, the diffusion and convection coefficients at $t=20\,{\rm{s}}$, as functions of the energy, are presented. The estimate of the coefficients is based on Eqs.\ (\ref{DifW}) and (\ref{SystW}), with $\Delta t$ small, whereto the energy of the electrons is monitored at a set of many regularly spaced monitoring times ${t}_{k}^{(M)}$, $k = 0, 1, ..., K$, with $K$ typically chosen as 200, where $t^{(M)}_0=0$ is the initial time and $t^{(M)}_K$ is the final time of interest. One then uses $t={t}_{k-1}^{(M)}$, ${\rm{\Delta }}t={t}_{k}^{(M)}-{t}_{k-1}^{(M)}$ in the estimates of the coefficients. Also, to account for the conditional averaging in Eqs.\ (\ref{DifW}) and (\ref{SystW}), the energies $W{\left({t}_{k-1}^{(M)}\right)}_{i}$ of the particles are divided into many logarithmically equispaced bins and the requested averages are performed separately for the particles in each bin. As Fig. \ref{pisf4} shows, both transport coefficients exhibit a power-law shape for energies above 1 keV, $D_W \sim  W^{1.51}$, and $F_W \sim W^{0.59}$. These results agree with the estimates reported from the hard-sphere model of \citet{Parker58} and \citet{Ramaty79}.

In order to validate the appropriateness of the Fokker-Planck equation, we insert the transport coefficients into the Fokker-Planck (FP) equation (Eq.\ (\ref{FokkerPVelLoss})) and solve it numerically, using the same initial condition and final time as in the random walk simulation. Additionally, we include the escape term, setting $t_{esc} = 8\,$s, which represents the median value from Fig.\ \ref{pisf21}b, as mentioned above. The transport coefficients are applied in the form of the power-law fit above $1\,$keV and set to a constant below $1\,$keV to ensure continuity at the transition. For the integration of the FP equation on the semi-infinite energy interval, we employ the pseudospectral method, based on an expansion in terms of rational Chebyshev polynomials in energy space, combined with the implicit backward Euler method for the time-stepping (see, e.g., \citet{Boyd2001}).

\begin{figure}[ht!]
\includegraphics[width=8cm]{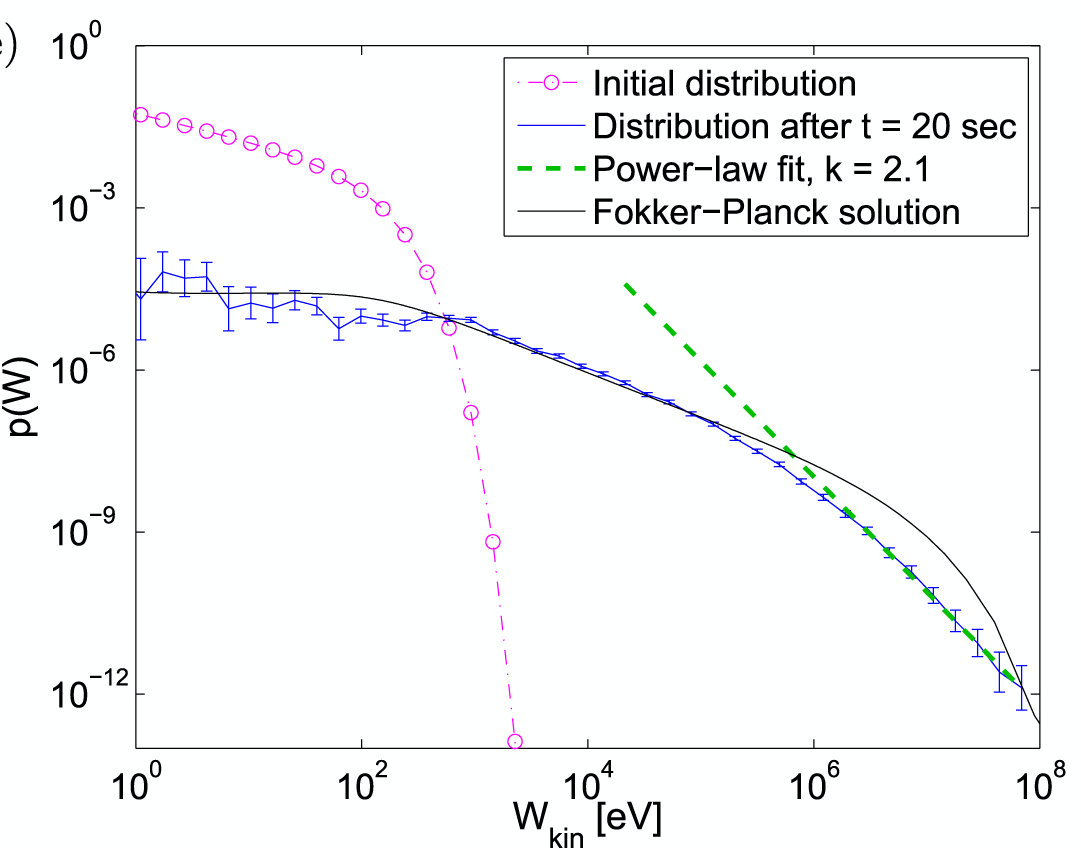}
	\caption{Energy distribution at $t=0$ and $t=20\,$s  for the electrons remaining inside the box, and the corresponding solution of the FP equation. Reproduced with permission from Pisokas et al., Astrop. J. {\bf 835}, 214 (2017), Copyrigt 2017 AAS.  }\label{pisf22}
\end{figure}

The resulting energy distribution at the final time is shown in Fig.\ \ref{pisf22}, which matches well with the distribution obtained from the electron simulation in the low and intermediate energy range, corresponding to the heating of the population (see also Eq.\ (\ref{SFHS4}), as derived from the analytical solution of the FP equation by \citet{Miller90}). The FP solution, however, cannot reproduce the power-law tail. The presence of the diffusive term in the FP equation does not allow a power-law tail to build up, in contrast to the purely convective case presented in Sec.\ \ref{FP_power_law}, Eq.\ (\ref{highenergyslope}).

The heating and the power-law tail in the energy distribution function form almost from the start of the simulation and reach their asymptotic shape on a timescale comparable to the acceleration time \cite{Pisokas16}. Initially, the power-law index is approximately 5, gradually decreasing to an asymptotic value of 2.1 over roughly 20 s, nearly twice the acceleration time.

With the considered parameters, the escape time is comparable to the acceleration time ($t_{acc} \sim t_{esc}$), indicating that the density of scatterers, which controls the mean free path, is a critical parameter in the setup used by \citet{Pisokas16}.  

In Eq.\ (\ref{FokkerPVelLoss}), the spatial transport is  represented only by  the mean escape time $t_{esc}(W)$. A careful analysis of the transport properties in a stochastic scattering process was conducted by \citet{Sioulas20}, who studied the spatial mean-square displacement inside a turbulent volume. They found that spatial transport is superdiffusive for high-energy particles (above $1\,$MeV), whereas  transport remains normal for low-energy, non-relativistic, only heated particles (below $10\,$keV). Similarly, the mean square displacement in energy reveals that transport is superdiffusive for high-energy particles and normal for non-relativistic particles. These results are consistent with  the findings presented just above in Fig.\ \ref{pisf22}, where the Fokker-Planck equation's solution agrees well with the numerically estimated energy distribution for the low-energy particles. However, it cannot reproduce the high-energy tail.

\citet{Greco10} conducted a numerical study of Fermi acceleration using a 2D model in which ions experience Fermi-like energization processes through interaction with a synthetic oscillating electromagnetic wave. This wave was carefully designed to simulate magnetic fluctuations (or magnetic clouds) that were randomly distributed within the $x$-$y$-plane (for more details on their numerical scheme, see \citet{Perri07a}). The free parameter in their system was the mean free path between magnetic clouds, $\lambda_{sc}$. Their results demonstrated efficient heating of ambient ions, particularly when the magnetic clouds were densely packed, resulting in a smaller mean free path. However, no attempt was made in their study to estimate transport coefficients, or  to analyze the stochastic Fermi acceleration characteristics for high-energy particles, such as the power-law index in the energy distribution, the acceleration time, or the evolution of the mean energy.

\subsubsection{Stochastic Fermi Energization within a Fractal Network of CoSs}\label{NumFractal1}

In the previous section, we followed Fermi's initial article \cite{Fermi49}, assuming that the scattering centers (SoCs) are uniformly distributed within the turbulent volume and that the mean free path, $\lambda_{sc}$, is constant. However, it is well known that large-scale magnetic disturbances and coherent structures in fully developed magnetohydrodynamic (MHD) turbulence exhibit a fractal or multifractal  structure, as observed in both space and laboratory plasmas \citep{Tu95, Marsch97, Shivamoggi97, Biskamp03, Dimitropoulou13, Leonardis13, Schaffner15, Isliker19}.

\begin{figure}[!ht]
     \begin{center}
         \includegraphics[width=0.4\textwidth]{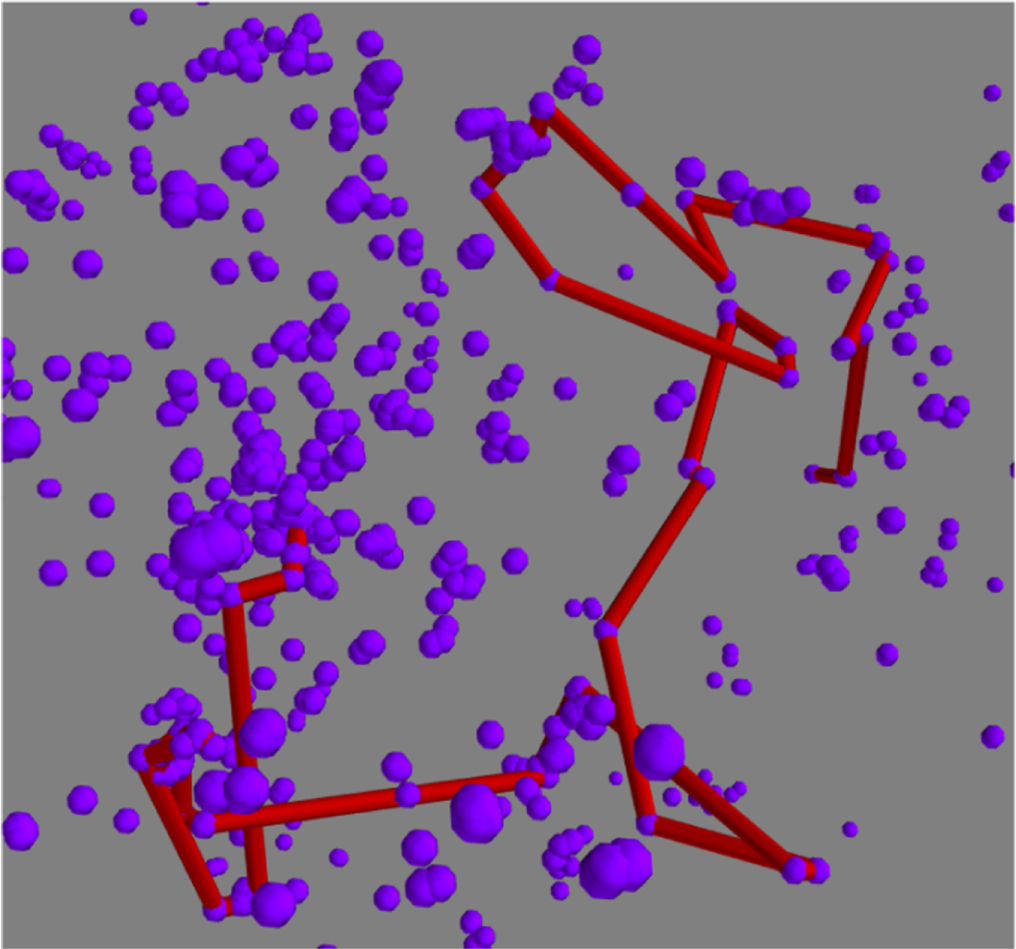}
     \caption {Illustration of the random walk through a fractal environment: Part of the fractal, with its constituent elementary volumes in blue color, and the orbit of a particle in red color, moving along straight paths and occasionally scattering off elementary volumes of the fractal. Reproduced with permission from Sioulas et al., Astrop. J. {\bf 895}, L14 (2020), Copyrigt 2020 AAS.  }\label{orbit2}
     \end{center}
     \end{figure}

In fully developed strong turbulence, coherent structures are distributed on a fractal set with a fractal dimension $D_F$, meaning that $\lambda_{sc}$ cannot be assumed constant \cite{Isliker03}. To address this, \citet{Isliker03} analyzed the behavior of particles undergoing a random walk in a 3D environment containing a natural fractal. Here, the fractal is not a purely mathematical abstraction but it is instead composed of finite, elementary volumes (the scattering centers). 
As considered by \citet{Isliker03}, natural fractals are finite (though typically large) and exhibit a clear fractal scaling from the overall  size down to its smallest constituent volumes.
Particles move freely through the space surrounding the fractal until they collide with its structure, where scattering occurs. Notably, the particles traverse the fractal rather than moving along it.
This dynamic is illustrated in Fig.\ \ref{orbit2}, depicting the random walk across such fractals.

\citet{Isliker03} derived the probability density function (PDF) $p_F(dr)$ for the distances $dr$ a particle travels between successive encounters with the fractal. The analysis assumes that a particle initially resides on a part of the fractal and then moves freely in a random direction until it collides with another fractal segment. For fractals with a dimension $D_F$ less than 2 (the case of interest here), the PDF is well approximated by a power-law:
\begin{equation} 
p_F(dr) = A dr^{D_F-3},
\label{fracstep} 
\end{equation}
where $A$ is a normalization constant determined by the fractal's overall size and the properties of its elementary volumes. For $D_F < 2$, the power-law index of $p_F(dr)$ lies in the range $-3 < D_F - 3 < -1$, indicating that the PDF shares the same asymptotic (large $dr$) behavior as stable Lévy distributions, see \ref{CTRW_Levy_real}. Consequently, particles occasionally undergo large spatial jumps or "Lévy flights", leading to anomalous spatial transport, as discussed in Sec.\ \ref{anomdef}. An earlier Monte Carlo simulation study that used Eq.\ (\ref{fracstep}) and investigated particle energization by fractally distributed electric fields was made in \citet{Vlahos04}.

Based on the results of \citet{Isliker03}, \citet{Sioulas20b} constructed a 3D simulation box with a linear size of $L = 10^{10}$cm and initialized it by uniformly distributing $10^{6}$ particles throughout the acceleration volume. At time $t=0$, the particle energy distribution was set to a Maxwellian with temperature $T$. Each particle was then allowed to perform a free flight of length $dr_i^{(j)}$, determined by the PDF in Eq.\ (\ref{fracstep}), before colliding with a scatterer (i.e., undergoing an energization event). During each collision, particles gained or lost energy stochastically, according to Eq.\ (\ref{Fermi1}). The scatterers were assumed to form a fractal set with a dimension $D_F = 1.8$ in the simulation, based on the analysis of \citet{Isliker19}. Using Eq.\ (\ref{fracstep}), the step length $dr_i^{(j)}$ for each particle followed the probability density $p_F(dr) \sim dr^{-\gamma}$, where $\gamma = 1.2$.
\begin{figure}[!ht]
     \begin{center}
         \includegraphics[width=0.4\textwidth]{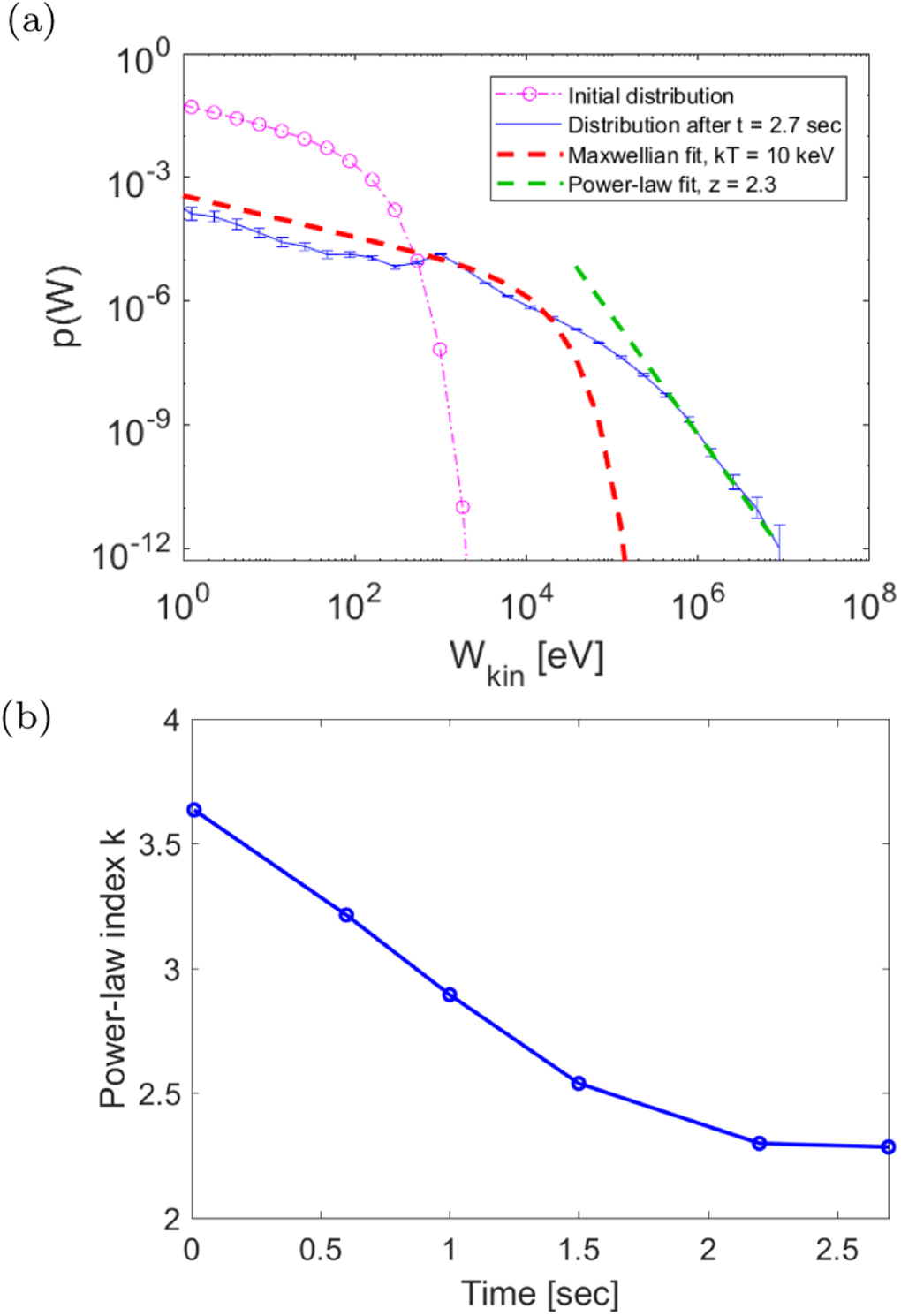}
     \caption { (a) Kinetic energy distribution at $t = 0$ and $t = 2.7 s$ (steady state) for the electrons remaining inside the box with size $L = 10^{10} cm$, together with a Maxwellian fit at low energies and a power-law fit at high energies. (b) Temporal evolution of the power-law index of the kinetic energy distribution's tail. Reproduced with permission from Sioulas et al., Astrop. J. {\bf 895}, L14 (2020), Copyrigt 2020 AAS. }\label{F6Sioulas}
     \end{center}
     \end{figure}
The spatial step lengths $dr_i^{(j)}$ ranged from a minimum value of ${\lambda_{sc, \text{min}}} = 10^{2}$ cm to a maximum of $\lambda_{sc, \text{max}} = 10^{10}$ cm, encompassing the full range of scales from the kinetic to the MHD regime. The lower limit corresponds to several ion gyroradii, while the upper limit matches the size of the acceleration box. Notably, the results were found to be insensitive to the exact values of $\lambda_{sc, \text{min}}$ and ${\lambda_{sc, \text{max}}}$, provided that the condition ${\lambda_{sc, \text{min}}} \ll {\lambda_{sc, \text{max}}}$ was satisfied. 
%\heinz{
It is important to stress here once again that CoSs in MHD turbulence form a well-defined fractal set, see e.g.\ the brief discussion in \citet{Biskamp03}, p.150, who otherwise mostly analyzes intermittency in turbulence.
%}

 As a result of the fractal scattering environment, particles occasionally undergo ``long flights," traveling large distances in a single step --- sometimes spanning almost the entire system --- before encountering a scatterer. Using standard parameters for the low solar corona (see \ref{stochasticFermi2}), simulations show that an initial Maxwellian electron distribution with a temperature of $100\,$eV can be rapidly heated to around $10\,\mathrm{keV}$. The high-energy tail of the distribution follows a power law with an index of approximately $-2.3$ (see Fig.\ \ref{F6Sioulas}a). This high-energy tail begins at around $100\, \mathrm{keV}$ and extends to about $10\, \mathrm{MeV}$. Notably, the index of the power-law tail depends on the system size and matches observed values in the low solar corona for realistic system sizes.

The heating and acceleration process occurs on a very short timescale, approximately $2\, \mathrm{s}$ (see Fig.\ \ref{F6Sioulas}b). This rapid acceleration is due to the trapping of particles within small-scale regions of the fractal environment, which significantly reduces the free path between scattering events. Small-scale activity facilitates the efficient extraction of particles from the thermal pool \cite{Pezzi22}.

The mean square displacement in space and energy is superdiffusive for high-energy particles, consistent with the anomalous transport properties expected in an energetically active fractal environment.

\subsubsection{Stochastic heating within a network of random Electric Fields}

\begin{figure}[!ht] \begin{center} \includegraphics[width=0.3\textwidth]{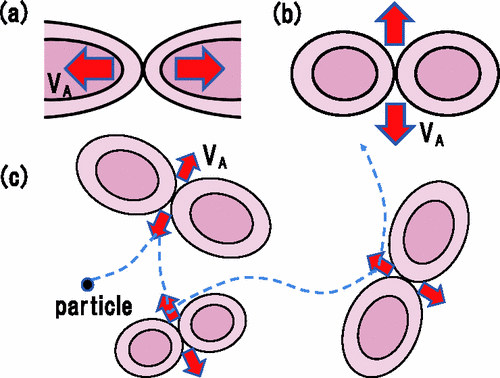} \caption{The interaction of an energetic particle with electric fields arising from the interaction of a network of magnetic islands. Black lines represent magnetic field lines, red arrows denote outflows traveling at the Alfvén speed $V_A$, and the blue dashed line illustrates the trajectory of the energetic particle. Reproduced with permission from Hoshino, Phys. Rev. Lett., {\bf 108}, 135003 (2012). Copyright 2012 APS.}\label{RandomE} \end{center} \end{figure}

The interaction between moving CoSs (e.g., eddies, filaments, or other structures) can generate localized convective electric fields that are randomly distributed in space (see Fig. \ref{RandomE}).

When a particle interacts with the randomly distributed effective electric field $E_{eff}$, the amount of energy change is expressed as
\begin{equation}\label{Eeff} 
	\delta W = |q| E_{eff}\ell_{eff} , 
\end{equation}
where $\ell_{eff}$ represents the effective acceleration length of the scatterer.

The energy change can result in either an energy gain or loss, depending on the particle’s velocity direction relative to the direction of the effective electric field. To model this stochastic acceleration process, two key probability density functions (PDFs) must be defined:
\begin{itemize}
   \item  $P(\ell_{eff})$: This PDF characterizes the distribution of the effective acceleration lengths, $\ell_{eff}$, of the scatterers. It is commonly assumed to have an extended power-law tail \citep{Zhdankin13}.
    
    \item $P(E_{eff})$: This PDF describes the effective electric field acting on a particle during its interaction with the scatterer. \\
\end{itemize}
As no prior studies have definitively established the exact form of the PDF $P(E_{eff})$, \citet{Sioulas22c} explored a range of distributions to assess their impact on the stochastic heating mechanism. They found that a Maxwellian energy distribution emerges under certain conditions when particles interact stochastically with the scatterers.

\begin{figure}[!ht] 
	\centering \includegraphics[width=0.95\columnwidth]{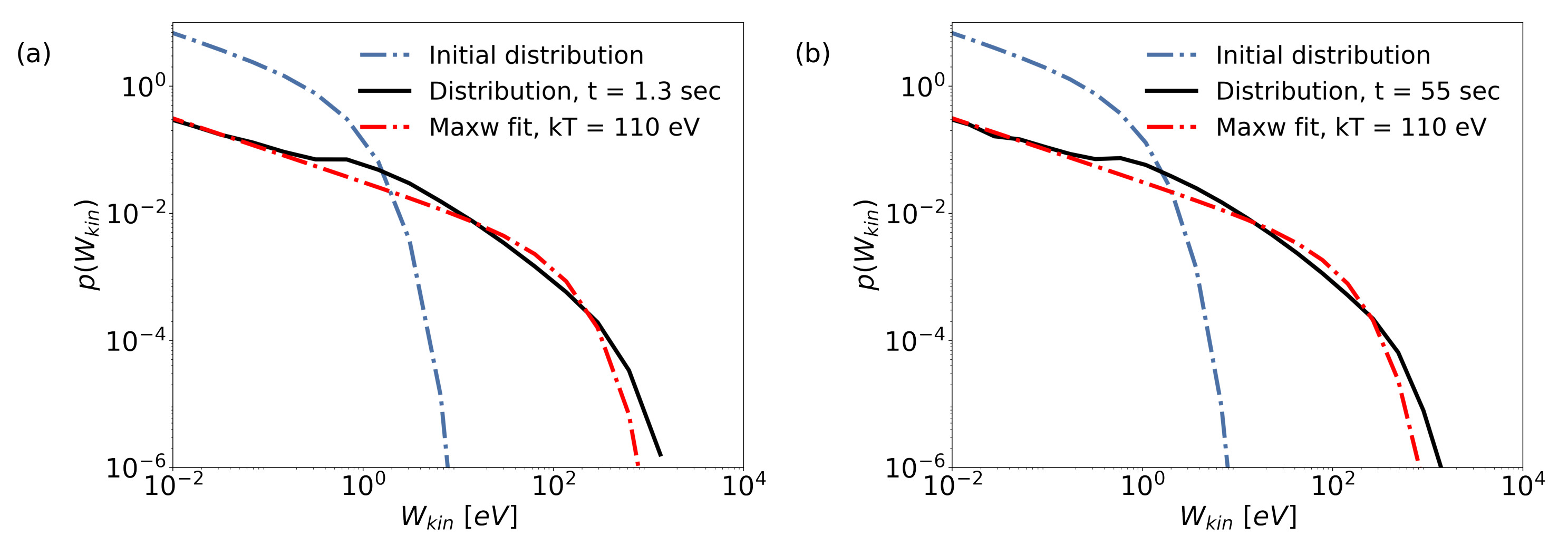} 
	\caption{ Steady-state kinetic energy distributions for non-reconnecting CSs, where $E_{eff}$ follows a Gaussian distribution with a mean of ${\mu\ {=}\ 10^{2} {\cdot} E_{D}}$ and a standard deviation of ${\sigma\ {=}\ E_{D}}$. (a) Electron and (b) proton distributions are shown alongside Maxwellian fits with a temperature $T \ = \ 110\ eV$ in both cases. Reproduced with permission from Sioulas et al., Astronomy and Astrophysics, {\bf 657}, A8 (2022). Copyright 2022 ESO.}\label{stochacc} 
\end{figure}

A significant distinction can be drawn between the stochastic interactions described here and the mechanisms explored  earlier  in stochastic Fermi energization, which  models energy gains  caused by moving magnetic disturbances. In contrast to Fermi acceleration, the interactions governed by Eq.\ (\ref{Eeff}) lack a systematic component (such as the term $V^2/c^2$ in Eq.\ (\ref{Fermi1})). This absence precludes the formation of a power-law tail in the energy distribution, as illustrated in Fig.\ \ref{stochacc}.

\subsection{Systematic energization by CoSs in a strongly turbulent plasma}

Let us assume that the CoSs (acting as scatterers) are not ``magnetized clouds" but rather Current Sheets (CSs). Numerous numerical studies have demonstrated that networks of CSs naturally arise in strongly turbulent systems \citep{Matthaeus86, Dmitruk04, Arzner04, Onofri04, Kowal11, Gordovsky11}. The statistical properties of CSs within 2D and 3D turbulent reconnection environments have been thoroughly analyzed in prior studies \citep{Uritsky10, Servidio11, Zhdankin13}.

It is crucial to distinguish between particle dynamics within evolving CSs and the broader behavior of particles traveling through the global environment, where they encounter a network of CSs acting as scatterers. This scenario is analogous to the magnetic clouds envisioned by Fermi (1949). Particles travel a characteristic distance, $\lambda_{sc}$, before interacting with a CS. During this interaction, they gain or lose an amount of energy, denoted as $\D W$.

\begin{figure}[!ht]
\centering
	\includegraphics[width=0.7\columnwidth]{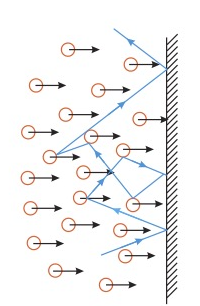}
	\caption{Systematic particle acceleration  throug a moving wall.}
    \label{FermiII}
\end{figure}

The energy increment, $\D W$, can become systematic if it is consistently positive ($\D W > 0$) during particle interactions with CSs or shocks \citep{Dahlin15, Guo15, Matsumoto15} (see Fig.\ \ref{FermiII}).

\subsubsection{Systematic acceleration within a network of   current sheets}

%\heinz{
The complex dynamics of particles in reconnecting current sheets (CSs) arises from the fragmentation process acting during the formation of the latter. It is essential to revisit the question: Why do current sheets fragment in a few Alfven times within highly turbulent plasma? As thoroughly explored in the review by \citet{Vlahos23}, the key reason lies in the current sheets' exposure to strong turbulence during their formation, which induces the emergence of Coherent Structures (CoSs) and subsequently drives them into turbulent reconnection \cite{Lazarian09}. A similar process occurs with shocks when they form within a highly turbulent plasma \cite{Pollock22}.
%}

\begin{figure}[!ht] \centering \includegraphics[width=0.7\columnwidth]{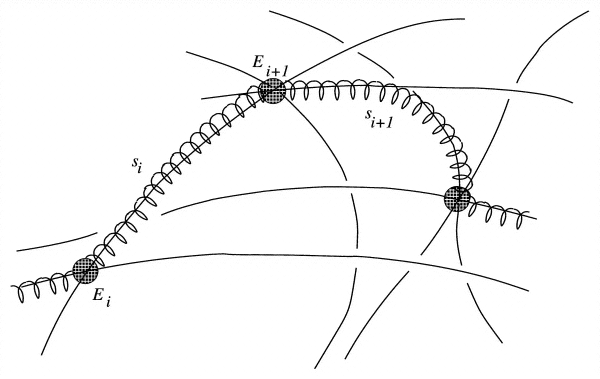} \caption{Sketch of the basic elements of the considered model. A particle (spiraling line) follows magnetic field lines (solid lines) while undergoing drifts and travels freely over a distance $s_i$ until it enters a CS (filled circles). Inside the CS, the associated effective DC electric field, $E_i$, accelerates the particle. After the acceleration event, the particle resumes free travel until it encounters another CS. Reproduced with permission from Vlahos et al., Astrophysical Journal, {\bf 608}, 540 (2004). Copyright 2004 AAS.}\label{Levywalk2} \end{figure}

Particles trapped inside a shortly lived CS can experience energy gain and loss on microscopic scales while interacting with CS fragments (see Fig.\ \ref{Levywalk2}).  Yet, particles systematically gain energy as they traverse a CS as a whole before eventually exiting, as illustrated in Figure 6(c) of \citet{Guo15} and discussed therein. This topic will be revisited in Section \ref{PIC simulations_CS}.

\begin{itemize}
    
\item The energy gain of a particle crossing a reconnecting CS can be approximated as
\begin{equation} \label{SystElec} 
\Delta W = |q| E_{eff} \ell_{eff}, 
\end{equation}
where $E_{eff}\sim (V/c)dB_{eff}$  (see Eq.\ (\ref{Econv})) represents the effective electric field of the CS, with $dB_{eff}$  the fluctuating magnetic field encountered by the particle. The characteristic interaction length $\ell_{eff}$ is proportional to $E_{eff}$, since a smaller $E_{eff}$ corresponds to smaller-scale CSs. Thus, a reconnecting CS acts as a dynamic structure that consistently accelerates charged particles as they cross it.

In the model considered here, particles approach scatterers (CSs) with an initial kinetic energy, $W_0$, and leave with a final energy, $W=W_0+\D W$. At the macroscopic level, the energy gain $\D W$ is systematic and depends on the statistical properties of the magnetic fluctuations $dB_{eff}$.

\item Systematic Fermi Acceleration at Contracting Islands

The systematic macroscopic energy gain caused by particle acceleration at contracting magnetic islands in a CS is a variant of classical Fermi acceleration (see Eq.\ (\ref{Fermi1})). For relativistic particles, considering only head-on collisions, the energy gain is always positive and can be expressed as
\begin{equation} \label{SystemColl1} 
\Delta W = 2 \Gamma^2 \left[ \left( \frac{V_A}{c} \right)^2 + \frac{V_A v_{||}}{c^2} \right] W . 
\end{equation}
$v_{||}$ is the component of the particle’s velocity parallel to the magnetic field, $V_A$ is the Alfv\'en speed, and $\Gamma$ is the Lorentz factor. For non-relativistic  Alfv\'en speeds $(V_A<<v<c$), the energy gain simplifies to
\begin{equation} \label{systemcol2} 
\frac{\Delta W}{W} \sim \left( \frac{2V_A}{c} \right) \left( \frac{v_{||}}{c} \right) \sim \left( \frac{2V_A}{c} \right) \left( \frac{v}{c} \right) \cos \phi, 
\end{equation}
where $\phi$ is a random angle in the range $-\pi/2<\phi < \pi/2$ \citep{DelPino05, Drake06, Hoshino12, Dahlin15, Guo15},
and the mean energy gain is
\begin{equation} \label{FF2} 
	\frac{\langle \Delta W \rangle}{W} = \frac{4}{3} \frac{V v}{c^2} . 
\end{equation} 

\end{itemize}

A critical distinction exists between energy gain through the effective electric field in CSs (Eq.\ (\ref{SystElec})) and Fermi acceleration at contracting islands (Eq.\ (\ref{FF2})). In the former, $\D W$  is independent of the particle's initial energy $W$, in contrast to the latter case.   Additionally, neither mechanism has significant energy dispersion, as both describe systematic energy gain without broadening (heating) the particle energy distribution.

\subsubsection{Numerical studies of systematic energization by CoSs}\label{systematicFermi1}

\begin{figure}[!ht]
     \begin{center}
         \includegraphics[width=0.5\textwidth]{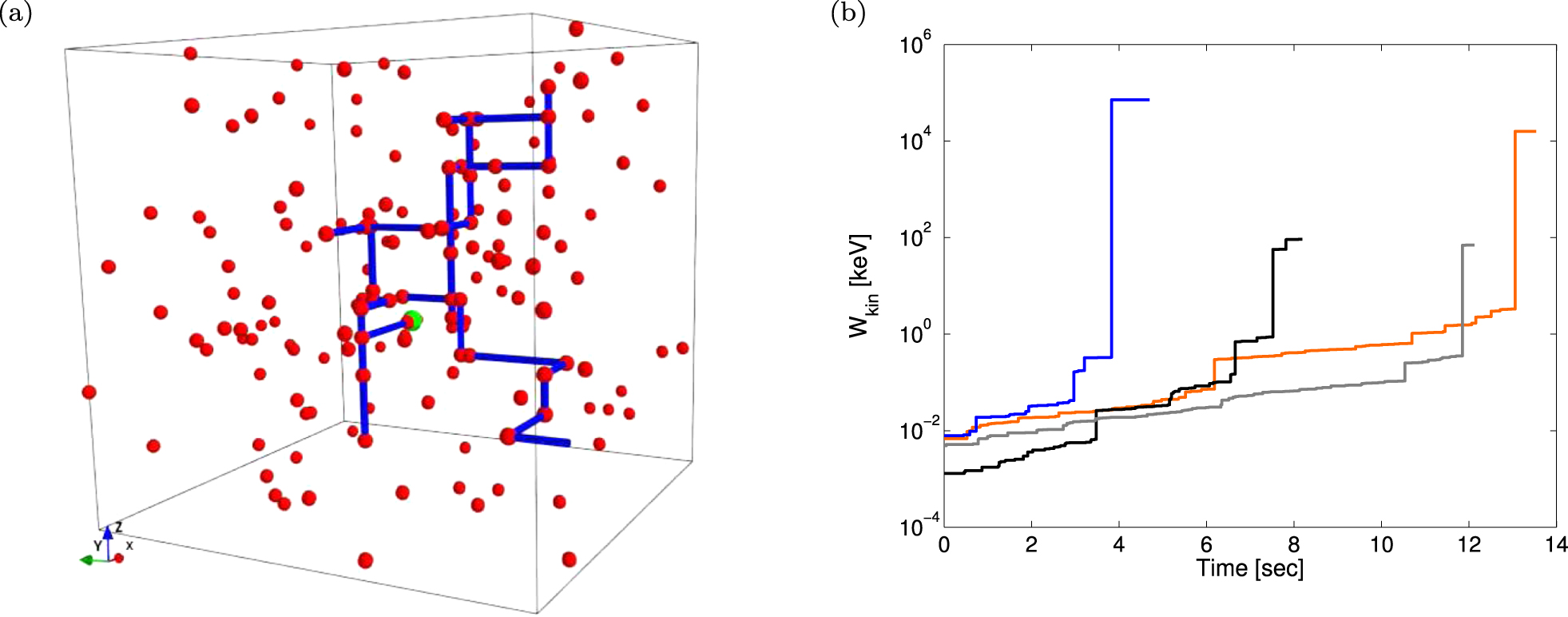}
     \caption { (a) Typical trajectory in the 3D simulation box (blue). The red spheres mark the randomly distributed CSs (active grid sites). (b) Energy as a function of time for a few selected particles.Reproduced with permission from Isliker et al., Astrop. J. {\bf 849}, 35 (2017), Copyrigt 2017 AAS. \cite{Isliker17}}\label{IslikerSystem1}
     \end{center}
     \end{figure}

\citet{Isliker17} employed a lattice gas model to simulate turbulent reconnection using a 3D grid of size $N \times N \times N$, with a grid spacing of $\ell=L/(N-1)$ and a total linear extent L. The setup mirrors the one used in Section \ref{stochasticFermi2} for stochastic energization (see Fig. \ref{IslikerSystem1}a).

According to Eq.\ (\ref{SystElec}), each scattering event at an active grid point increases the particle's energy by $\D W=|q|[(V_A/c)\delta B]\ell_{eff}$. The parameters chosen in the study correspond to plasma conditions in the low solar corona (for further details, see \citet{Isliker17}).

The grid in \citet{Isliker17}'s model is open, allowing particles to escape the acceleration region upon reaching any boundary. In position space, the particles perform a classical random walk on the grid. However, in energy space, the dynamics take the form of a systematic random walk, as energy increments $\D W$ are always positive (see Fig. \ref{IslikerSystem1}b). For a detailed discussion of the energy gain mechanism within the CS, refer to \cite{Dahlin15, Guo15, Matsumoto15, Isliker17}.

\begin{figure}[!ht] 
     \begin{center}
         \includegraphics[width=0.45\textwidth]{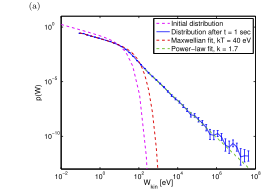}\\
         \includegraphics[width=0.45\textwidth]{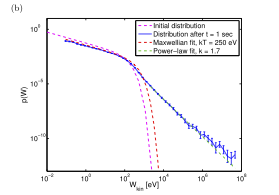}     \caption {Energy distribution at $t = 0$ and $t=1\,{\rm{s}}$ of the electrons that have remained inside the box, for an initial temperature of (a) $10\,$eV and (b) $100\,$eV. Reproduced with permission from Isliker et al., Astrop. J. {\bf 849}, 35 (2017), Copyrigt 2017 AAS. \cite{Isliker17}}\label{Isliker17}
     \end{center}
     \end{figure}

\citet{Isliker17} monitor the electron population, injected at t = 0, until half of it has escaped, which happens at ${t}_{1/2}=4.5\,{\rm{s}}$. The electrons inside the box at $t=1\,$s are distributed as shown in Fig. \ref{Isliker17}. The energization process heats the low-energy particles below one keV, where the initial distribution follows a Maxwellian with a temperature of $40\,$eV or $250\,$eV, respectively, depending on the initial temperature. The high-energy part of the distribution exhibits a clear power-law tail with an index of $k\approx 1.7$, which extends from approximately 1 keV to 100 MeV. The power-law tail is formed in a few milliseconds and persists even when more electrons escape from the acceleration volume. This implies that the acceleration process is extremely rapid, and most of the dissipated energy will be transferred to the accelerated tail. 

%\heinz{
At this point, it is worthwhile to comment on the limitations of test particle and Monte Carlo simulations. 
First, we should clearly distinguish between Monte Carlo (random walk) simulations and test-particle simulations. In test-particle simulations, the particles move in the electromagnetic fields of an MHD simulation, without giving feed-back to the electromagnetic fields of the MHD simulation. In this case, the available MHD energies are known, and the test-particles can and should be stopped before they exceed the potentially available MHD energy. Still though, the absence of self-consistency implies that the results for the maximum energy reached and the power-law tail index are only qualitatively correct but not very accurate \cite{Pugliese25}. An interesting discussion of this problem is given in  Onofri et al. \cite{Onofri06}, who conclude that an excessive energy gain of test-particles in an MHD simulation could also be interpreted as the breakdown of the MHD approach, such that kinetic modeling would ultimately be the appropriate tool, if technically feasible. 

In random walk models, the situation is less clear, and particles can gain more energy than they would in a more complete modeling approach. Random walk models thus are more explorative and indicative for a certain energization mechanism than strictly quantitative.
In this review, we present both, random walk and test-particle simulations, and the case of the simulation presented here, a lattice gas model, belongs to the random walk approach.
%}

\subsubsection{Classical and Fractional approach to model transport when particles are systematically energized \label{systematic_fractional}}

As discussed in Sec.\ \ref{normalDiff1}, the validity of the Fokker-Planck (FP) equation relies on the assumption that particles execute Brownian motion in energy space as they interact with localized CSs. This model presumes small random-walk steps with finite mean and variance. However, the systematic acceleration observed in Fig.\ \ref{Isliker17}b deviates from this framework, raising questions about the validity of the FP equation. Notably, Lévy walks in energy, which lead to anomalous transport, are observed (see Sec.\ \ref{anomdef}). Using Eqs.\ (\ref{DifW}) and (\ref{SystW}), \citet{Isliker17} estimate the transport coefficients $F(W,t)$ and $D(W,t)$ based on the trajectories obtained from the lattice gas model simulations, they insert the coefficients into the FP equation (Eq.\ (\ref{FokkerPVelLoss})) and solve the latter numerically, as previously outlined in Sec.\ \ref{stochasticFermi2}.

\begin{figure}[!ht] \begin{center} 
		\includegraphics[width=0.4\textwidth]{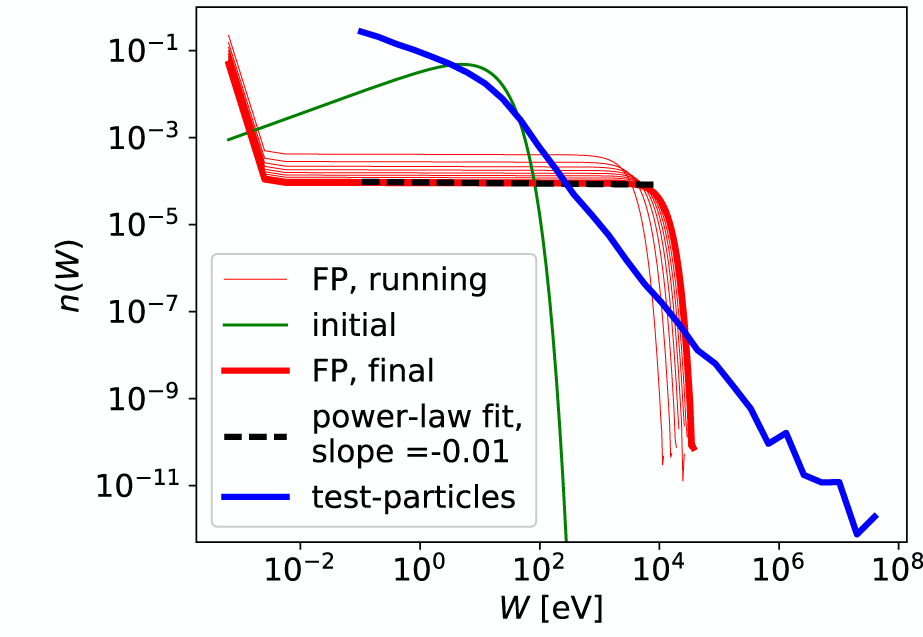} \includegraphics[width=0.4\textwidth]{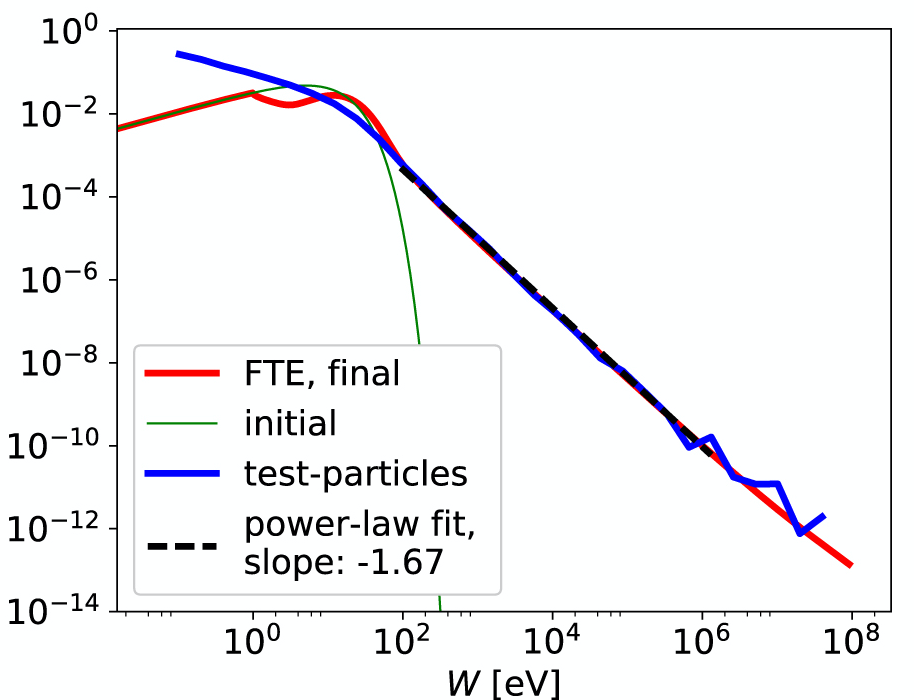} 
		\caption{(Top) Solution of the FP equation compared with the energy distribution from the lattice gas simulation. (Bottom) The fractional transport equation (FTE) solution compared with the energy distribution from the lattice gas simulation. Reproduced with permission from Isliker et al., Astrophysical Journal, {\bf 849}, 35 (2017). Copyright 2017 AAS \cite{Isliker17}.}\label{IslikerSystem3} \end{center} \end{figure}

The results are shown in Fig.\ \ref{IslikerSystem3} (top), where the energy distribution from the lattice gas model is compared to the FP solution. A significant discrepancy emerges: the FP solution is flat across the entire energy range, whereas the particle distribution exhibits a decaying, extended power-law tail. This mismatch suggests that the FP equation fails to  describe this system's transport dynamics.

In Sec.\  \ref{anodi}, we discussed anomalous transport and the derivation of the Fractional Transport Equation (FTE), a powerful tool for describing non-local and anomalous transport phenomena. This motivates whether the FTE can serve as a more accurate model for acceleration at reconnecting CSs or contracting islands. The order of the fractional derivative $\alpha$ in the FTE (Eq.\ (\ref{fraceq}), Sec.\ \ref{fracdiffeq}) can be determined from the power-law index $z$ of $p_w(w)$, the distribution of energy increments $w$ in the random walk (see Fig.\ 7(c) and 8(c) in \citet{Isliker17}), $\alpha=-z-1$. Alternatively, $\alpha$ can be calculated with the characteristic function approach\cite{Koutrouvelis1980,Borak2005}, which also allows determining the parameter $a$. The temporal derivative is assumed to be of ordinary first order, as in Eq.\ (\ref{fraceqexam}), since the energy increments are monitored over fixed and constant time-intervals $\D t$, which implies that $b=\D t$, see Sec.\ \ref{symmLevy}. We note that all the parameters of the FTE, Eq.\ (\ref{fraceqexam}), are determined from the lattice gas simulation data, no hypotheses are introduced. The FTE is solved numerically 
as described in Sec.\ \ref{stron_turb} below
%by using the Gr\"unwald–Letnikov definition of fractional derivatives (e.g., \citet{Kilbas2006}), implemented in the matrix form of \citet{Podlubny2009} 
(see \citet{Isliker17} for more details).

Fig.\ \ref{IslikerSystem3} (bottom) shows the numerical solution of the FTE at $t=0.7\,$s, alongside the particle energy distribution. Unlike the FP equation, the FTE successfully reproduces the power-law  part of the particle energy distribution.

In summary, systematic acceleration is an efficient process for forming power-law energy distributions in laboratory, space, and astrophysical plasmas. In these systems, heating is a minor effect, and transport is anomalous, necessitating using a FTE to describe the dynamics at intermediate to high energies.

\subsection{Synergy of Stochastic and Systematic Energization within a network of  CoSs} \label{SecSynergy}

\begin{figure}[!ht] 
\begin{center} 
\includegraphics[width=0.5\textwidth]{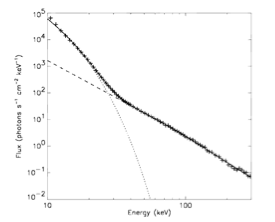} 
\caption{RHESSI X-ray spectrum from a solar flare, showing both plasma heating and acceleration in the high energy tail, which together form a Kappa distribution. Reproduced with permission from Lin et al., Astroph. J.,   {\bf 595}, L69 (2003), Copyrigt 2003 AAS.}\label{Kappa1} 
\end{center} 
\end{figure}

The defining feature of strong turbulence is the combination of intense large-magnitude magnetic disturbances $(\delta B/B \sim 1)$  with randomly distributed current sheets (CSs), which form due to colliding filaments and shocks (see \citet{Vlahos23}). In most explosive laboratory and astrophysical systems, plasma heating and particle acceleration are strongly correlated. For example, plasma heating accompanies particle acceleration in solar flares, as shown by RHESSI X-ray spectra \citep{Lin03}, see Fig.\ \ref{Kappa1}.

Despite this correlation, current studies often treat plasma heating and particle acceleration as separate processes. While particle acceleration is attributed to reconnection or shocks, both these mechanisms have no established process for impulsive heating. The combined effect of stochastic and systematic  energization, on the other hand, produces particle distributions that show very clear features of heating and acceleration, which can also be described as Kappa distributions (for more details see Sec.\ \ref{SecKappa}).

\subsubsection{Numerical studies of a fractal distribution of stochastic and systematic scatterers}

The physics of charged particle energization within a strongly turbulent plasma, where current sheets (CSs) naturally emerge due to large-scale magnetic topologies' evolution, was analyzed by \citet{Sioulas22c}.  Intense magnetic disturbances and CSs are fractally distributed in such environments, forming a complex and dynamic energization framework, see e.g.\ \citet{Vlahos23}. \citet{Sioulas22c} implemented a Monte Carlo simulation to investigate the effects of this fractality on particle energization. Specifically, they examined how the interplay between energization at strong magnetic disturbances and systematic particle acceleration in reconnecting CSs influences key processes, such as plasma heating, power-law high-energy tails, and the characteristic acceleration and escape times of electrons and ions. Their findings highlight the crucial role of current sheets in systematically accelerating particles, thereby contributing significantly to the development of power-law tails in energy distributions. Meanwhile, stochastic particle energization is identified as a primary mechanism for heating, relevant to both the solar corona's steady heating and the impulsive heating observed during solar flares (see Fig.\ \ref{Kappa1}).

The simulation setup in \citet{Sioulas22c} closely follows the one described in Sec.\ \ref{NumFractal1}, with parameters chosen to match those relevant to the solar corona. In this setup, the turbulent plasma environment is populated by two types of scatterers: stochastic and systematic ones. The scatterers' fraction $P$ ($0\leq P \leq 1$) are intense magnetic disturbances, where particles undergo stochastic acceleration. The remaining fraction, $1-P$, are reconnecting CSs, which impart systematic energy kicks to the particles.

At $t=0$, a population of $10^6$ particles is injected into  the environment, with an initial Maxwellian velocity distribution corresponding to a temperature of $T=100\,$eV. The study focuses on the heating and acceleration dynamics of electrons and ions, though the results presented here pertain primarily to electrons. It is noted that while the overall behavior of electrons and ions in this environment is similar, the time scales required to reach steady-state conditions differ due to the mass disparity.  
\begin{figure}
\begin{center}
         \includegraphics[width=0.5\textwidth]{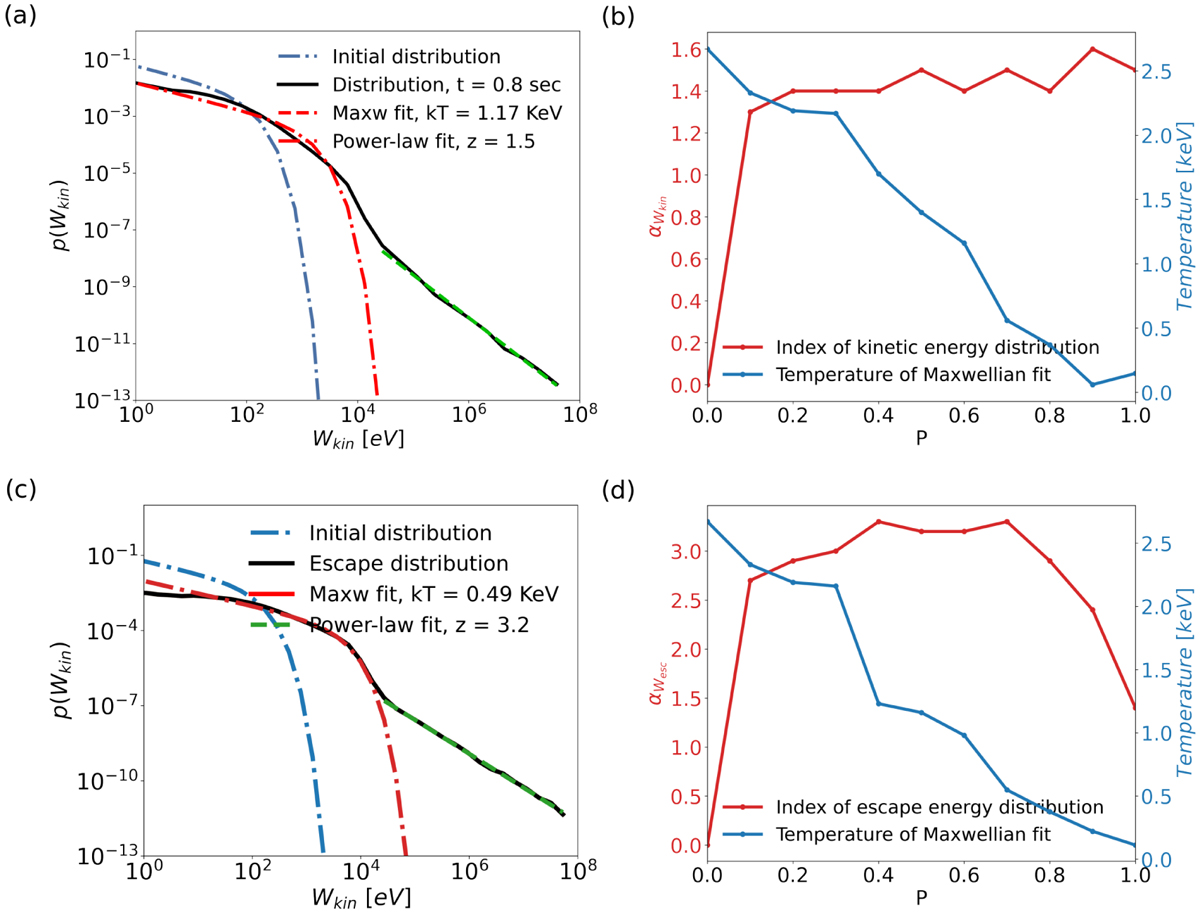}
     \caption {(a) Steady-state kinetic energy distribution, along with a Maxwellian fit of temperature $T \sim 1\,$keV and a power-law fit, for $P=0.5$. (b) Power-law index of the steady-state kinetic energy distribution (blue), and the temperature of the Maxwellian fit (red), for different ratios $P$ of the two kinds of scatterers. (c) Kinetic energy distribution for the escaped particles, for $P=0.5$. (d) Power-law index of the kinetic energy distribution of the escaped particles (blue), and the temperature of the Maxwellian fit (red), for different ratios $P$ of the two kinds of scatterers. Reproduced with permission from Sioulas et al., Astron. Astrophys.  {\bf 657}, A8 (2022), Copyright 2022 ESO. }\label{SioulasAA2}
     \end{center}
     \end{figure}

Since the fraction of scatterers acting as systematic or stochastic in a turbulent environment cannot be expected to remain constant (as it may vary with the system's evolution over time), it becomes crucial to understand how different combinations of these two scatterer types influence the overall energization process. Consequently, in their study, \citet{Sioulas22c} explore a range of cases, starting with $P=0$ (where only stochastic scatterers are present) and gradually increasing the fraction of CSs up to $P=1$ (where  systematic scatterers dominate the system).

Examining the particles' steady-state kinetic energy distributions, \citet{Sioulas22c} observe striking differences between the two extremes. When $P=1$, where the turbulent environment consists solely of reconnecting CSs, the particle energy distribution exhibits significant acceleration, resulting in extended power-law tails with only minimal heating of the lower-energy particles. On the other hand, for $P=0$, where only stochastic scatterers are present, the system produces nearly perfect Maxwellian energy distributions with  increased temperatures, and no discernible power-law tails form. These contrasting behaviors are illustrated in Fig.\ \ref{SioulasAA2}b. 

For intermediate values of $P$, particularly $P=0.5$, the energy distribution reveals a hybrid behavior. As shown in Fig.\ \ref{SioulasAA2}a, the steady-state kinetic energy distribution of the particles remaining inside the turbulent volume at time $t \sim 0.8\,$s features a low-energy Maxwellian component alongside a high-energy power-law tail. The Maxwellian component, corresponding to a temperature $T \approx 1\,$keV, represents the bulk heating of the low-energy particles due to stochastic processes. The high-energy portion, however, deviates from the Maxwellian and forms a power-law tail with index $k=1.5$, extending from approximately $10\,$keV to $100\,$MeV for particles confined within the simulation box. 

The particles that escape from the acceleration volume display a steeper power-law distribution than the confined particles, with an index $k=3.2$, as shown in Fig.\ \ref{SioulasAA2}c. Notably, this power-law tail emerges after only a few milliseconds of evolution and persists even as more particles escape from the turbulent volume.

The combination of stochastic and systematic scatterers, therefore, transitions the behavior of a turbulent environment from a pure systematic accelerator ($P=1$) or an efficient heating mechanism ($P=0$) to a highly effective hybrid mechanism that simultaneously heats and accelerates particles (e.g., $P=0.5$). This dual functionality underscores the importance of the interplay between these two energization processes in turbulent plasma environments. 
A similar pattern emerges when analyzing the energy distributions of the escaping particles, as shown in Figs.\ \ref{SioulasAA2}c,d.

\subsubsection{Diffusive shock acceleration in the presence of strong turbulence}

Diffusive shock acceleration (DSA) alone cannot efficiently accelerate particles without the presence of strong turbulence $(\delta B/B \geq 1)$ in the vicinity of the shock, whether self-generated or pre-existing \cite{Perri22}. \citet{Garrel18} investigated the scenario where large-amplitude magnetic disturbances are present upstream and downstream of a shock, establishing a state where current sheets (CSs) can influence particle dynamics. They showed that these magnetic disturbances and CSs, which spontaneously form in strong turbulence near a shock \cite{Zank15, Matsumoto15, leRoux16}, can facilitate particle acceleration as effectively as the typical shock-crossing mechanism described by DSA, especially in large-scale systems over extended periods.

\citet{Garrel18} start their analysis by assuming the presence of 'elastic' scatterers on both sides of the shock, estimating the energy distribution of particles, with only the classical DSA mechanism at the shock front actively energizing the particles. This initial analysis successfully replicates the expected results for systematic acceleration, as described by classical DSA theory. Then, they consider the additional effect of active scatterers, specifically significant magnetic disturbances and reconnecting CSs, which coexist with the shock discontinuity in its neighborhood. Through this combined interaction, they demonstrate that the resulting energy distribution of accelerated particles closely resembles that predicted by DSA alone. However, the combination of DSA and CSs leads to significantly enhanced acceleration efficiency: acceleration times are reduced by an order of magnitude, and the maximum attainable energy is increased by a factor of two.
\begin{figure}
\begin{center}
         \includegraphics[width=0.5\textwidth]{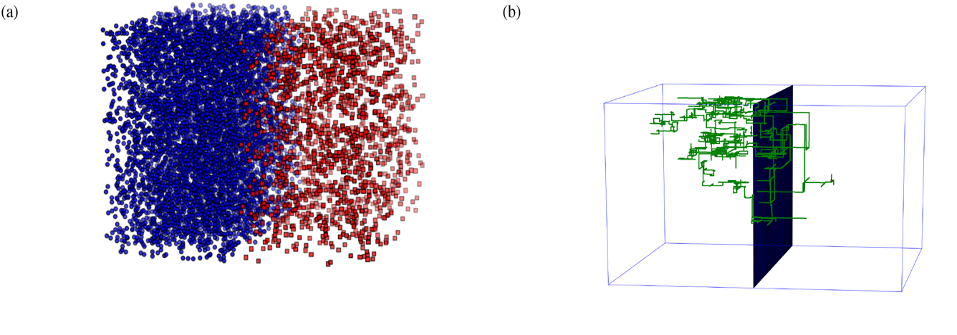}
     \caption {(a) Small version of the simulation box with the planar shock wave in the middle.(b) Trajectory of a typical electron inside the simulation box. Reproduced with permission from Garrel et al., MNRAS {\bf 478}, 2018 (2018), Copyrigt 2018 RAS.}\label{Garrel1}
     \end{center}
     \end{figure}

Figure \ref{Garrel1} illustrates the simulation setup used by \citet{Garrel18}, which consists of a 3D grid $(N \times N \times N)$ with linear size $L$, representing a region split into downstream and upstream areas relative to the shock discontinuity located in the middle plane. The downstream and upstream regions correspond to areas behind and in front of the shock, respectively. The grid is populated with scatterers, with $N_{sc,dwn}$ scatterers downstream and $N_{sc,up} = N_{sc,dwn}/r$ scatterers upstream, where $r$ is the compression ratio across the shock. The fraction of grid points occupied by scatterers is defined as $R_{up/dwn} = N_{sc,up/dwn} / [N^2 \times (N-1)/2]$, and this fraction determines the mean free path ($\lambda_{sc,up/dwn}$) of the particles, which is different on each side of the shock.

The particles injected into the system at $t = 0$ follow a Maxwellian distribution characterized by a temperature $T$, and the subsequent time evolution of the distribution is analyzed. The numerical experiments by \citet{Garrel18} were based on parameters  representing plasma conditions in the upper solar corona, providing realistic insights into shock-particle interactions in space environments.
\begin{figure}
\begin{center}
         \includegraphics[width=0.5\textwidth]{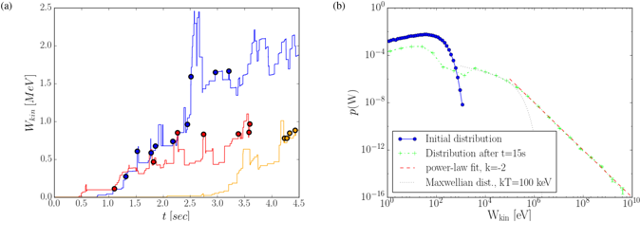}
     \caption {Acceleration of electrons by the synergy of CoSs and DSA: (a) The kinetic energy evolution of typical electrons interacting with the three accelerators. The circles mark the crossings of the shock surface. (b) Initial kinetic energy distribution (blue) and distribution after $15s$ for electrons escaping from the box (green), with a power-law fit  index $k=2$, red), and a Maxwellian distribution fit with temperature $T= 100$ keV (grey). Reproduced with permission from Garrel et al., MNRAS {\bf 478}, 2018 (2018), Copyrigt 2018 RAS.}\label{Garrel3}
     \end{center}
     \end{figure}

 \citet{Garrel18} extend the framework developed by \citet{Pisokas18} by incorporating both DSA and strong turbulence near the shock into their simulations. Scatterers are divided into two classes in these modified simulations: a fraction $P$ are traditional stochastic scatterers, while the remainder $(1 - P)$ are reconnecting CSs. The choice of $P = 0.5$ allows for a balanced investigation of the roles played by each type of scatterer. The resulting energization forms a heated energy distribution with a power law tail (a Kappa distribution, see Sec.\ \ref{SecKappa}) from the injected Maxwellian distribution, as depicted in Figure \ref{Garrel3}, which showcases the kinetic energy evolution of individual particles and the resulting energy distribution after interacting with the shock and its turbulent environment. The synergy between stochastic and systematic acceleration thus becomes evident when the plasma environment reaches a strongly turbulent state \cite{Zank15, leRoux16}.

The results presented by \citet{Garrel18} demonstrate that DSA is initially the dominant mechanism for particle acceleration. However, once CoSs (reconnecting current sheets and large-scale disturbances) are established, strong turbulence is the primary driver of heating and acceleration. The shock triggers a suitable environment for turbulence to take hold \cite{Vlahos23}. In such a strongly turbulent environment, the acceleration process becomes a cooperative mechanism involving DSA and stochastic and systematic components from CSs, leading to enhanced efficiency compared to traditional DSA alone.

\subsubsection{The interplay of stochastic and systematic acceleration and the Kappa distribution}\label{SecKappa}

\begin{figure}[!ht]
\centering
\includegraphics[width=0.8\columnwidth]{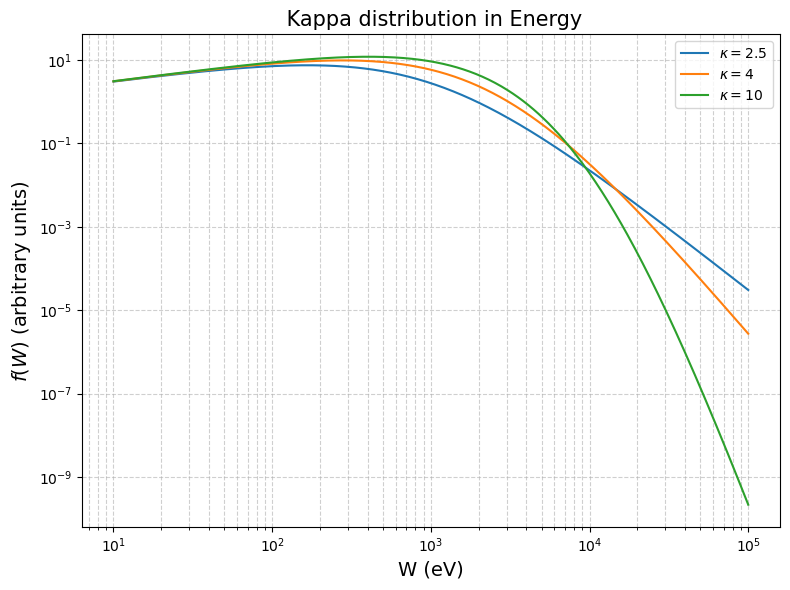}%
\caption{The Kappa distribution as a function of energy $W$, for three different parameters $\kappa$. }
\label{f2}
\end{figure}

The Kappa distribution (or $\kappa$-distribution) is a widely used probability density function in space plasma physics and astrophysics, describing particles' velocity or energy distributions analytically in environments that deviate from thermal equilibrium. Unlike the Maxwell-Boltzmann distribution, which assumes thermal equilibrium and  is derived from the Gaussian distribution of the velocity components, the Kappa distribution accounts for the non-thermal, high-energy tails observed in space plasmas  (see Fig. \ref{f2}) \citep{Summers91, Burlaga06, Decker08, Zouganelis08, Mace09, Pierrard10, Yoon12, Livadiotis13, Bian14, Livadiotis23}.

The Kappa distribution for the velocity $\mb v$ is given as 
	\begin{equation}\label{Kappa_mbv} 
		f_{\kappa}(\mb v) = A_{\mb v} \left( 1 + \frac{m \mb v^2}{(\kappa -3/2) 2 k_B T} \right)^{-(\kappa + 1)} ,
	\end{equation}
	with normalization constant
	\begin{equation}
		A_{\mb v} = \frac{1}{[\pi (\kappa -3/2)]^{3/2}} \cdot \frac{1}{[2k_BT/m]^{3/2}} \cdot \frac{\Gamma(\kappa+1)}{\Gamma(\kappa-1/2)} ,
	\end{equation}
	where $k_B$ is the Boltzmann constant, $T$ the temperature, $m$ the particle mass, $\Gamma$ the Gamma function, and $\kappa > 3/2$ the kappa index that determines the strength of the suprathermal tail.
	
	From the velocity distribution the distribution for the speed $v=|\mb v|$ can be derived as
	\begin{equation}\label{Kappa_v} 
		f_{\kappa}(v) = A_v \left( 1 + \frac{m v^2}{(\kappa -3/2) 2 k_B T} \right)^{-(\kappa + 1)} v^2 ,
	\end{equation}
	with
	\begin{equation}
		A_v = \frac{2}{((\kappa -3/2))^{3/2}} \cdot \frac{1}{[2k_BT/m]^{3}} \cdot \frac{\Gamma(\kappa+1)}{\Gamma(\kappa-1/2) \Gamma{(3/2)}} ,
	\end{equation}
	as well as the distribution for the kinetic energy $W=(1/2)m\mb v^2$,
	\begin{equation}\label{Kappa_W} 
		f_{\kappa}(W) = A_W \left( 1 + \frac{W}{(\kappa -3/2) k_B T} \right)^{-(\kappa + 1)} W^{1/2} ,
	\end{equation}
	with
	\begin{equation}
		A_W = \frac{1}{((\kappa -3/2))^{3/2}} \cdot \frac{1}{[k_BT]^{3/2}} \cdot \frac{\Gamma(\kappa+1)}{\Gamma(\kappa-1/2) \Gamma{(3/2)}}
	\end{equation}
	(see \citet{Livadiotis13}, Eqs.\ (3.8), (3.9), (3.10)). The normalization constants $A_.$ ensure that the total probability integrates to one.

The Kappa  distribution  provides a statistical model for the particle energy distribution in systems with significant suprathermal particle populations. 
 One can define a characteristic energy $W_{\kappa}$ that is related to the system's temperature $T$ by
\[
W_{\kappa} = (\kappa - 3/2) {k_B T},
\]
where, as stated above,  $\kappa > 1.5$, and Eq.\ (\ref{Kappa_W}) can be written as
\begin{equation}\label{Kappa} 
	f_{\kappa}(W) = A_W \left( 1 + \frac{W}{W_\kappa} \right)^{-(\kappa + 1)} W^{1/2} .
\end{equation}

As $\kappa \to \infty$, the term 
${m \mb v^2}/[(\kappa -3/2) 2 k_B T]$ in Eq.\ (\ref{Kappa_mbv}) becomes small. Using 
the limit definition of the exponential function, one finds
 
\begin{equation} 
		\left( 1 + \frac{m \mb v^2}{(\kappa -3/2) 2 k_B T} \right)^{-(\kappa + 1)} 
		\approx \exp\left( -\frac{m \mb v^2}{2 k_B T} \right)
\end{equation}

This approximation shows that, in the limit of large $\kappa$, the Kappa distribution in energy reduces to a Maxwell-Boltzmann distribution,  thus converging to thermal equilibrium conditions.

Stochastic and systematic acceleration inside a strongly turbulent plasma will lead to various distribution functions, ranging from pure acceleration of the high-energy tail to pure heating without a tail. One of the most common energy distributions arising under these conditions is the Kappa distribution \citep{Vasiliunas68, Lin03}, where both, the heating and the acceleration mechanisms, can be represented.

\section{Particle Heating and Acceleration in 3D Coherent Structures from MHD and Kinetic simulations}

The heating and acceleration of particles in space and astrophysical plasmas is a complex, multi-scale phenomenon that necessitates an integrated approach to address both large- and small-scale dynamics. Based on the current literature, this review examines two complementary methodologies widely employed to tackle these extremes: magnetohydrodynamics (MHD) for large-scale plasma behavior and Particle-In-Cell (PIC) simulations for small-scale kinetic processes.

The MHD framework offers valuable insights into the evolution of large-scale coherent structures (CoSs), critical to plasma dynamics. However, its approximations come with notable limitations: MHD cannot adequately resolve the kinetic evolution of energetic particles, nor can it fully account for energy dissipation mechanisms at smaller scales. On the other hand, PIC simulations excel in capturing the kinetic effects within CoSs, accurately modeling particle behavior and local energy transfer processes. Yet, these simulations are constrained by practical  limitations, including simplified initial conditions, the use of periodic boundaries, and limitations in capturing long-term particle evolution and escape dynamics.

Although both, the  MHD and the PIC approach, have demonstrated significant strengths, their limitations highlight a major gap in our understanding of plasma processes across scales. Bridging this gap through directly coupling the two methods remains a substantial challenge. While progress has been made in developing feedback mechanisms between resistive MHD and kinetic-scale fields, these efforts are still in their early stages \cite{Drake19}. Consequently, a unified framework seamlessly integrating behavior across all relevant scales is yet to be realized.

This review discusses results from both methodologies, highlighting their respective contributions and current limitations. Initially, we investigate particle energization within large-scale structures by analyzing the behavior of test particles in MHD simulations. Subsequently, we delve into findings from PIC simulations, emphasizing small-scale dynamics, in particular particle heating and acceleration mechanisms. This analysis includes a detailed examination of the formation and evolution of 3D coherent structures—such as large-amplitude magnetic  structures, current sheets, and shocks—and their pivotal roles in driving energy dissipation and particle energization.

\subsection{Combining 3D MHD and test particle simulations}

\subsubsection{Strong Turbulence driven by large amplitude MHD waves \label{stron_turb}}

 The initiation of strong turbulence in plasma systems has been extensively studied using various modeling approaches through both two-dimensional (2D) and three-dimensional (3D) numerical MHD simulations \cite{Biskamp89,Dmitruk03,Arzner06,Servidio09,Servidio10,Servidio11,Zhdankin13,Isliker17a,Rueda21,Pezzi22, Richard22}. In this subsection, we employ the methodology initially developed by Dmitruk et al. \cite{Dmitruk04} and later refined by Arzner et al. \cite{Arzner06}, Zhdankin et al. \cite{Zhdankin13}, Isliker et al. \cite{Isliker17a}, and Pezzi et al. \cite{Pezzi22}. Unlike approaches that impose predefined reconnection geometries or use specific wave ensembles as proxies for turbulence \cite{Arzner04}, this  approach allows the magnetohydrodynamic (MHD) equations to evolve naturally, producing intrinsically correlated field structures.

This self-organized evolution leads to turbulence dominated by coherent structures, such as dense current filaments or current sheets (CSs), rather than random, uncorrelated noise. By avoiding externally imposed turbulence characteristics, this framework enables the MHD system to generate a realistic turbulent environment self-consistently, capturing correlated structures and regions of localized intense current density. Such a dynamic approach is essential for advancing our understanding of turbulent reconnection processes and energy dissipation mechanisms in astrophysical and laboratory plasma systems.

Ambrosiano et al. \cite{Ambrosiano88} analyzed the evolution of test particles within turbulent magnetic reconnection. Many years later, several studies revisited this problem, investigating the evolution of particle distributions within snapshots of 3D MHD  simulations that model turbulence under various aspects \cite{Dmitruk03, Dmitruk04, Arzner06, Dalena14, Gonzalez16, Gonzalez17, Isliker17a}. 

The 3D, resistive, compressible, and normalized MHD equations used in \citet{Isliker17a} are
\beq
\p_t \rho = -\nabla \cdot \mathbf{p} ,
\eeq
\beq
\p_t \mathbf{p} =
- \mathbf{\nabla}  \cdot
\left( \mathbf{p} \mathbf{u} - \mathbf{B} \mathbf{B}\right)
-\nabla P - \nabla B^2/2 ,
%+ \rho \bar{\nu} \nabla^2 \mb{v} 
\eeq
\beq
\p_t \mathbf{B} =
-  \nabla \times \mathbf{E} ,
\eeq
\beq
\p_t (S\rho) = -\mathbf{\nabla} \cdot \left[S\rho \mathbf{u}\right] ,
\eeq
with $\rho$ the mass density, $\mathbf{p}$ the momentum density,
$\mathbf{u} = \mathbf{p}/\rho$,
$P$ the thermal pressure,
$\mathbf{B}$ the magnetic field,
\beq \label{ohmsaLaw}
\mathbf{E}   = -  \mathbf{u}\times \mathbf{B} + \eta \mathbf{J}\eeq
the electric field,
$\mathbf{J} =  \mathbf{\nabla}\times\mathbf{B}$
the current density, $\eta$ the resistivity,
$S=P/\rho^\Gamma$ the entropy,
and $\Gamma=5/3$ the adiabatic index.

In these MHD simulations, test particles are tracked within a fixed snapshot of the MHD configuration that exhibits fully developed strong turbulence. The particles evolve over short timescales, focusing not on wave-particle scattering but on their interactions with electric fields. In this particular experiment, anomalous resistivity effects were included in the model.

Physical units are defined by setting the box size $L=10^5\,$m, the Alfvén speed $ v_A=2\times 10^6\,$m/s, and the background magnetic field $B_0=0.01\,$T. To accurately map the fields to the particle positions, \citet{Isliker17a} applied a tri-cubic interpolation of the field values at the grid points.

%\paragraph{Test-particle simulations.}
The relativistic guiding center equations (without collisions) are used
for the evolution of the position $\mb{r}$ and  the parallel component $u_{||}$ of the relativistic 4-velocity of the particles  \cite{Tao2007}, 
\beq \label{gc:r}
\frac{d\mb{r}}{dt}= \frac{1}{B_{||}^*} 
\left[ \frac{u_{||}}{\gamma} \mb{B}^* + \hat{\mb{b}}\times 
\left(\frac{\mu}{q\gamma}\nabla B -\mb{E}^* \right) \right]
\eeq
\beq \label{gc:u}
\frac{du_{||}}{dt} = - \frac{q}{m_0 B_{||}^*}\mb{B}^* 
\cdot\left(\frac{\mu}{q\gamma} \nabla B -\mb{E}^* \right)
\eeq
where %$\mb{B}^*$ and $\mb{E}^*$ are the modified fields defined by
$\mb{B}^*=\mb{B} +\frac{m_0}{q} u_{||}\nabla\times\hat{\mb{b}}$, 
$\mb{E}^*=\mb{E} -\frac{m_0}{q} u_{||} \frac{\p\hat{\mb{b}}}{\p t}$, 
$\mu   = \frac{m_0 u_\perp^2}{2 B}  $
is the magnetic moment, 
$\gamma=\sqrt{1+\frac{u^2}{c^2}}$,
$B=|\mb{B}|$, $\hat{\mb{b}}=\mb{B}/B$, 
$u_\perp$ is the perpendicular component of the relativistic 4-velocity,
and $q$, $m_0$ are the particle charge and rest-mass, respectively. 

The test particles that \citet{Isliker17a} consider throughout are electrons.
Initially, all particles are located at random positions; they obey a 
Maxwellian distribution 
$n(W, t=0)$ 
with temperature $T=100\,$eV. The simulation box is open; the particles can escape from it when they reach any of its boundaries.
\begin{figure}[h!]
\centering
\includegraphics[width=0.5\columnwidth]{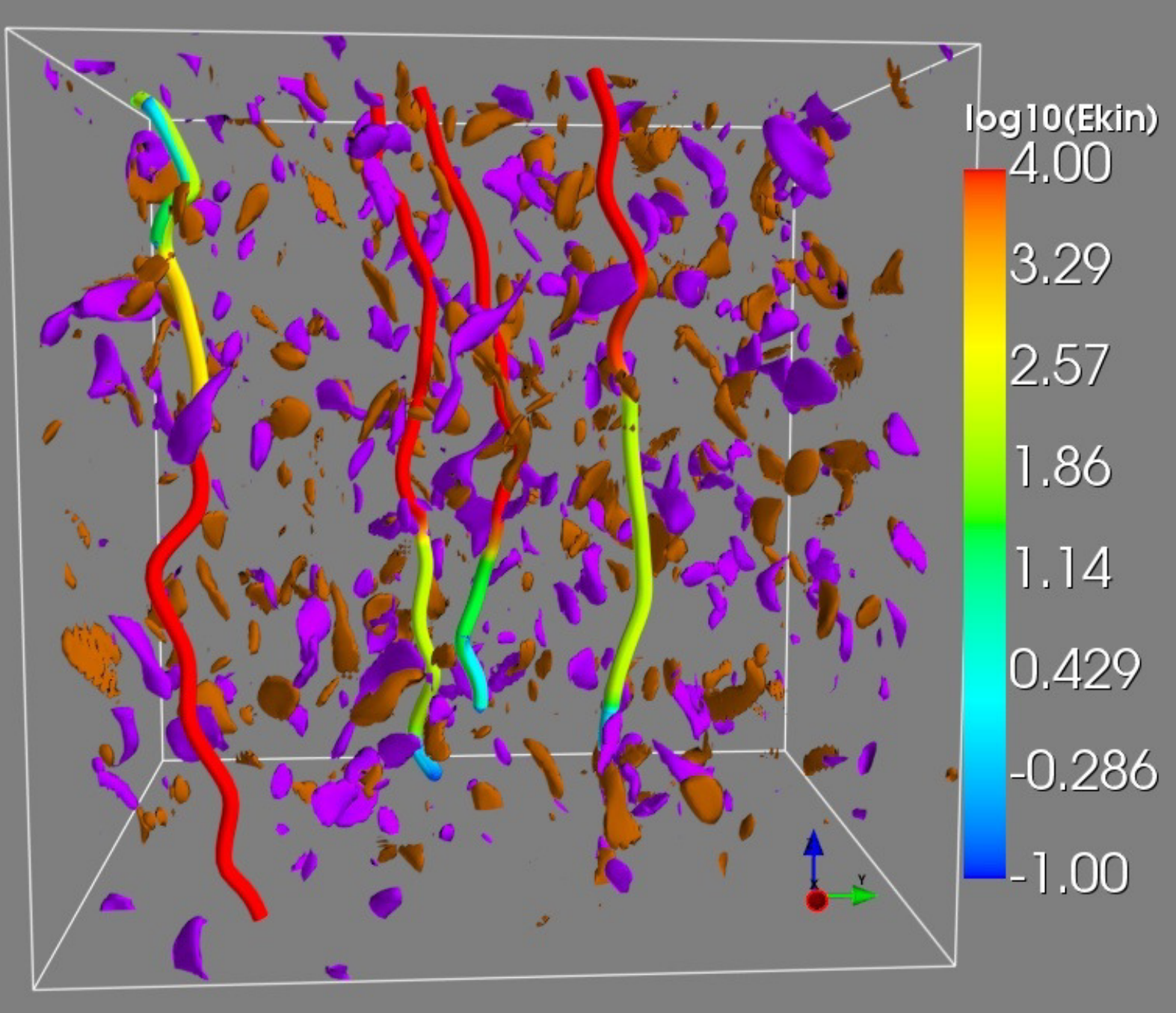}\\
\includegraphics[width=0.6\columnwidth]{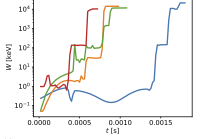}
\caption{(a) Isocontours of the supercritical current density component $J_z$ (positive in brown, negative in violet), and a few orbits of energetic particles, colored according to the logarithm of their kinetic energy in keV (see color bar). (b) The energy evolution of four energetic particles (marked with different colors) is shown. Reproduced with permission from Isliker et al., Phys. Rev. Lett. {\bf 119}, 045101 (2018), Copyrigt 2018 APS. 
 }
\label{Snapshot}
\end{figure}

The particle acceleration process is highly efficient, reaching an asymptotic state within just $0.002\,$s, equivalent to $7\times 10^5$ gyration periods. The test-particle simulation is run until a final time of $t=0.8\,$s, or until a particle escapes the simulation box. Figure \ref{Snapshot}a illustrates the $J_z$ component of the current density in regions exceeding the critical threshold. These regions are fragmented into numerous small-scale current filaments, or current sheets, representing coherent structures in the nonlinear, super-Alfvénic MHD environment.

\begin{figure}[!ht]
\centering
\includegraphics[width=0.70\columnwidth]{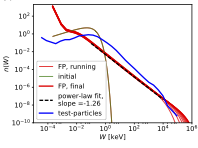}\\
\includegraphics[width=0.70\columnwidth]{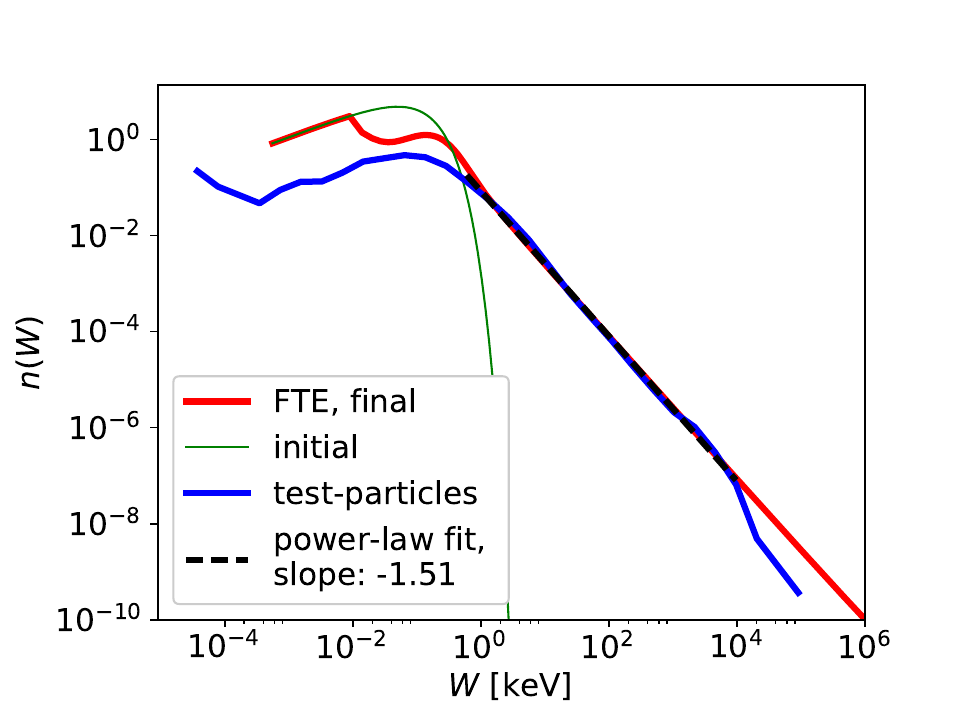}
\caption{(a) Solution of the classical FP equation up to a final time of 0.002 s, together with the solution at a few intermediate times, and the energy distribution from the test-particle simulations at $t= 0.002\,$ s.
(b) Initial and final (at $t
			= 0.002\,$sec) kinetic energy distribution from the test-particle simulations, together with a power-law fit, and the solution of the fractional transport equation at the final time. Reproduced with permission from Isliker et al., Phys. Rev. Lett. {\bf 119}, 045101 (2018), Copyrigt 2018 APS.
}
\label{Ekin}
\end{figure}

Figure \ref{Ekin}b shows the energy distribution at the initial and final times. The acceleration process is systematic, with particles experiencing rapid energy gains predominantly when traversing reconnecting current sheets formed by colliding current filaments (see Fig.\ \ref{Snapshot}b). The resulting energy distribution develops a clear power-law tail in the intermediate-to-high energy range, with a spectral index of $-1.51$ and a slight turnover at the highest energies. This systematic acceleration is accompanied by moderate particle heating, with the initial temperature approximately doubling, consistent with previous observations in both test-particle simulations \cite{Arzner04} and particle-in-cell (PIC) simulations \cite{Dahlin15,Guo15}.

The particle energization process is characterized by localized, abrupt jumps in energy, occurring repeatedly at different current filaments (see Fig.\ \ref{Snapshot}b). These sudden energy increments contrast with the continuous evolution assumed in the classical Fokker-Planck (FP) framework. \citet{Isliker17a} demonstrated this discrepancy using the method outlined in Sec. \ref{stochasticFermi2}. By estimating the transport coefficients directly from the test particle dynamics \cite{Isliker17a} (see Eqs.\ (\ref{SystW}) and (\ref{DifW})), they solved the FP equation. They compared the result with the energy distribution from the test particle simulations. As shown in Fig.\ \ref{Ekin}a, the FP solution produces a power-law tail in the energy distribution, albeit much flatter than observed in the test-particle simulations. This divergence persists even with longer integration times, underscoring the limitations of the FP approach.

The failure of the classical FP equation can be attributed to the statistical nature of the energy increments, $ \delta W_j=W_j(t+\D t)-W_j(t)$ (where j is the particle index). These increments follow a power-law distribution at intermediate-to-high energies, resulting in occasional large jumps in energy space --- so-called Lévy flights (see Fig.\ \ref{Snapshot}b, and also Sec.\ \ref{symmLevy}). This behavior has profound implications: (1) The mean value of the transport coefficients $F$ (Eq.\ (\ref{SystW})) and variance $D$ (Eq.\ (\ref{DifW})) theoretically diverge for a scale-free power-law distribution, complicating their practical computation. (2) A mean value is inherently unrepresentative of the underlying distribution. (3) As discussed in Sec.\ \ref{fokkpla}, the prerequisites for deriving an FP equation are fundamentally violated.

\citet{Isliker17a}, following the steps outlined in Sec. \ref{fracdiffeq} and Sec.\ \ref{systematic_fractional}, constructed a Fractional Transport Equation (FTE) based on insights gathered from test-particle trajectories. Specifically, they probed the energy increments over a fixed time interval, $\Delta t$, and assumed the waiting time distribution to be $p_{\tau}(\tau) = \delta(t - \Delta t)$. It thus follows that $\beta = 1$ and $b = \Delta t$ in Eq.\ (\ref{fraceq}), see Sec.\ \ref{systematic_fractional}. This approach is  appropriate when test-particle data are  given as time series, where no direct information is available about the waiting times between scattering events.

As a result, the fractional transport equation they proposed includes an ordinary first-order time derivative and a fractional derivative in energy. Following Eq.\ (\ref{fraceqexam}), the used equation takes the form 
\beq \label{FTE} 
\p_t n = \frac{a}{\Delta t} D^{\alpha}_{|W|} n - \frac{n}{t_{esc}} , 
\eeq
where an additional escape term is incorporated. The parameters $\alpha$ and $a$ are determined as described in Sec.\ \ref{systematic_fractional}.

For the numerical solution of the fractional transport equation, the authors used the Grünwald-Letnikov definition of fractional derivatives (see, e.g., \citet{Kilbas2006}) implemented in the matrix formulation of \citet{Podlubny2009}. The derivative scheme detailed in \cite{Podlubny2013} was applied to non-equidistant grid points, which allows to use grid-points that are equi-spaced in the logarithm of the energy. This is well suited to model power-law tails of arbitrarily large extent, and it also ensures compatibility with the grid used for solving the classical Fokker-Planck equation.  The fractional derivative was applied only for energies above 10 eV, as the focus was on the evolution of the high-energy tail of the energy distribution. The authors noted that the FTE, Eq.\ (\ref{FTE}), in its current form, is unsuitable for modeling low-energy phenomena or heating processes; it is specifically derived to model the long tails observed at the high-energy end of the distribution.

Figure \ref{Ekin}b shows the solution of the FTE at $t = 0.002\,$s, showing excellent agreement with the power-law tail yielded by the test-particle simulations across its entire extent. Varying the anomalous resistivity $\eta_{an}$ from $10\eta$ to $10^4\eta$, \citet{Isliker17a} demonstrated that the FTE consistently reproduces the simulation data, which further validates the appropriateness of the FTE for capturing the high-energy dynamics of the system.
They also observed that for $\eta_{an} = \eta$, no power-law tail is formed.

\begin{figure}[!ht]
\centering
\includegraphics[width=0.70\columnwidth]{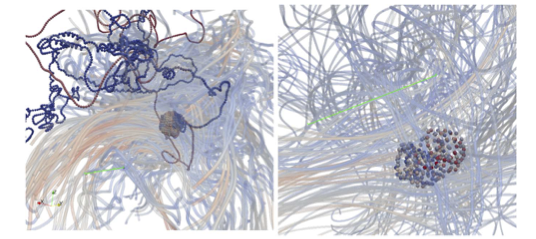}\\
\includegraphics[width=0.70\columnwidth]{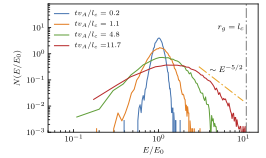}
\caption{(a) Particle trajectory in the 3D domain, with the points colored with the particle energy, where the color scale goes from blue to red as the particle energy increases. Magnetic field lines, colored with the magnitude of the magnetic field itself (again from blue to red as the magnetic field magnitude increases), indicate that the particle is trapped in a flux tube and that it is accelerated when the flux tube is feeling the gradients associated with the interaction with another large-scale structure. The right panel shows an inset of the left plot zoomed in the trapping region and limited in time to a few particle gyrations. (b) Particle PDFs at different time instants to show the energization process. Reproduced with permission from Pezzi et al., Astrop. J. {\bf 928}, 25 (2022), Copyrigt 2022 AAS.
 }
\label{Pezzi}
\end{figure}

\citet{Pezzi22} also discuss the energization of relativistic charged particles in three-dimensional incompressible MHD turbulence and the diffusive properties of the motion of the particles. They show that the random electric field induced by turbulent plasma motion leads test particles moving in a simulated box to be stochastically energized. A small fraction of these particles are trapped in large-scale structures, most likely formed due to the interaction of current filaments in the turbulence (see Fig. \ref{Pezzi}a). 

These particles are systematically accelerated, provided their pitch angle satisfies some conditions. They discuss the characterization of the accelerating structure and the physical processes responsible for rapid acceleration. It is evident from Fig. \ref{Pezzi}b that stochastic energization dominates, and the power law tail caused by systematic energization at the very few CSs present in their simulation box is less prominent. 

\subsubsection{Strong turbulence driven by reconnecting current sheets}

\citet{Onofri06} investigates the fragmentation of a {\sl single} current sheet (see also \cite{WangYulei25}) its role in particle acceleration, as simulated using resistive magnetohydrodynamics (MHD) for three-dimensional magnetic reconnection. Their initial equilibrium magnetic field includes a longitudinal component parallel to the current sheet, a key factor in understanding particle acceleration. By examining the evolution of the electromagnetic fields generated by this current sheet, they highlight the limitations of the MHD approach and stress the importance of kinetic effects.

The simulation solves the incompressible, dissipative and resistive MHD equations in a 3D Cartesian domain, using dimensionless units with kinetic and magnetic Reynolds numbers $ R_v = 5000 $ and $R_M = 5000 $. The initial conditions feature a plasma at rest, permeated by an equilibrium magnetic field $ \mathbf{B}_0 $, sheared along the $ \hat{x}$-direction, and with a central current sheet. Nonlinear evolution leads to forming small-scale structures in the lateral regions and coalescing magnetic islands in the center, disrupting the initial equilibrium. This process is reflected in the 3D structure of the electric field, where current filaments fragment the current sheet after $ t = 50 \tau_A$ (Alfv\'en time). 
\begin{figure}[!ht]
\includegraphics[width=0.60\columnwidth]{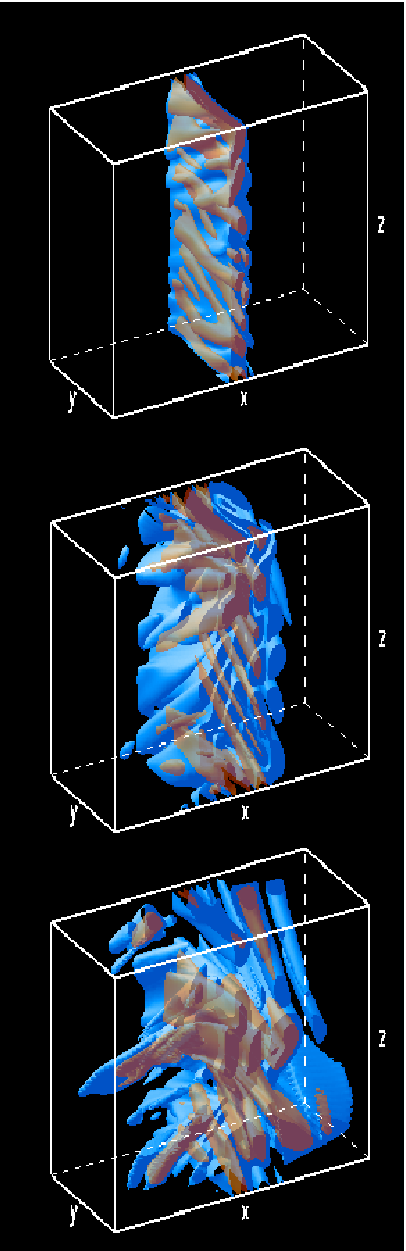}
\caption{Electric field isosurfaces at $t=50\tau_A, t=200 \tau_A, t=500 \tau_A$. Reproduced with permission from Onofri et al., Phys. Rev. Lett. {\bf 96}, 151102 (2006), Copyrigt 2006 APS.
}
\label{Onofri1}
\end{figure}
Figure \ref{Onofri1} illustrates the electric field's evolution, showing isosurfaces for high (red) and low (blue) field intensities. Initially localized, strong electric field regions fragment further by \( t = 400 \tau_A \), destroying the initial current sheet. These regions, identified as particle acceleration sites, expand throughout the simulation box, increasing the probability of particle acceleration.

\citet{Onofri06} quantify the electric field by constructing its distribution function, separating the resistive and convective components. Although the resistive component is weaker, it is more critical for particle acceleration, as confirmed by simulations where each component is tested in isolation.  For the results presented here, both components of the electric field have been taken into account.

Protons and electrons are injected into the simulation box to study the particle dynamics under the influence of the electromagnetic fields that remain static during particle motion. This assumption is justified by the faster time scale of particle acceleration compared to field evolution. Particle trajectories are computed by solving the relativistic equations of motion, using a fourth-order Runge-Kutta scheme, with the fields being interpolated between grid points by using local 3D interpolation (see Sec.\ \ref{stron_turb}, Eqs.\ (\ref{gc:r}), Eqs.\ (\ref{gc:u})).

\begin{figure}[!ht]
\includegraphics[width=0.95\columnwidth]{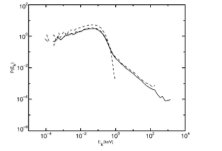}
\caption{ Distribution function of electron kinetic energy at $t = 8 \times 10^{-5}$ s (solid line), $t = 3 \times 10^{-5}$ s (dot-dashed line), and the initial distribution (dashed line). Reproduced with permission from Onofri et al., Phys. Rev. Lett. {\bf 96}, 151102 (2006), Copyrigt 2006 APS.
 }
\label{Onofri2}
\end{figure}

For electrons, the energy distribution evolves rapidly (Fig. \ref{Onofri2}). Some particles are quickly accelerated, forming a power-law tail. The kinetic energy of the electrons increases until it equals the magnetic field energy. Since the particles exert no feedback on the fields in the test-particle approach, the particles' energy can grow indefinitely until they exit the simulation box. The maximum energy reached by electrons at the final simulation time $ t_{pe} \sim 8 \times 10^{-5}\,$s is approximately $ 1 \, \text{MeV} $. The simulations exclude collisions due to the much longer collisional time scale compared to $ t_{pe} $. The final particle energy distribution develops a power-law tail with a logarithmic slope of $ \sim 1 $, beginning at energies around $ 1 \, \text{keV}$. Acceleration occurs so rapidly that electrons primarily interact with the systematically acting electric field of the fragmented current sheet, with minimal contribution from stochastic effects (Fig. \ref{Onofri2}).

\citet{Kowal2012} and \citet{Pezzi24} provide additional perspectives on particle acceleration in MHD turbulence and flux rope perturbations. \citet{Kowal2012} studied particle energization in an evolving current sheet, ignoring the resistive electric field term and retaining only the convective component $( \mathbf{E} = -\mathbf{v} \times \mathbf{B}/c $). Their use of periodic boundary conditions and exclusion of the resistive term prevented the formation of a power-law tail in the particle energy distribution. Instead, stochastic energization led to particle heating. Similarly, \citet{Pezzi24} analyzed ideal MHD simulations of perturbed solar wind flux ropes, neglecting the resistive electric field. While the particles were rapidly heated, the lack of systematic acceleration mechanisms and periodic boundary conditions hindered the development of a power-law tail.

\begin{figure}[!ht]
\includegraphics[width=0.70\columnwidth]{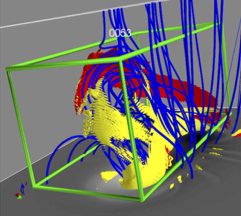}
\caption{MHD simulations, zoomed into the coronal part: Magnetic field lines are shown (blue), and an isocontour plot of the parallel electric field (red to yellow 3D surfaces). The green cube outlines the region from which the initial conditions of the particles are chosen. Reproduced with permission from Isliker et al., Astroph. J. {\bf 882}, 52 (2019), Copyrigt 2019 AAS. }
\label{IslAr1}
\end{figure}

In their work, \citet{Isliker19} analyzed the interaction between emerging and pre-existing magnetic fields in the solar atmosphere, using a 3D resistive MHD code. Their findings demonstrate that such interactions can initiate various dynamic processes, such as eruptions and jets. A key feature of these interactions is the formation of large-scale current sheets, which eventually fragment, creating a strongly turbulent environment. Their study focuses on understanding the kinetic behavior of particles in response to the fragmented magnetic fields (see Fig. \ref{IslAr1}).

Their research uses electrons treated as test particles, with an  integration time of 0.1 seconds. In each simulation, 100'000 particles are traced, using the relativistic guiding-center approximation as equations of motion (see Sec.\ \ref{stron_turb}, Eqs.\ (\ref{gc:r}), Eqs.\ (\ref{gc:u})). The initial spatial distribution of particles is chosen randomly within the region surrounding the main reconnection area, marked by the green cube in Fig.\ \ref{IslAr1}, which encompasses the entire current sheet and its fragments. The initial velocities of the particles are also randomized, following a Maxwellian distribution with a temperature of approximately $10^6\,$K. At 100 pre-defined monitoring times, including the final time, the positions and velocities of all particles are recorded for statistical analysis. Additionally, particles that leave the simulation box before the final time are tracked separately.

\begin{figure}[!ht]
\includegraphics[width=0.95\columnwidth]{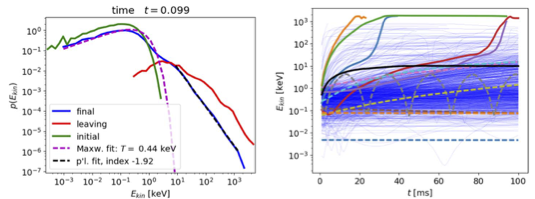}
\caption{ Left: Kinetic energy distribution of electrons after 0.1 s, without collisions, with fits to the low-energy Maxwellian and high-energy power-law components. The plot also shows the initial distribution and the distribution of particles that escaped during the simulation. Right: Time evolution of kinetic energy for the test particles (thin blue lines), highlighting several high-energy (solid) and low-energy (dashed) particles, along with the evolution of the mean energy (solid black). Reproduced with permission from Isliker et al., Astroph. J. {\bf 882}, 52 (2019). Copyright 2019 AAS. 
 }
\label{IslAr2}
\end{figure}

The statistical properties of the turbulent, fragmented electric fields drive efficient heating through stochastic particle interactions,  and lead to systematic acceleration of the  particles. This interaction forms a power-law tail at high energies on sub-second timescales, as can be seen from Fig.\ \ref{IslAr2} (left), which shows the energy distribution of the particles at $0.1\,$s. The distribution thus approaches the well-known Kappa distribution (see Sec.\ \ref{SecKappa}). A fraction of the accelerated particles escape from the reconnection region,  representing a super-hot component with a temperature close to $150\,$MK and a high-energy tail with a power-law index ranging from $-2$ to $-3$ (also shown in Fig.\ \ref{IslAr2} (left)). In Fig.\ \ref{IslAr2} (right), the evolution of the kinetic energy over time is shown for individual particles (thin blue lines), with several high-energy (solid lines) and low-energy (dashed lines) particles highlighted in different colors, which clearly reveals that the high-energy particles experience systematic acceleration. The overall  evolution of the mean kinetic energy is shown as solid black line.

\citet{Isliker19} estimate the transport coefficients in energy space from the dynamics of the charged particles inside the fragmented, fractal electric fields (as described in Sec.\ \ref{stochasticFermi2}). 
They find that the solution of a fractional transport equation (see Sec.\ \ref{stron_turb}),
as appropriate for a strongly turbulent plasma, agrees well with the test-particle simulations. 
By contrast, they show that the conventional Fokker–Planck equation fails to provide an adequate transport model for the systematic acceleration in these conditions (see Sec. \ref{stochasticFermi2}).

In a related study, \citet{Baumann13} investigated particle acceleration mechanisms within three-dimensional magnetic reconnection null-point regions in the solar corona. Starting from a potential field extrapolation of a Solar and Heliospheric Observatory (SOHO) magnetogram taken on November 16, 2002, their work utilized MHD simulations driven by observed photospheric horizontal motions to study electric current build-up in the fan plane of a null point. A sub-section of the MHD results was used as initial and boundary conditions for a kinetic particle-in-cell model of the plasma. They discovered that a systematic  electric field accelerates sub-relativistic electrons within the current sheet. Their simulations produced a non-thermal electron population with a power-law energy distribution featuring a spectral index of approximately $-1.78$. Their study successfully bridges macroscopic scales, on the order of hundreds of Mm, with kinetic scales, down to centimeters, providing insight into such a cross-scale coupling by effectively utilizing either physical modifications or equivalent modifications of the constants of nature. These high-resolution simulations --- up to 135 billion particles and 3.5 billion grid cells of $17.5\,$km size --- represent a significant advance in understanding particle acceleration in solar-like settings.

\begin{figure}[!ht]
\includegraphics[width=0.90\columnwidth]{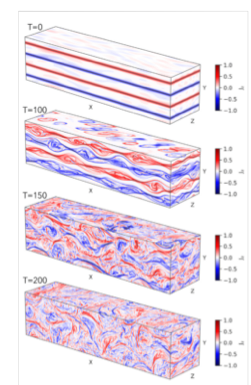}
\caption{Time evolution (t = 0, 100, 150, 200) of the current density $J_z$. Reproduced with permission from Nakanotani et al., Frontiers Astr. Sp. Scien. {\bf 9}, 151102 (2022). Copyright Frontiers 2022.
 }
\label{Nakano1}
\end{figure}

\begin{figure}[!ht]
\includegraphics[width=0.90\columnwidth]{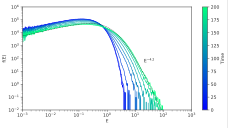}
\caption{Time evolution of the energy distribution of test particles. The black dashed line represents a Kappa distribution with a temperature $T_{\kappa} = 0.8$ and a Kappa index of $\kappa = 3.2$. Reproduced with permission from Nakanotani et al., Frontiers Astr. Sp. Scien. {\bf 9}, 151102 (2022). Copyright Frontiers 2022.
 }
\label{Nakano2}
\end{figure}

\citet{Nakanotani22} extended the investigation of particle acceleration mechanisms by examining an MHD-scale system consisting of multiple current sheets through combined 3D MHD and test-particle simulations. Due to tearing-mode instabilities, magnetic reconnection leads to the formation of magnetic filaments, and the subsequent interactions between these filaments drive the system toward a turbulent state (see Fig.\ \ref{Nakano1}). Their 3D simulations revealed spatial power-law spectra of index $-11/3$ for magnetic field fluctuations and $-7/3$ for velocity fluctuations (see Fig.\ 8 in \citet{Nakanotani22}). The resulting turbulence efficiently energizes the particles, creating Kappa-like energy distribution functions (see Fig.\ \ref{Nakano2}). They found that more energetic particles tend to move between magnetic filaments than being confined within them, emphasizing the role of inter-filament regions in particle energization.

In summary, these studies highlight the intricate dynamics involved in particle acceleration during magnetic reconnection and fragmentation events. The formation of turbulent magnetic and electric fields plays a critical role in energizing particles and shaping their distributions into power-law or Kappa forms. The systematic acceleration observed in these environments emphasizes the complex interplay of magnetic topology, turbulence, and  their interaction with particles, which drive the solar corona's dynamic behavior.

\subsubsection{Strong turbulence driven by the convection zone on the sun}

\citet{Turkmani05, Turkmani06} based their analysis on the earlier work of \citet{Galsgaard96} and \citet{Galsgaard05}, who conducted three-dimensional MHD simulations to explore how small-scale coronal structures, such as current sheets (CSs), emerge from simple photospheric boundary motions. These simulations assumed a coronal magnetic field anchored at both ends in the photosphere and "straightened out", meaning that large-scale magnetic field curvature was neglected. The initial magnetic field was uniform, while the density varied by a factor of 1000 between the photosphere and the corona. The atmosphere was initialized in hydrostatic equilibrium, and boundary motions in the photosphere were imposed using sinusoidal, incompressible shear patterns that varied randomly over time. These boundaries were impenetrable and perfectly conducting, ensuring that the imposed motions advected the magnetic footpoints and applied stress to the magnetic system.

\begin{figure}[!ht] 
\includegraphics[width=0.80\columnwidth]{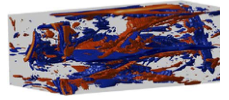} \includegraphics[width=0.80\columnwidth]{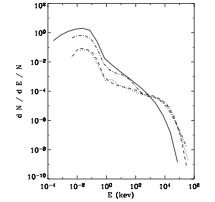} \caption{ (a) Snapshots of the resistive electric field configuration in the coronal volume, as calculated from the global MHD model. Blue and red regions correspond to electric fields pointing toward the left and right footpoints, respectively. Details of the model are described in the main text.
(b) Distribution functions (dotted, dashed, and dashed-dotted lines) for particles leaving from the right footpoint, the left footpoint, and the domain sides, respectively. The solid curve represents particles remaining inside the domain. Reproduced with permission from Turkmani et al., Astron. Astroph. {\bf 449}, 749 (2006). Copyright ESO 2006. 
 } 
\label{Turk1} 
\end{figure}

The simulations followed the system’s time-dependent evolution by numerically solving the three-dimensional non-ideal MHD equations in Cartesian geometry \cite{Galassi17}. These equations incorporated heat conduction and optically thin radiation. After several driving cycles, the magnetic field developed a highly structured topology characterized by scattered current concentrations throughout the simulation domain. Specific details of these numerical experiments can be found in \citet{Galsgaard96}.

\citet{Turkmani05, Turkmani06} used a snapshot of the simulated coronal magnetic and electric fields to investigate particle acceleration (see Fig.\ \ref{Turk1}). They tracked test particles by solving the relativistic equations of motion, using adaptive time-stepping and linear interpolation to calculate field values at arbitrary positions (see Sec.\ \ref{stron_turb}). The electric field, fully three-dimensional, consisted of two components (see Eq.\ (\ref{ohmsaLaw})):
(a) an inductive field and  (b) a resistive field, originating from finite resistivity, which localizes diffusion to regions of steep gradients using hyper-resistivity \cite{Galsgaard97a}.

The highly fragmented structure of these field components correlates with magnetic field disturbances, which reach magnitudes of $5\%$ – $10\%$ of the ambient field. The inductive term $\mathbf{V}\times \mathbf{B}/c$ in Eq.\ (\ref{ohmsaLaw}) dominates outside current sheets (CSs), leading primarily to particle drifts. Conversely, the resistive term $\eta \mathbf{J}$ becomes significant near reconnection sites, where it drives particle acceleration parallel to the ambient field (see Fig.\ \ref{Turk1}). The study focused on the resistive field, as it provided systematic energy gain. Still, results were consistent across three MHD simulations \cite{Turkmani06, Onofri06, Isliker19}, confirming the robustness of the statistical properties of resistive and convective electric fields.

\begin{figure}[!ht] 
\includegraphics[width=0.70\columnwidth]{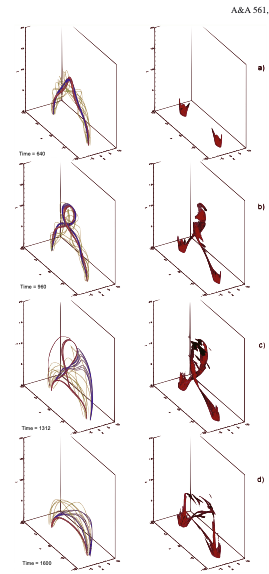} \includegraphics[width=0.80\columnwidth]{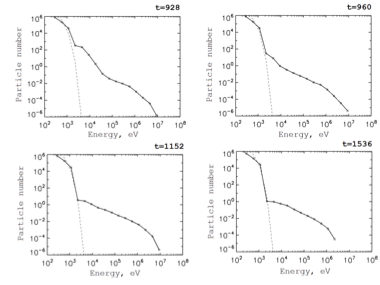} \caption{ (a) Selected magnetic field lines (left panels) and current density isosurfaces (right panels) during magnetic reconnection. Different colors indicate changes in connectivity: blue and red lines originate near the footpoint center, while green lines belong to the twisted fluxtube. Times are shown in the lower-left corners.
(b) Electron energy spectra at different times. Reproduced with permission from Gordovskyy, Astroph. J., {\bf 729}, 101 (2011). Copyright AAS 2011.
 } 
\label{Gordovsky1} 
\end{figure}

\citet{Gordovsky11, Gordovskyy14} expanded on these ideas by modeling energy release in an unstable, twisted coronal loop (see Fig.\  \ref{Gordovsky1}a), focusing on thermal and non-thermal components in the energization process. Their model combined 3D MHD simulations with test-particle tracking of electron and proton trajectories under the influence of Coulomb collisions.

Two key features distinguish their model:
\begin{itemize}
    \item {\it Large-Scale Magnetic Curvature:} The curved geometry of the twisted fluxtube influenced both reconnection and particle trajectories.
    \item {\it Atmospheric Stratification:} Density-dependent anomalous resistivity affected the MHD evolution, while Coulomb collisions influenced particle acceleration to a lesser extent.
\end{itemize}
Their simulations also calculated hard X-ray bremsstrahlung emission, allowing direct comparison with observations. Results on energy distributions (Fig.\  \ref{Gordovsky1}b) aligned with the findings of \citet{Turkmani05}, highlighting the systematic acceleration of particles in current sheets formed inside a driven coronal loop.

\begin{figure}[!ht] \includegraphics[width=0.90\columnwidth]{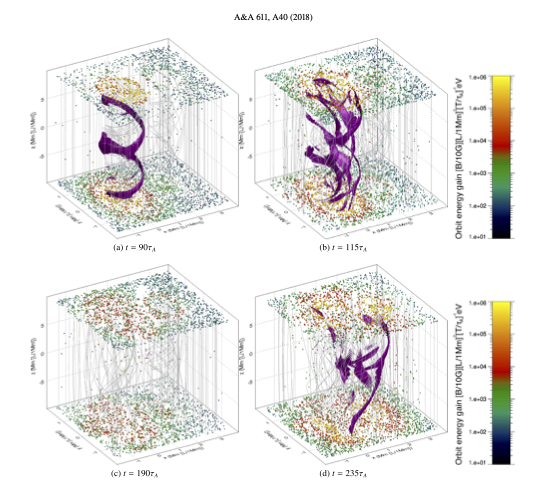}\ \includegraphics[width=0.90\columnwidth]{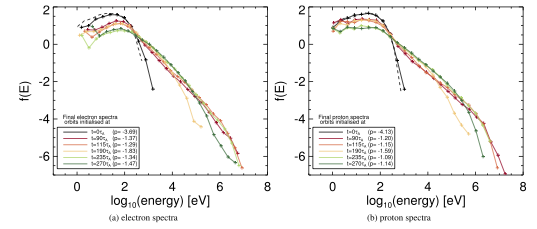} \caption{ (a) 3D map of final particle positions (colored orbits) and energies, based on four MHD snapshots during loop interactions. Current density isosurfaces above a critical threshold are shown.
(b) Temporal evolution of energy spectra for electrons (left) and protons (right). Dashed histograms indicate initial spectra, while colored spectra correspond to final energies at different initiation times. Power-law fits are shown, with spectral indices and times listed in the legend. Reproduced with permission from Threlfall et al., Astr. Astroph. {\bf 611}, A40 (2018). Copyright ESO 2018.
 } 
\label{Threlfall1} 
\end{figure}

\citet{Threlfall18} extended this work to unstable multi-threaded flaring coronal loops, analyzing non-thermal particle dynamics in high-resolution 3D MHD simulations. They employed snapshots from simulations involving two interacting loop threads (Fig.\ \ref{Threlfall1}a).

They distinguished the effects of uniform resistivity from anomalous resistivity and studied energy gains in simulations where one or both threads became unstable and eventually merged. The evolving energy spectra (Fig.\ \ref{Threlfall1}b) showed power-law behavior, with spectral indices dependent on the timing of particle injection into the system. This approach provided a detailed view of how particle energization depends on the dynamic interplay between interacting loop structures and evolving magnetic topologies.

\subsubsection{Strong turbulence driven in the edge of a Tokamak}

\citet{Isliker2022} conducted test-particle simulations to investigate the behavior of electrons during a nonlinear MHD simulation of a type-I Edge Localized Mode (ELM). Their study explored the kinetic effects of an eruptive plasma filament, revealing that electrons are moderately heated and accelerated over a short timescale of approximately $0.5\,$ms during the filamentary eruption. A non-thermal tail forms in the kinetic energy distribution, adopting a power-law shape with some particles reaching energies as high as 90 keV. The acceleration is observed exclusively in the direction parallel to the magnetic field, with a preference for the countercurrent direction, and the parallel electric field is identified as the primary driver of this acceleration.

High-energy particles escaping the system predominantly leave along a distinct strike-line in the outer divertor leg, typically during their energization phase. The escaping electrons  from the tail of the energy distribution are unaffected by collisions and exhibit characteristics of runaway electrons. The mean square displacement analysis shows that transport in energy space is superdiffusive. When the acceleration process is interpreted as a random walk, it is found that the distributions of energy increments display exponential tails, indicating a transport mechanism that combines convective (systematic) and diffusive (stochastic) elements. Analyzing the underlying MHD simulation, it turns out that the histograms of the parallel electric field in the edge region exhibit power-law shape, and this non-Gaussian statistics  is one of the reasons for the moderately anomalous particle transport observed in energy space.

\begin{figure}[!ht] \includegraphics[width=0.70\columnwidth]{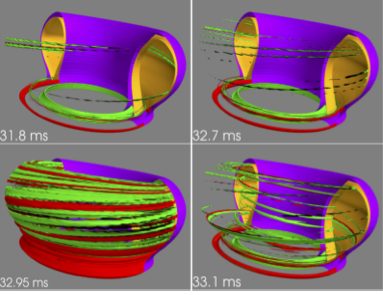}\ \includegraphics[width=0.70\columnwidth]{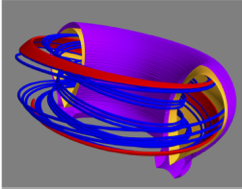}\ \includegraphics[width=0.90\columnwidth]{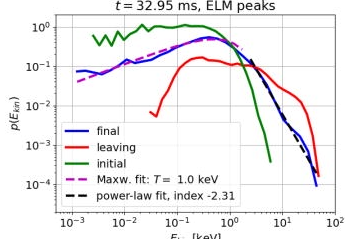} \caption{(a) Iso-contours of $E_\parallel$ (red: positive at $10\,$V/m, green: negative at $-10\,$V/m) at selected times, alongside the separatrix (orange) and plasma boundary (violet), with half of the surfaces cut away for clarity. (b) Separatrix (yellow), domain boundary (violet), and examples of two-particle orbits: a passing particle (blue; first $2\%$ of integration time) and a trapped particle (red; first $20\%$ of integration time). (c) Kinetic energy distributions  at the time where the ELM peaks: initial distribution, distribution of confined particles, and distribution of particles that have escaped by the time of interest. A Maxwellian fit (low energies) and a power-law fit (high energies) are shown for the confined particles. Reproduced with permission from Isliker et al. , Phys. Plas. {\bf 29}, 112306 (2022). Copyright ESO 2022.
	} 
\label{IslikerTok} 
\end{figure}

Figure \ref{IslikerTok} displays key aspects of the parallel electric field ($E_\parallel$) and the particle behavior during the ELM simulation. Iso-contours of $E_\parallel$ (Fig.\ \ref{IslikerTok}a) reveal band-like helical structures forming on the low-field side (LFS) and high-field side (HFS) of the plasma near the peak ELM activity. These structures, particularly prominent near the separatrix, become densely packed and more pronounced as the parallel electric field intensifies.  Increased parallel electric fields are also observed in the divertor leg regions, reaching magnitudes of $10\,$V/m at the base of the divertor legs during the ELM peak.
Fig. \ref{IslikerTok}b shows typical particle orbits, highlighting the spatial dynamics of passing and trapped particles. 

Fig.\ \ref{IslikerTok}c shows the kinetic energy distribution of particles  at the time where the ELM peaks. Initially, the distribution follows a Maxwellian shape. Over time, a non-Maxwellian high-energy tail emerges, forming a clear power-law shape during the ELM’s peak and  nearby stages. The tail extends beyond $20\,$keV  when the ELM intensifies, reaches approximately $50\,$keV  when the ELM peaks, and approaches $90\,$keV  when the ELM ends. The power-law tail steepens as time progresses, with the power-law index evolving from -1.9 when the ELM intensifies, to -2.3  when the ELM peaks, and -2.9  when the ELM ends. This steepening indicates a progressive reduction in the fraction of highly accelerated particles as the system evolves.

Escaping particles leave the system at different times, and their distributions are asynchronous, based on their energy at the time of escape. As Fig.\ \ref{IslikerTok}c shows, the escaping particles also form a non-thermal population, with  a clear departure from the Maxwellian baseline.

\subsection{3D Particle in Cell simulations of particle energization in strong turbulence}

\subsubsection{Strong turbulence driven by large amplitude waves}

Comisso et al. \cite{Comisso19, Comisso22} employed large-scale particle-in-cell (PIC) simulations to investigate plasma turbulence, focusing on scenarios where turbulence arises from strong magnetic field fluctuations that gradually decay over time. They solve the complete Vlasov-Maxwell system of equations using the PIC method \cite{Birdsal05}, which evolves electromagnetic fields through Maxwell's equations while computing particle trajectories via the Lorentz force. For this purpose, they utilize the fully relativistic electromagnetic PIC code TRISTAN-MP \cite{Spitkovsky05} to perform comprehensive three-dimensional (3D) simulations of plasma turbulence.
\begin{figure}[!ht]
\includegraphics[width=0.70\columnwidth]{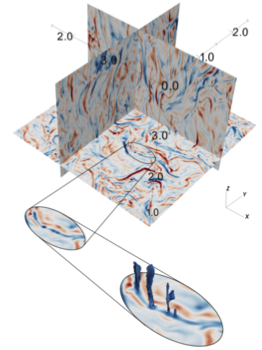}
\caption{Chain of flux ropes formed in a reconnecting current sheet that self-consistently develops in 3D turbulence. Isosurfaces of the current density $J_z$ are blue in the zoomed-in region, highlighting four flux ropes (3D plasmoids/filaments) elongated along $z$, i.e., the direction of the mean magnetic field. The color scheme for the shaded isocontours is such that blue indicates regions with $J_z<0$, while red indicates regions with $J_z>0$. Reproduced with permission from Comisso et al., Astroph. J. {\bf 886}, 122 (2019), Copyrigt 2019 AAS.
}
\label{Comisso1}
\end{figure}
Their computational setup is a cubic domain, with periodic boundary conditions applied in all three dimensions. This setup  allows to follow the evolution of all particle momenta and electromagnetic field components over time. The simulations begin with a uniform electron - positron plasma with  the momentum distribution following a Maxwell-J\"uttner distribution. A uniform magnetic field is initialized along the $z$-direction, and the initial equilibrium is perturbed by strong magnetic fluctuations (as detailed in \citet{Comisso19}).

Fig.\ \ref{Comisso1} illustrates a chain of flux ropes formed in a reconnecting current sheet that develops self-consistently within the 3D turbulent environment. These flux ropes (3D plasmoids or filaments) are elongated along the $z$-direction, coinciding with the direction of the mean magnetic field. 

\begin{figure}[!ht]
\includegraphics[width=0.70\columnwidth]{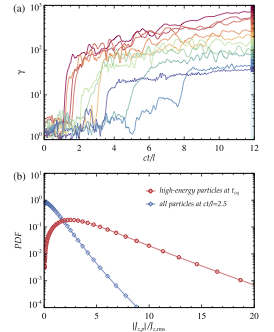}\\
\caption{   (a) Time evolution of the Lorentz factor for ten representative particles selected to end up in different energy bins  (matching the different colors in the color bar on the right). (b) pdfs of $|J_{z,p}| /J_{z,rms}$ experienced by the high-energy particles at their $t_{inj}$ (red circles) and by all our tracked particles at $ct/l=2.5$ (blue diamonds). About $80\%$ of the high-energy particles are injected at regions with $|J_{z,p}|>  2J_{z,rms}$. Reproduced with permission from Comisso et al., Astroph. J. {\bf 886}, 122 (2019), Copyright 2019 AAS.
 }
\label{Comisso2}
\end{figure}

\citet{Comisso19} extensively discuss the interplay between systematic and stochastic particle energization in their analysis. During the initial stages, the appearance of numerous filaments and 3D reconnecting current sheets (CSs) dominates the particle acceleration process, which they term as ``injection" or the ``first stage" of acceleration. As the magnetic fluctuations decay over time, the dominance of systematic acceleration associated with reconnecting current sheets diminishes. At this stage, stochastic scattering of particles by turbulent fluctuations or other coherent structures (CoSs) within the strong turbulence takes over, leading to continued particle energization. 

The study also demonstrates that elongated current sheets are susceptible to rapid filamentation instabilities, breaking into flux ropes separated by secondary current sheets. This process leads to fast magnetic reconnection and efficient particle  acceleration.

Fig.\ \ref{Comisso2} highlights the relationship between particle injection and electric current density. The time evolution of the Lorentz factor is presented for ten representative particles, each ending up in a different energy bin (Fig.\ \ref{Comisso2}a). Approximately $80\%$ of the high-energy particles are injected in regions with intense current densities, indicating that the particle energization strongly correlates with localized intense current structures (Fig.\ \ref{Comisso2}b).
\begin{figure}[!ht]
\includegraphics[width=0.70\columnwidth]{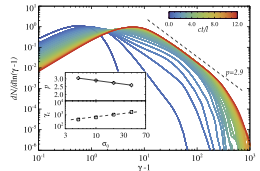}
\caption{Time evolution of the particle energy spectrum. The spectrum displays the formation of a Kappa distribution (heating and a power-law tail with index $\sim 2.9$). About $16\%$ of the particles have $\gamma >15$ (twice the particle energy spectrum's peak), indicating the percentage of nonthermal particles. Reproduced with permission from Comisso et al., Astroph. J. {\bf 886}, 122 (2019), Copyrigt 2019 AAS. 
 }
\label{Comisso3}
\end{figure}
The initial systematic energization of particles is primarily driven by the work done by the electric field component parallel to the local magnetic field ($\mathbf{E}_\parallel$), which is nonzero at reconnecting current sheets. After this phase of fast systematic energization, the perpendicular electric field component ($\mathbf{V}\times \mathbf{B}$ ) takes over, contributing to particle heating. The final energy distribution thus reflects the combined effect of systematic and stochastic processes. Notably, the particle pitch-angle distribution retains signatures of these different energization mechanisms.
Fig.\ \ref{Comisso3} shows the time evolution of the particle energy spectrum, which evolves into a Kappa distribution with a power-law tail of index $\sim 2.9$. Notably, about $16\%$ of the particles attain Lorentz factors greater than 15, indicating significant nonthermal particle acceleration.

\citet{Comisso19} also determine the energy diffusion coefficient associated with stochastic acceleration driven by turbulent fluctuations. Following the initial systematic acceleration phase, they show that the stochastic evolution of particle energy transport can be effectively described using the Fokker-Planck equation.

It is crucial to note that their simulation box employs periodic boundary conditions, preventing energetic particles from escaping the acceleration region. This confinement influences the final velocity distribution and the relative contributions of the systematic and the stochastic acceleration mechanism. Moreover, other coherent structures, such as eddies or reflections from stochastic field lines, also play a significant role in shaping the final velocity distribution of particles.

%\heinz{
A recent study by \citet{Comisso24} analyzed the acceleration mechanism of ultra-high-energy cosmic rays driven by magnetically dominated turbulence through comprehensive kinetic particle-in-cell simulations. This type of turbulence accelerates particles on a short time-scale, resulting in a power-law energy distribution characterized by a well-defined cutoff. Notably, the significance of this simulation lies in the novel inclusion of particle escape from the turbulent acceleration zone within the PIC framework, yielding,  in particular, the energy dependence of particle escape.
%}

\subsubsection{Strong turbulence driven by fragmented current sheets}
\label{PIC simulations_CS}

\begin{figure}[!ht]
\includegraphics[width=0.90\columnwidth]{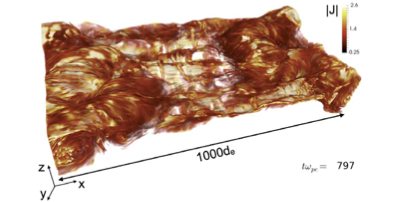}
\caption{The distribution (volume rendering) of the magnitude of the current density in the reconnection region for the standard 3D run.  Reproduced with permission from Guo et al., Astroph. J. {\bf 806}, 167 (2015), Copyrigt 2015 AAS.
 }
\label{Guo1}
\end{figure}

\citet{Guo21} presented a series of three-dimensional, fully kinetic simulations of relativistic turbulent magnetic reconnection (RTMR) in positron-electron plasmas with system domains much larger than the kinetic scales. These simulations begin with a force-free current sheet subject to several long-wavelength magnetic field perturbation modes, which drive additional turbulence within the reconnection region. Consequently, the current layer breaks up, and the reconnection region quickly evolves into a turbulent layer.  Fig.\ \ref{Guo1} presents the distribution (volume rendering) of the current density magnitude in the reconnection region for the standard 3D run. Under the influence of injected turbulence, the current sheet fragments into a turbulent reconnection region filled with various coherent structures, such as flux ropes and current sheets.

\begin{figure}[!ht]

\includegraphics[width=0.50\columnwidth]{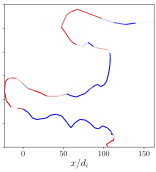}\\

\includegraphics[width=0.90\columnwidth]{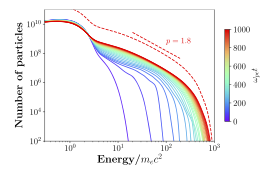}

\caption{(a) Particle trajectory showing energy versus x-position. The energy gain is systematic. (b) Particle energy distribution at different simulation times. The spectrum at the last time step is replotted and shifted by a factor of 10. The spectral index is $p = 1.8$. Reproduced with permission from Guo et al., Astroph. J. {\bf 806}, 167 (2015), Copyrigt 2015 AAS.
 }
\label{Guo2}
\end{figure}

The flux ropes evolve rapidly after their formation and can be disrupted entirely by secondary kink instabilities. This turbulent evolution leads to the superdiffusive behavior of magnetic field lines, similar to that observed in magnetohydrodynamic (MHD) studies of turbulent reconnection. Meanwhile, nonthermal particle energization and the timescale for energy release are very fast and do not depend strongly on the turbulence amplitude. The acceleration mechanism is due to the interplay of stochastic and systematic acceleration processes. The stochastic energization, supported by the perpendicular convective electric field ($(\mathbf{E}_{\perp}=\mathbf{V}\times \mathbf{B}$), plays a dominant role, resulting in a final energy distribution (Fig.\ \ref{Guo2}b) that approximates a superthermal distribution due to stochastic heating. In contrast, systematic acceleration due to the parallel electric field (termed ``nonideal" in this article) plays a subdominant role, and the high-energy tail is less pronounced. In Fig.\ \ref{Guo2}a, the systematic energization of a particle can be observed, contributing to the formation of the moderate energy tail.   The absence of multiple reconnecting current sheets in the fragmented initial current sheet is critical to this simulation's lack of a prominent power-law tail.

\citet{Dahlin15} also analyzed the fragmentation of an isolated large-scale current sheet and explored particle energization, using simulations with the massively parallel 3D PIC code p3d \cite{Zeiler02}. Particle trajectories are calculated using the relativistic Newton-Lorentz equation, and the electromagnetic fields are advanced using Maxwell's equations.  These simulations reduce the computational expense by using an artificial proton-to-electron mass ratio of  $m_i/m_e=25$. All simulations are initialized with a force-free configuration of a current sheet and use periodic boundary conditions. 

The spectra from these simulations reveal significant electron  energization in 3D. The stochastic structure of the magnetic field in 3D allows particles following the field lines to wander throughout the chaotic reconnecting region. Surfaces of section from field line tracing reveal that the stochastic magnetic field area in the 3D simulation roughly matches the spatial distribution of energetic electrons.

The mechanism of particle energization here also involves an interplay between stochastic interaction with the convective electric field ($(\mathbf{E}_{\perp}$) and systematic acceleration from the electric field component parallel to the local ambient magnetic field ($(\mathbf{E}_{||}$).

The electron spectra from these simulations \cite{Dahlin15} do not exhibit a power-law distribution. This is partly due to the limited energy gain possible within the modestly sized 3D simulation, and the limited number of systematic interactions with the fragments of the initial current sheet is also a contributing factor. Additionally, these simulations employ periodic boundary conditions, meaning that no particles are lost from the system, which is not a realistic set-up for most astrophysical systems.

\subsubsection{Strong turbulence driven by a large scale shock}

\citet{Matsumoto15} present results from computer simulations of a high Alfvén-Mach number collisionless shock. They investigate the shock's evolution in two dimensions through ab initio particle-in-cell (PIC) simulations,  conducting large-scale simulations to explore electron energization in multiscale shock structures.

To simulate a collisionless shock, ions and electrons were injected continuously from the boundary on the right-hand side of the simulation domain at supersonic (super-Alfvénic) speeds, moving toward the opposite boundary, where they were reflected. The simulations were, therefore, carried out in the shock downstream frame, with the shock front propagating upstream. Specifically, they modeled a perpendicular shock in which the upstream magnetic field is orthogonal to the shock normal. In the later stages of the simulation, approximately $10^{10}$ particle movements were tracked across $24'000 \times 2'048$ computational cells.

\begin{figure}[!ht]
\includegraphics[width=0.90\columnwidth]{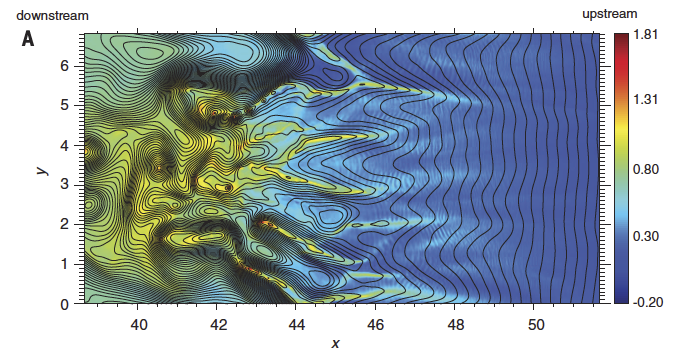}\
\includegraphics[width=0.40\columnwidth]{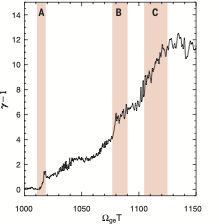}
\includegraphics[width=0.40\columnwidth]{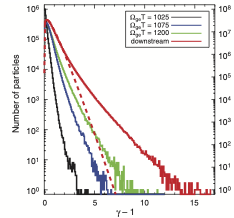}
\caption{(a) Supercomputer simulations of a strong collisionless shock revealing spontaneous turbulent reconnection. (b) Electron acceleration from multiple reconnection sites and magnetic islands. (c) Energy distribution of $10^6$ sampled electrons (black, blue, and green) and in the downstream region (red). Reproduced with permission from Matsumoto et al., Science \textbf{347}, 974 (2015), Copyright 2015 AAAS.
 }
\label{Matsumoto1}
\end{figure}

The domain features a transition region, spanning between the upstream region ($x > 51$) and the downstream region ($x < 39$), characterized by tangled magnetic field lines (see Fig. \ref{Matsumoto1}). The magnetic field in this transition region is highly turbulent, making it challenging to identify the shock front visually. Additionally, filamentary structures develop in the density profile within this region. These filaments are associated with folded magnetic field lines, and the enhanced density regions contain current sheets (CSs). Multiple instances of magnetic reconnection are observed in these current sheets, forming magnetic islands along the CSs. Similar evidence of magnetic reconnection is also seen in other CSs within the transition and downstream regions, which form spontaneously.

The energy history of the particles (see Fig. \ref{Matsumoto1}b) indicates that the acceleration mechanism has two components: rapid systematic acceleration when a particle encounters a reconnecting CS, and slower acceleration from magnetic fluctuations.  Notably, the accelerated relativistic electrons constitute a moderate high-energy tail in the energy distribution, similar to what was observed in the simulations by \citet{Dahlin15}. Due to the short period of these simulations, there is insufficient time for a clear power-law tail to form in the energy distribution.

\citet{Caprioli14a} conducted three-dimensional hybrid simulations (kinetic ions - fluid electrons) to investigate particle acceleration and magnetic field amplification in non-relativistic astrophysical shocks. The particle acceleration scenario explored by \citet{Caprioli14a} contrasts with that of \citet{Matsumoto15}. Caprioli and Spitkovsky did not analyze the formation of CSs in a turbulent environment or their role in particle acceleration. Instead, they interpret the presence of high-amplitude magnetic fluctuations downstream as "magnetic field amplification," which facilitates elastic scattering of relativistic particles, forcing them to cross the shock surface multiple times and gain energy. Their study did not discuss the potential role of systematic acceleration by CSs.

\begin{figure}[!ht]
\includegraphics[width=0.90\columnwidth]{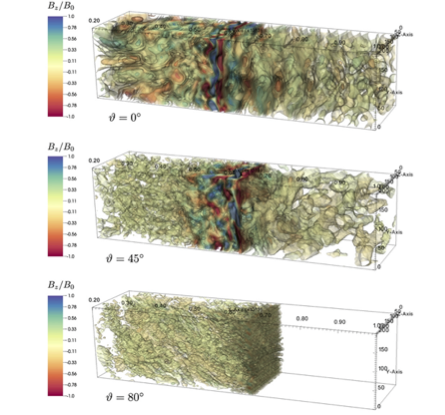}\
\includegraphics[width=0.90\columnwidth]{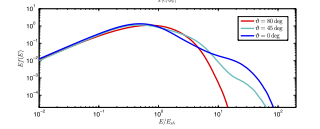}
\caption{(a) Self-generated component of the magnetic field, $B_z$, in units of the initial field $B_0$, which lies in the $xy$-plane. The panels correspond to $t = 200 \omega^{-1}_c$ for different 3D simulations with inclinations of $\theta = 0^\circ, 45^\circ, 80^\circ$ (top to bottom). The intensity of magnetic field fluctuations varies significantly between cases, with $\theta = 0^\circ$ showing several regions where $B_z \approx B_0$, while for $\theta = 80^\circ$, $B_z \approx 0.1B_0$. (b) The downstream spectrum integrated for the three cases above. The non-thermal power-law tail develops only at low-inclination shocks, while for quasi-perpendicular shocks, ions are only moderately heated. Reproduced with permission from Caprioli and Spitkovsky, Astroph. J. \textbf{783}, 91 (2014), Copyright 2014 AAS.
 }
\label{Capri1}
\end{figure}

As shown in Fig.\ \ref{Capri1}, there is a transition from stochastic and systematic acceleration at quasi-parallel shocks to predominantly stochastic acceleration and heating as the inclination becomes more perpendicular. The coupling between shocks and turbulence has also been highlighted in several other studies, e.g., \citet{Karimabadi2014, Trotta21, Bessho23}.

All of the above PIC simulations share similar limitations: the computational box is periodic, which means that particles do not escape, the escape times are not estimated and thus play no role for the final particle energization; the system's evolution is only modeled over a short time; and the analysis is carried out using either electron-positron pair plasmas or an ion-to-electron mass ratio that is relatively small compared to realistic conditions.

The central idea in the articles referred to above was that a shock acts as an isolated structure generating its turbulence, which then contains local scatterers. This scenario is very similar to the one described earlier in this review, where an isolated current sheet generates its turbulence as part of its evolution. In realistic astrophysical settings, however,  shocks propagate through pre-existing turbulence (see \citet{Trotta21,Guo21B}). Such interactions characterize spectacular, high-energy events, such as supernova explosions propagating through the turbulent interstellar medium, coronal mass ejections passing through the turbulent solar wind, and complex environments like Earth's bow shock. In many of these scenarios, oblique shocks generate coherent field-aligned beams (FABs), as observed at Earth's bow shock \cite{Burgess12}. These FABs are an essential source of free energy throughout the interplanetary medium. Turbulence-generated coherent structures and waves may interact with the shock discontinuity, likely playing a crucial role in particle acceleration and plasma heating.

%\heinz{
Recent initiatives have been launched to compare the energization mechanism between a compressible Hall magnetohydrodynamic model for test particle simulations and a corresponding hybrid particle-in-cell method for a  self-consistent approach, with simulations run in both 2D and 3D \cite{Pugliese25}. According to \citet{Pugliese25}, although test particles can replicate some qualitative features of self-consistent models, they often overlook more nuanced phenomena and tend to exaggerate energization. It remains uncertain whether these discrepancies originate from the inherent limitations of the test particle method or from scale-related approximations within the two different simulation approaches. To ensure an accurate comparison, the CoSs considered in different models must exhibit similar electromagnetic properties and scales. Thus, this question awaits further research. 

We just briefly mention that for test-particle simulations in the frame of turbulent MHD simulations, \citet{Pugliese22} studied the effect of different charge
to mass ratios in particle energization of heavy ions, and \citet{Pugliese23} analyzed the relation between wave modes and coherent structures in turbulence and their effect on the consequent energization of test particles.
%} 

\section{Summary}
This review begins with the established understanding that strong turbulence in three-dimensional (3D) magnetized plasmas spontaneously generates a variety of coherent structures (CoSs). These include current sheets, filaments, shocks, switchbacks, and large amplitude magnetic disturbances. The interactions of these CoSs with particles at the kinetic level give rise to two fundamental particle energization mechanisms: systematic and stochastic energization \cite{Fermi49, Fermi54}.

We explore these universal mechanisms from multiple perspectives, with a focus on highly turbulent, 3D magnetized systems. Our analysis demonstrates that particle energization emerges from the interplay between perpendicular electric fields, which drive turbulent heating through randomly moving scatterers, and parallel electric fields, which enable systematic acceleration. This coupling of perpendicular heating and parallel acceleration often produces Kappa distributions (heating at low energies and a power-law tail at high energies) --- energy distributions widely observed in laboratory, space, and astrophysical plasmas that experience strong turbulence.

The transport properties associated with these energization mechanisms differ fundamentally:
\begin{itemize} 
    \item Stochastic energization involves small, random energy jumps that follow Gaussian statistics, effectively modeled by the Fokker-Planck equation.
    \item Systematic acceleration, by contrast, causes consistent, positive energy increments that align with Lévy flight statistics, necessitating the modeling with a fractional transport equation.
\end{itemize}

Furthermore, the fractal spatial distribution of CoSs indicates that spatial transport --- controlling e.g.\  particle escape times from turbulent regions --- also deviates from a traditional Fokker-Planck descriptions.

Systematic acceleration proves particularly efficient during the initial and most energetic phase of strong turbulence, often called the ``injection phase" or ``first stage." In contrast, stochastic heating progresses gradually, characterizing a ``second phase" of particle energization.

An intriguing aspect of CoSs is their diversity and dynamics. Observations and simulations suggest that reconnecting current sheets (CSs) constitute only a tiny fraction of CoSs \cite{Hou21, Vinogradov24}. While reconnecting CSs play a dominant role during the explosive onset of turbulence, their prevalence diminishes rapidly. Unfortunately, the statistical properties and energy dissipation characteristics of other CoSs remain poorly understood. The development of AI tools may provide a breakthrough in identifying and characterizing these lesser-known CoSs, shedding light on their contributions to energy dissipation \cite{Sisti21, Trung24}.

Strong turbulence governs plasma heating and acceleration across a wide array of astrophysical environments, including magnetic field lines in the solar corona driven by turbulent convection, solar eruptions from unstable magnetic topologies, the turbulent solar wind, interactions between the solar wind and Earth's bow shock, Earth's magnetotail, astrophysical jets, supernova remnants, and others.
In these contexts, plasma heating and acceleration typically result in Kappa distributions. 

The specific energization mechanism depends on the dominant type of CoSs: (a) Systems with predominantly stochastic scatterers exhibit plasma heating, (b) systems with mainly systematic scatterers favor particle acceleration, (c) systems with a balanced mix of CoSs exhibit both, heating and acceleration. 

This review also identifies several pressing questions: (1) {\it Kinetic theory for diverse CoSs:} A detailed kinetic theory describing particle interactions with commonly occurring CoSs, beyond reconnecting current sheets or shocks, is lacking. These underexplored CoSs may be pivotal for mechanisms of heating and acceleration that generate  Kappa distributions in various plasma environments. (2) {\it Particle escape times:} Estimating particle escape times from turbulent volumes remains an open challenge, crucial for understanding energy transport and dissipation in turbulent systems.

By addressing these questions, we aim to advance the understanding of particle energization processes in laboratory, space, and astrophysical plasmas and to unify further theoretical and observational insights into 3D large-amplitude magnetized plasma turbulence.

\begin{acknowledgements}
	We want to express our gratitude to our colleagues, Profs. Dennis Papadopoulos, Peter Cargill, Vasilis Archontis, Fabio Lebreti, Cristos Tsironis, Eva Ntormousi, and Drs. Anastasios Anastasiadis, Manolis Georgoulis, Marco Onofri, Rene Kluven, Kaspar Arzner, Rim Turkmani, Theofilos Pisokas, Vasilis Tsiolis, Nikos Sioulas, and Argiris Koumtzis, who collaborated with us in developing some of the concepts discussed in this review.
\end{acknowledgements}

\section{References}

\bibliographystyle{plainnat}
%\nocite{*}
\bibliography{vlahosturb3}% Produces the bibliography via BibTeX.

\end{document}